\documentclass[a4paper,12pt]{elsarticle}

\usepackage[utf8]{inputenc}

\usepackage[margin=2.5cm]{geometry}

\usepackage{setspace}
\usepackage[shortlabels]{enumitem}

\usepackage{graphicx}
\graphicspath{{./figures/}}

\usepackage[font=footnotesize,labelfont=bf,subrefformat=parens]{subcaption}

\usepackage{comment}

\usepackage{xcolor}

\usepackage{hyperref}
\hypersetup{
    colorlinks,
    linkcolor={red!50!black},
    citecolor={blue!50!black},
    urlcolor={blue!80!black}
}

\usepackage{amsmath}

\usepackage{amssymb}

\usepackage[lighttt]{lmodern}
\ttfamily
\DeclareFontShape{OT1}{lmtt}{m}{it}
     {<->sub*lmtt/m/sl}{}

\usepackage{listings}
\definecolor{codegreen}{rgb}{0,0.6,0}
\definecolor{codegray}{rgb}{0.5,0.5,0.5}
\definecolor{codepurple}{rgb}{0.58,0,0.82}
\definecolor{backcolour}{rgb}{0.95,0.95,0.92}
\lstdefinestyle{mystyle}{
    keywordstyle=\color{black}\bfseries, % bold black keywords
    numberstyle=\tiny\color{codegray},
    stringstyle=\color{black}\slshape,  % \itshape also works due to the above workarround
    basicstyle=\ttfamily\footnotesize,
    numbers=left,
    showspaces=false,
    showstringspaces=false,
}
\usepackage[british]{babel}
\usepackage{booktabs}

\usepackage{lineno}

\usepackage{afterpage}
\usepackage{placeins}

\title{A revisit of the development of viscoplastic flow in pipes and channels}

\begin{document}

\newcommand{\vf}[1]{\underline{#1}}
\newcommand{\tf}[1]{\underline{\underline{#1}}}
\newcommand{\pd}[2]{\frac{\partial #1}{\partial #2}}
\newcommand{\ucd}[1]{\overset{\scriptscriptstyle \triangledown}{\tf{#1}}}
\newcommand{\gd}{\dot{\gamma}}
% Dimensionless numbers
\newcommand\Rey{\operatorname{\mathit{Re}}}
\newcommand\Str{\operatorname{\mathit{Sr}}}
\newcommand\Wei{\operatorname{\mathit{Wi}}}
\newcommand\Deb{\operatorname{\mathit{De}}}
\newcommand\Bin{\operatorname{\mathit{Bn}}}
\newcommand{\ry}{\ensuremath{\tilde{r}_0}}
\newcommand{\yy}{\ensuremath{\tilde{y}_0}}

\begin{frontmatter}

\author[MME]{Alexandros Syrakos\corref{cor1}}
\ead{syrakos.alexandros@ucy.ac.cy}

\author[MME,MAS]{Evgenios Gryparis}
\ead{gryparis.evgenios@ucy.ac.cy}

\author[MAS]{Georgios C. Georgiou}
\ead{georgiou-x.georgios@ucy.ac.cy}

\cortext[cor1]{Corresponding author}

\address[MME]{Department of Mechanical and Manufacturing Engineering, University of Cyprus, P.O.
Box 20537, 1678 Nicosia, Cyprus}

\address[MAS]{Department of Mathematics and Statistics, University of Cyprus, PO Box 20537, 1678,
Nicosia, Cyprus}

% \maketitle

\begin{abstract}
This study revisits the development of viscoplastic flow in pipes and channels, focusing on the
flow of a Bingham plastic. Using finite element simulations and the Papanastasiou regularisation,
results are obtained across a range of Reynolds and Bingham numbers. The novel contributions of this
work include: (a) investigating a definition of the development length based on wall shear stress, a
critical parameter in numerous applications; (b) considering alternative definitions of the Reynolds
number in an effort to collapse the development length curves into a single master curve,
independent of the Bingham number; (c) examining the patterns of yielded and unyielded regions
within the flow domain; and (d) assessing the impact of the regularisation parameter on the accuracy
of the results. The findings enhance the existing literature, providing a more comprehensive
understanding of this classic flow problem.
\end{abstract}


\end{frontmatter}

\journal{JNNFM}

% \begin{linenumbers}

\section{Introduction}
\label{sec: introduction}

The flow in pipes and channels is a simple yet very important case of flow. If the pipe or channel
has uniform cross-section and is straight and long enough then the flow can be approximated as fully
developed (Poiseuille) throughout, which greatly simplifies calculations. However, in reality, close
to the entrance of the conduit or wherever there is a change in cross-section, a bifurcation, a
bend, or any other cause of disturbance, the flow will deviate from its fully-developed state and
the fluid will have to travel some distance until it resettles. Knowledge of this development length
is very desirable in many applications, for example in microfluidics \cite{Haase_2017,
Pasias_2020}, haemodynamics \cite{Grotberg_2021, Clark_2015}, or the petroleum industry
\cite{Kfuri_2011}.

Due to the simplicity and importance of pipe flow development, it has been studied for over a
century (at least the Newtonian case); a comprehensive list of publications is provided in
\cite{Durst_2005}. Unless the momentum of the flow is very small, the development length is
determined by the relative strength of the viscous forces compared to inertia, with the former
tending to decrease this length and the latter tending to increase it; eventually, the viscous
forces harness the momentum and pin the flow to its fully developed state. Under these conditions,
the development length should be proportional to the ratio of inertia to viscous forces, i.e.\ to
the Reynolds number. Most past studies mainly focused on obtaining the constant of proportionality.
Nevertheless, it was relatively recently that Durst et al.\ \cite{Durst_2005} pointed out that when
inertia is very small the development length is actually determined by a competition between the
viscous forces themselves, and hence when the Reynolds number tends to zero the development length
tends to a finite, non-zero value.

Flow development is more complex when the fluid is non-Newtonian. There being different kinds of
non-Newtonian behaviour, the flow development of each class of non-Newtonian fluids exhibits its own
distinct characteristics. The cases of generalised Newtonian \cite{Ookawara_2000, Poole_2007,
Fernandes_2018, Lambride_2023} and viscoplastic flows \cite{Soto_1976, Ookawara_2000, Poole_2010,
Philippou_2016, Panaseti_2017} have been studied more extensively compared to the viscoelastic case
\cite{Yapici_2012, Bertoco_2021}. One complication compared to the Newtonian case is that the
standard definition of the development length as the location along the centreline (or midplane)
where the velocity reaches 99\% of its fully developed value may no longer be an accurate indicator
of the overall development of the flow. This is particularly evident in the case of viscoplastic
flows, as shown in \cite{Ookawara_2000, Philippou_2016}. The arbitrary 99\% mark is set because, in
the absence of a yield stress, the flow approaches its fully developed velocity asymptotically. But
in viscoplastic flow an unyielded plug forms in the central region of the pipe or channel, whose
velocity is everywhere fully developed and uniform. This, in combination with the fact that
viscoplasticity makes the fully developed velocity profile more blunt and hence reduces the
difference in velocity between the inlet and the plug, causes the standard centreline development
length to decrease as the Bingham number increases. Yet in the yielded region between the plug and
the walls the flow may continue to develop for a significant distance downstream of the point where
the plug starts to form. Using indicators other than the centreline development length reveals that
it is in fact not true that increased viscoplasticity results in earlier development. The
similarities between shear-thinning and viscoplasticity raises the suspicion that the standard
development length definition may be inadequate also for shear-thinning flows, which was recently
confirmed in \cite{Lambride_2023}.

What other indicators can be used, that give a more accurate picture? Ookawara et al.\
\cite{Ookawara_2000}, who were the first to note that in Bingham flow the velocity develops faster
at the centreline than elsewhere, proposed an alternative definition of the development length as
the axial distance required for the velocity to reach 99\% of its fully developed value at a radial
location of 95\% of the plug radius, a definition adopted also by Poole and Chhabra
\cite{Poole_2010}. Philippou et al.\ \cite{Philippou_2016} examined the development lengths at all
radial (or spanwise) locations, and defined the ``global development length'' as their maximum (this
length will be defined more precisely later, as it is among those that we shall use). They also
examined a version of this length restricted to radial or spanwise locations where the fully
developed velocity is larger than the inlet velocity. Recently, Lambride et al.\
\cite{Lambride_2023} proposed the wall shear stress development length, which is the distance from
the entrance beyond which the shear stress at the channel or pipe walls has stabilised within 1\% of
its fully developed value. This happens to be equivalent to the slip velocity development length in
cases with Navier slip \cite{Kountouriotis_2016}. The motivation for defining this development
length is twofold: on one hand, for certain flows (e.g.\ haemodynamics \cite{Zhou_2023}) the wall
shear stress is one of the most important quantities of interest; and on the other hand, shear
stress is one of the major factors that govern the flow development.

An additional complication is that whereas Newtonian flow is determined by a single dimensionless
number -- the Reynolds number -- the constitutive equations of non-Newtonian fluids give rise to
additional dimensionless numbers upon which the development length is also dependent. So, at first
glance, the Newtonian convenience of having a single indicator of flow development is lost. However,
some deeper consideration suggests that, if attention is restricted to generalised Newtonian and
viscoplastic fluids, it may be possible to correlate the flow development to a single dimensionless
parameter that has the significance of a Reynolds number; indeed, as mentioned above, provided that
inertia is not very small, the flow development of such fluids is determined by the relative
strength of viscous/viscoplastic forces compared to inertia, which is precisely what the Reynolds
number quantifies. The challenge is to find definitions of the Reynolds number that employ general
representative inertial fluxes and viscous/viscoplastic forces that govern the flow development
irrespective of the particularities of each non-Newtonian constitutive model -- for example, by
postulating that the development length is governed by the ratio of the average momentum flux to the
wall shear stress, irrespective of the kind of fluid. If such a definition is successful, it will
collapse the development length curves of the various non-Newtonian flows onto the Newtonian one.
The downside is that proceeding to actually calculate the values of such Reynolds numbers requires
knowledge of the flow field (e.g.\ in order to calculate the wall shear stress in the aforementioned
example). Therefore, the resulting mathematical expressions will be more complicated and will
require more effort to derive compared to the standard definitions of the Reynolds number.

Efforts to find such Reynolds numbers have been reported in the literature. For power-law fluids,
Poole and Ridley \cite{Poole_2007} tested three different Reynolds numbers, comparing the average
momentum flux to: (a) a stress calculated from a characteristic shear rate equal to the mean
velocity divided by the pipe diameter; (b) a stress calculated similarly, but using the wall
viscosity instead of the characteristic shear rate viscosity; and (c) the wall shear stress. They
referred to the latter as the ``Metzner-Reed'' Reynolds number because it was proposed by Metzner
and Reed \cite{Metzner_1955} in order to maintain for power-law flow the same relationship between
the Reynolds number and the friction factor as for Newtonian flow. The Metzner-Reed Reynolds number
proved to be the most successful in \cite{Poole_2007}, and has been used also by other researchers
in the past (see references in \cite{Poole_2007}). It is one of the numbers examined in the present
study, adapted for viscoplastic flow. Ookawara et al.\ \cite{Ookawara_2000} proposed Reynolds
numbers for power-law and Bingham fluids, respectively, which were relatively successful in
collapsing the respective development length curves onto the Newtonian one, thus providing a single
curve for all three classes of fluids. The derivation of these Reynolds numbers in
\cite{Ookawara_2000} is not entirely clear; Poole and Chhabra \cite{Poole_2010} assumed that the
viscoplastic variant is the Metzner-Reed equivalent for viscoplastic flow, but this does not appear
to be entirely accurate (see the present discussion in Sec.\ \ref{ssec: Re MRC}). Nevertheless,
Ookawara's viscoplastic Reynolds number was adopted in \cite{Poole_2010, Philippou_2016}, and both
studies verified that it does a good job collapsing the development length curves onto the Newtonian
one.

These approaches are based on the assumption that the development length is determined by the
competition between inertia and viscous (viscoplastic) resistance, which is invalid when the
Reynolds number is too small. In the low inertia regime, what governs the development length is
competition between the stresses themselves, which depends on the constitutive equation and its
parameters. Hence, in this regime the development length curves cannot be made to collapse no matter
what definition of the Reynolds number is chosen. This was noted in \cite{Poole_2007} for power-law
fluids and in \cite{Poole_2010, Philippou_2016} for viscoplastic fluids whereas it seems to have
been overlooked previously, when the low inertia regime was overlooked (just like for Newtonian
flows).

Although beyond the scope of the present paper, it is perhaps useful to mention that it seems
unlikely that such a single flow development indicator can be found for non-Newtonian flows that are
governed by constitutive equations with additional time derivatives, such as viscoelastic and
thixotropic flows. Such flows are also determined by other time scales (e.g.\ the relaxation time of
viscoelastic flows) in addition to the viscous time scale embodied nondimensionally in the Reynolds
number, and unless these additional time scales are sufficiently smaller than the viscous one, the
Reynolds number cannot inform us of the overall time scale of flow development. For viscoelastic
flows, it has been found that elasticity causes the flow to develop more slowly than a corresponding
Newtonian one (e.g.\ \cite{Poole_2007, Oliveira_2007, Li_2015, Bertoco_2021}) or a generalised
Newtonian one that exhibits the same degree of shear-thinning \cite{Syrakos_2018}. The
thixotropic case has been studied relatively little, with only a few experimental
\cite{Corvisier_2001} and computational \cite{Zhang_2023} investigations available.

In the present paper we revisit the viscoplastic entrance flow problem, comparing several
definitions of the Reynolds number in their ability to provide development length correlations that
are independent of the degree of viscoplasticity of the flow. Different definitions of the
development length are employed, including the standard (centreline), global, and wall shear stress
lengths. To keep things simple, we only consider a Bingham fluid (no shear thinning or thickening).
The flow is computed numerically, with a finite element method, using Papanastasiou regularisation
\cite{Papanastasiou_1987} as with most previous studies. Compared to previous studies, though, we
examine in detail also the pattern of yielded/unyielded zones, something that was previously missing
from the literature. The effect of the regularisation parameter on the solution accuracy is also
investigated. We begin with a description of the problem in Section \ref{sec: problem description},
including the definitions of the various development lengths in subsection \ref{ssec: development
lengths}. The alternative definitions of the Reynolds number are presented in Section \ref{sec:
Renolds numbers}. Our results are presented in Section \ref{sec: results}, and the paper concludes
with Section \ref{sec: conclusions}.

\section{Problem description and governing equations}
\label{sec: problem description}

\begin{figure}[tb]
  \centering
  \includegraphics[width=12cm]{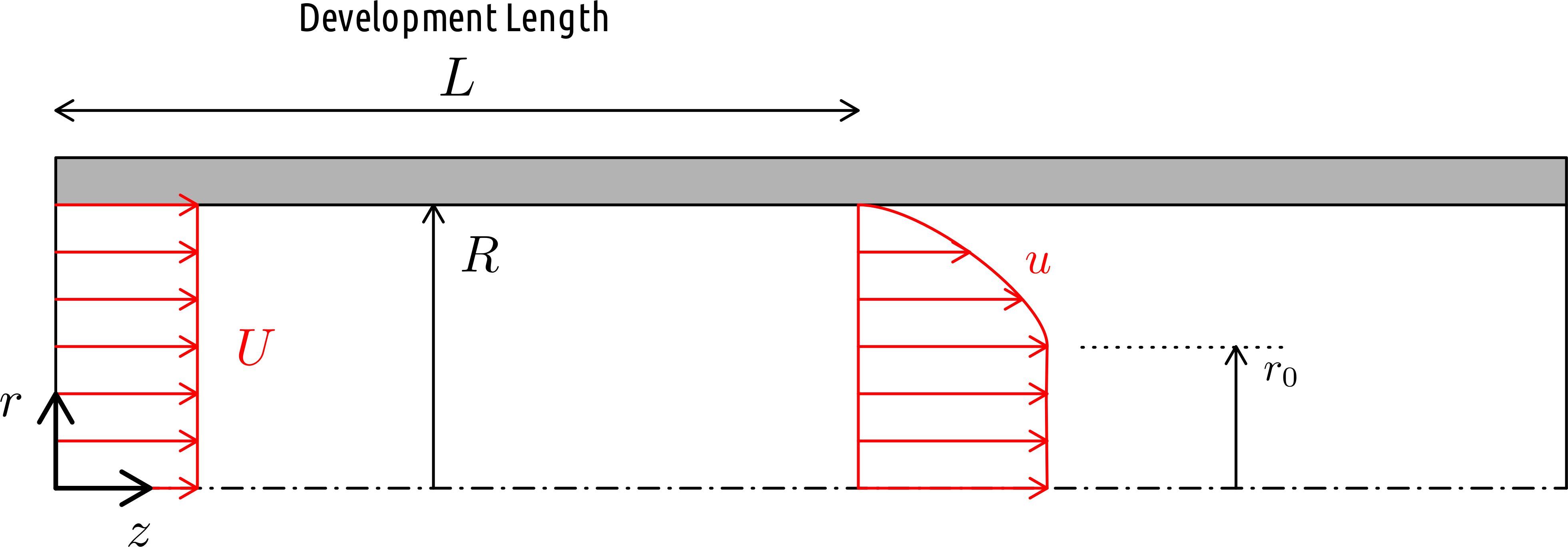}
  \caption{Development of axisymmetric viscoplastic flow in a pipe.}
  \label{fig: geometry}
\end{figure}

Consider the isothermal flow of a Bingham fluid entering a long horizontal cylindrical tube of
radius $R = D/2$, where $D$ is the diameter, or a long horizontal infinite-width planar channel of
semi-height $H$. The first geometry is axisymmetric, with $(r,z)$ denoting the radial and axial
coordinates, while the second is described by a Cartesian coordinate system with $(x,y)$ denoting
the directions parallel and perpendicular to the channel walls. Figure \ref{fig: geometry} shows a
sketch of the axisymmetric arrangement, but the channel case is similar. At the inlet, the axial
velocity is uniform, $u_z = U$, and the radial velocity is zero, $u_r = 0$. At the outlet, located
sufficiently far from the inlet, the flow is assumed to be fully developed, so that $u_r = 0$
and $\partial u_z / \partial z = 0$. The no-slip condition is imposed at the wall. Along the axis
(or plane) of symmetry, the standard symmetry conditions of zero radial velocity and shear stress
are imposed. Due to the similarity between the axisymmetric and planar cases, for the sake of
conciseness we will describe mainly the axisymmetric case, noting any important differences between
the two cases.

\subsection{Governing equations}
\label{ssec: governing equations}

The governing equations for steady-state flow are the continuity equation,
\begin{equation}
 \label{eq: continuity}
 \nabla \cdot \vf{u} \;=\; 0
\end{equation}
the momentum equation,
\begin{equation}
\label{eq: momentum}
\rho \vf{u} \cdot \nabla \vf{u} \;=\; - \nabla p \;+\; \nabla \cdot \tf{\tau}
\end{equation}
and the Bingham constitutive equation,
\begin{equation}
 \label{eq: constitutive}
 \left\{
 \begin{array}{ll}
   \tf{\dot{\gamma}} \;=\; 0  &  \tau \leq \tau_0
   \\[0.25cm]
   \tf{\tau} \;=\; \left( \dfrac{\tau_0}{\dot{\gamma}} + \mu \right) \, \tf{\dot{\gamma}} \quad
               &  \tau > \tau_0
 \end{array}
 \right.
\end{equation}
where $\rho$ is the (constant) fluid density, $\vf{u}$ is the velocity vector, $p$ is the pressure,
$\tf{\tau}$ is the extra stress tensor and $\tau = \sqrt{(\tf{\tau}:\tf{\tau})/2}$ is its
magnitude, $\tf{\dot{\gamma}} = \nabla \vf{u} + (\nabla \vf{u})^{\mathrm{T}}$ is the rate-of-strain
tensor and $\dot{\gamma}$ is its magnitude, $\tau_0$ is the yield stress, and the constant $\mu$ is
called the ``plastic viscosity''. When the yield stress is zero, the Bingham constitutive equation
reduces to the Newtonian one.

\subsection{Development lengths}
\label{ssec: development lengths}

As in \cite{Philippou_2016}, let us define the function $L(r)$ (or $L(y)$ in planar flow) as the
minimum axial distance from the entrance beyond which the velocity $u_z(r,z)$ differs from the fully
developed profile $\bar{u}_z(r)$ by no more than 1\%. The standard development length concerns the
velocity $u_z(0,z)$ at the centreline (midplane), and hence it is equal to $L_c = L(0)$. As
mentioned in Section \ref{sec: introduction}, this length may not be representative of the overall
flow development, hence we can also define the global development length $L_g$ \cite{Philippou_2016}
as the maximum value of $L(r)$ over all radii, $L_g = \max\limits_{0 \leq r \leq R} L(r)$. Finally,
as also mentioned in Section \ref{sec: introduction}, it is useful to define also the wall shear
stress development length $L_{\tau w}$, as the minimum length beyond which the shear stress at the
wall remains within a $\pm 1$\% margin of its fully-developed value.

\subsection{Dimensionless equations and numbers}
\label{ssec: dimensionless}

Scaling lengths by the pipe radius $R$ or the channel semi-height $H$, velocities by the inlet
velocity $U$, and the pressure and stress by $\mu U/R$ (or $\mu U/ H$), the continuity, momentum,
and constitutive equations are converted to the following non-dimensional forms, respectively:

\begin{equation}
 \label{eq: continuity ND}
 \tilde{\nabla} \cdot \tilde{\vf{u}} \;=\; 0
\end{equation}

\begin{equation}
 \label{eq: momentum ND}
 \frac{1}{2} \Rey \tilde{\vf{u}} \cdot \tilde{\nabla} \tilde{\vf{u}}
 \;=\;
 -\tilde{\nabla} \tilde{p}
 \;+\;
 \tilde{\nabla} \cdot \tilde{\tf{\tau}}
\end{equation}

\begin{equation}
 \label{eq: constitutive ND}
 \left\{
 \begin{array}{ll}
   \tilde{\tf{\dot{\gamma}}} \;=\; 0  &  \tilde{\tau} \leq \Bin/2
   \\[0.25cm]
   \tilde{\tf{\tau}} \;=\;
     \left( \dfrac{\Bin}{2\tilde{\dot{\gamma}}} + 1 \right) \, \tf{\dot{\gamma}} \quad
     &  \tilde{\tau} > \Bin / 2
 \end{array}
 \right.
\end{equation}
The dimensionless numbers appearing in the above equations are the standard Reynolds number,
\begin{equation}
 \label{eq: Re}
 \Rey \;\equiv\; \frac{2 \rho U R}{\mu}
\end{equation}
and the Bingham number,
\begin{equation}
 \label{eq: Bn}
 \Bin \;\equiv\; \frac{2 \tau_0 R}{\mu U}
\end{equation}
(with $R$ replaced by $H$ in the planar case). To be consistent with the rest of the literature,
these numbers are based on the pipe diameter $2R$ or channel height $2H$, hence the $1/2$ factors
appearing in the governing equations. The dimensionless yield stress is equal to $\tilde{\tau}_0 =
\Bin/2$.

\subsection{Fully developed flow}

A brief description of the fully developed (Poiseuille) flow will facilitate the discussion in the
subsequent sections. Let us denote the pressure gradient by $G = -dp/dz$. In fully developed flow,
the pressure gradient is balanced by the shear stress, whose magnitude is zero at the centre and
increases linearly towards the walls:
\begin{equation}
 \label{eq: FD stress}
 \tau_{rz} \;=\; \frac{1}{2} G r
\end{equation}
This means that for any finite value of yield stress, $\tau_0 > 0$, there will be an unyielded
cylindrical plug at the centre of the pipe with radius
\begin{equation}
 \label{eq: r0}
 r_0 \;=\; \frac{2 \tau_0}{G}
\end{equation}
Note that there will be no flow if $r_0 \geq R$. The fully developed velocity profile is:
\begin{equation}
 \label{eq: FD velocity}
 \bar{u}_z(r) \;=\; \frac{G}{4\mu}
   \left\{
   \begin{array}{ll}
     (R - r_0)^2, & 0 \leq r \leq r_0
     \\
     \left[ (R - r_0)^2 \;-\; (r - r_0)^2 \right], \quad & r_0 \leq r \leq R
   \end{array}
   \right.
\end{equation}
and the mean velocity is related to the pressure gradient by:
\begin{equation}
 \label{eq: FD mean velocity}
 U \;=\; G \frac{R^2}{24\mu} \left( 3 - 4 \ry + \ry^4 \right)
\end{equation}
where $\ry = r_0 / R$ is the non-dimensional plug radius, which is a convenient indicator of the
viscoplasticity of the flow, alternative to the Bingham number. Their relationship can be found by
substituting $\tau_0$ from \eqref{eq: r0} and $U$ from \eqref{eq: FD mean velocity} into the Bingham
number definition \eqref{eq: Bn}:
\begin{equation}
 \label{eq: Bn vs r0}
 \Bin \;=\; \frac{24 \ry}{3 - 4 \ry + \ry^4}
\end{equation}
This is a one-to-one relationship between them, so they convey the same information.

A useful expression for the wall shear stress $\tau_w$ is obtained by setting $r = R$ in Eq.\
\eqref{eq: FD stress} and using Eq.\ \eqref{eq: FD mean velocity} to substitute for the pressure
gradient:
\begin{equation}
 \label{eq: FD wall shear stress}
 \tau_w \;=\; \frac{12 \mu U}{R} \, \frac{1}{3 - 4 \ry + \ry^4}
\end{equation}

\subsubsection{Planar flow}
\label{sssec: fully developed planar}

In planar Poiseuille Bingham flow, Eqs.\ \eqref{eq: FD stress} -- \eqref{eq: FD wall shear stress}
are replaced by:

\begin{equation}
 \label{eq: FDP stress}
 \tau_{yx} \;=\; G y
\end{equation}

\begin{equation}
 \label{eq: y0}
 y_0 \;=\; \frac{\tau_0}{G}
\end{equation}

\begin{equation}
 \label{eq: FDP velocity}
 \bar{u}_x(y) \;=\; \frac{G}{2\mu}
   \left\{
   \begin{array}{ll}
     (H - y_0)^2, & 0 \leq y \leq y_0
     \\
     \left[ (H - y_0)^2 \;-\; (y - y_0)^2 \right], \quad & y_0 \leq y \leq H
   \end{array}
   \right.
\end{equation}

\begin{equation}
 \label{eq: FDP mean velocity}
 U \;=\; G \, \frac{H^2}{6\mu} \left( 2 - 3 \yy + \yy^3 \right)
\end{equation}

\begin{equation}
 \label{eq: Bn vs y0}
 \Bin \;=\; \frac{12 \yy}{2 - 3 \yy + \yy^3}
\end{equation}

\begin{equation}
 \label{eq: FDP wall shear stress}
 \tau_w \;=\; \frac{6 \mu U}{H} \, \frac{1}{2 - 3 \yy + \yy^3}
\end{equation}
where $\yy = y_0/H$ is the dimensionless half-thickness of the plug.

\subsection{Solution method}
\label{ssec: solution}

The equations were solved in dimensionless form. For the numerical solution, the constitutive
equation \eqref{eq: constitutive ND} was regularised according to the popular method of
Papanastasiou \cite{Papanastasiou_1987}:
\begin{equation}
 \label{eq: constitutive Papanastasiou}
 \tilde{\tf{\tau}} \;=\; \left[ \frac{\Bin \left( 1 - \mathrm{e}^{-M \tilde{\dot{\gamma}}}
   \right)}{2 \tilde{\dot{\gamma}}}
\;+\; 1 \right] \, \tilde{\tf{\dot{\gamma}}}
\end{equation}
where $M = mU/R$, $m$ being the dimensional regularisation parameter. This method has been used also
in previous studies \cite{Poole_2010, Philippou_2016} in order to avoid the difficulties associated
with the discontinuous nature of the original Bingham equation. The introduced parameter $M$ must
have a sufficiently large value in order for Eq.\ \eqref{eq: constitutive Papanastasiou} to mimick
the original equation \eqref{eq: constitutive ND} accurately. Values of $M \geq 500$ are generally
considered adequate \cite{Papanastasiou_1987, Mitsoulis_2007, Poole_2010, Syrakos_2013,
Syrakos_2014} for most purposes, unless one wants to capture the yield surfaces with high accuracy.
However, for the present flow we will see later that higher values of $M$ may be required at higher
Reynolds numbers.

The governing equations were solved with a standard Galerkin finite element method employing
biquadratic and bilinear basis functions for the velocity and pressure, respectively. The
simulations were performed using the open-source finite element package FEniCS \cite{FEniCS_2015}
(\href{https://fenicsproject.org}{fenicsproject.org}); in selected cases the results were validated
against predictions by an in-house code \cite{Philippou_2016}. There was no noticeable difference
between the predictions of the two codes.

\section{Alternative Reynolds numbers}
\label{sec: Renolds numbers}

\subsection{The effective Reynolds number}
\label{ssec: Re effective}

Recognising that in viscoplastic flows it is more illuminating to compare inertia to a
characteristic
\textit{viscoplastic} force rather than to its viscous component only, various authors have replaced
the plastic viscosity $\mu$ in Eq.\ \eqref{eq: Re} with an effective viscosity of
$\mu_{\mathrm{eff}} = \tau_0/\dot{\gamma}_c + \mu$, where $\dot{\gamma}_c$ is a characteristic
strain rate of the flow. The resulting Reynolds number is referred to as ``effective'' or
``modified'' Reynolds number,
\begin{equation}
 \label{eq: Re effective}
 \Rey^* \;=\; \frac{\rho U^2}{\tau_0 + \mu \dot{\gamma}_c}
\end{equation}
and has been shown to be much more useful as a flow regime indicator than the standard Reynolds
number \cite{Nirmalkar_2013, Syrakos_2016, Thompson_2016, Ferrari_2024}; it conveys information
about the flow even on its own, whereas the standard Reynolds number tells us nothing about the flow
unless it is accompanied by the Bingham number. Thompson and Soares \cite{Thompson_2016} strongly
advocate for the adoption of this Reynolds number, even proposing that the definition \eqref{eq: Re
effective} should be referred to simply as the ``Reynolds number'' without further qualification.
The reasoning is that the definition \eqref{eq: Re effective} conveys analogous information for
viscoplastic flows as the standard Reynolds number does for Newtonian flows, quantifying the ratio
between inertia and the internal forces that resist the development of velocity gradients.

The ``effectiveness'' of the effective Reynolds number depends on the selection of a truly
characteristic rate of strain $\dot{\gamma}_c$, which is art as much as it is science. The choice
$\dot{\gamma}_c = U/D$ affords the analogue of the ``Collins-Schowalter'' Reynolds number for
power-law flow \cite{Poole_2007}:
\begin{equation}
 \label{eq: Re effective D}
 \Rey^{*_D} \;=\; \frac{\rho U^2}{\tau_0 + \mu U/D}
 \;=\;
 \frac{\Rey}{\Bin + 1}
 \;=\;
 \Rey \, \frac{3 - 4 \ry + \ry^4}{3 + 20\ry + \ry^4}
\end{equation}
However, intuitively, it seems that $U/D$ rather under-represents the shear rates occurring in the
pipe, whereas $U/R$ seems more typical of a mean shear rate between the wall (where the velocity is
zero) and the centre of the pipe (where the velocity is actually greater than $U$). Therefore, in
order to investigate the effect of the choice of $\dot{\gamma}_c$, we define also the effective
Reynolds number based on the radius:
\begin{equation}
 \label{eq: Re effective R}
 \Rey^{*_R} \;=\; 2 \frac{\rho U^2}{\tau_0 + \mu U/R}
 \;=\;
 2 \frac{\Rey}{\Bin + 2}
\end{equation}
where the factor $2$ has been introduced so that for Newtonian flow ($\Bin = 0$) $\Rey^{*R}$ and
$\Rey$ become identical.

\subsubsection*{Planar case}
% \label{sssec: Re effective planar}

In the planar case, the respective Reynolds numbers are given by the same formulae as for the
axisymmetric one:
\begin{equation}
 \label{eq: Re effective planar}
 \Rey^{*2H} = \frac{\Rey}{\Bin + 1}
 \quad , \quad
 \Rey^{*H} = 2 \frac{\Rey}{\Bin + 2}
\end{equation}

\subsection{The Metzner-Reed Reynolds number}
\label{ssec: Re MR}

Arguably, the fully developed wall shear stress $\tau_w$ is much more crucial for driving the
dynamics of this flow than a shear stress based on $\dot{\gamma}_c = U/D$ or even $\dot{\gamma}_c =
U/R$. So, let us replace the latter with the former in \eqref{eq: Re effective D} to define the
following Reynolds number:
\begin{equation}
 \label{eq: Re MR 0}
 \Rey_{MR} \;=\; 8 \, \frac{\rho U^2}{\tau_w}
\end{equation}
The factor $8$ has been introduced so that $\Rey_{MR}$ reduces to $\Rey$ when the flow is Newtonian.
Using Eq.\ \eqref{eq: FD wall shear stress}, $\Rey_{MR}$ can be expressed as:
\begin{equation}
 \label{eq: Re MR}
 \Rey_{MR} \;=\; \Rey \, \frac{3 - 4\ry + \ry^4}{3}
\end{equation}

We will call $\Rey_{MR}$ the ``Metzner-Reed'' Reynolds number because the reasoning behind it is
essentially the same as for the analogous Reynolds number that Metzner and Reed \cite{Metzner_1955}
proposed for power-law fluids. Metzner and Reed designed their Reynolds number so that, for
power-law fluids, it would relate to the friction factor in the same way that the standard Reynolds
number relates to it in laminar Newtonian flows: $\Rey = 16/f$, where $f = \tau_w / (\frac{1}{2}
\rho U^2)$ is the Fanning friction factor \cite{Metzner_1955}.

Note that, just like the relationship $\Rey = 16/f$ does not hold for turbulent Newtonian flow, the
expressions \eqref{eq: Re MR 0} and \eqref{eq: Re MR} are no longer equivalent for turbulent Bingham
flow for which the wall shear stress is not given by Eq.\ \eqref{eq: FD wall shear stress} (the
present study is limited to laminar flows). Expression \eqref{eq: Re MR 0} is more general, applying
across different fluid types and flow regimes. In fact, it has been used for viscoplastic pipe flow
by Kfuri et al. \cite{Kfuri_2011}. $\Rey_{MR}$ can be straightforwardly calculated from
measurements, since the average velocity $U$ is readily obtained from the flow rate $Q$ as $U =
Q/A$, where $A$ is the pipe's cross-sectional area, and the wall shear stress can be obtained from
the pressure gradient $G$ from Eq.\ \eqref{eq: FD stress} at $r = R$. If the pipe is not circular,
then the corresponding expression is $\tau_w = G (A/S)$ where $S$ is the periphery of the pipe's
cross-section; for arbitrary cross-sectional shapes, this would return an average value of $\tau_w$
around the wall.

\subsubsection*{Planar case}
% \label{sssec: Re MR planar}

In the planar case, instead of \eqref{eq: Re MR} we obtain:
\begin{equation}
 \label{eq: Re MR planar}
 \Rey_{MR} \;=\; \Rey \, \frac{2 - 3\yy + \yy^3}{2}
\end{equation}

\subsection{Applying a momentum correction}
\label{ssec: Re MRC}

As mentioned in Section \ref{sec: introduction}, it was Ookawara et al.\ \cite{Ookawara_2000} who
first proposed a Reynolds number definition for the development of Bingham flow in a pipe that
results in a correlation that is close to that for Newtonian flow. Their definition was later used
by other authors as well \cite{Poole_2010, Philippou_2016}. Ookawara et al.\ did not provide a
detailed derivation of their Reynolds number, but it appears to be based on Eq.\ \eqref{eq: Re MR 0}
with a correction factor applied to the momentum flux in the numerator. Let us follow such an
approach, and define a ``momentum-corrected'' version of Eq.\ \eqref{eq: Re MR 0} as:
\begin{equation}
 \label{eq: Re MRC 0}
 \Rey_{MRC} \;=\; \alpha \, \frac{\frac{1}{A} \int_A \rho \bar{u}^2 \, \mathrm{d}\!A}{\tau_w}
\end{equation}
We have therefore replaced the numerator $\rho U^2$ of Eq.\ \eqref{eq: Re MR 0} by the more accurate
momentum flux $(1/A) \int_A \rho \bar{u}^2 \mathrm{d}\!A$ where $A$ denotes the pipe cross-section
or its area, and $\bar{u}(r)$ is the fully-developed velocity. The number $\alpha$ in the expression
is a scaling factor, playing the same role as ``$8$'' in Eq.\ \eqref{eq: Re MR 0} in ensuring that
the Reynolds number \eqref{eq: Re MRC 0} reduces to the standard one (Eq.\ \eqref{eq: Re}) in
Newtonian flow.

$\Rey_{MRC}$ requires more detailed knowledge about the flow compared to the previous Reynolds
numbers; in particular, it requires knowledge of the fully-developed velocity profile. In the
present case the velocity profile is given by Eq.\ \eqref{eq: FD velocity}. Integrating, and using
$\tau_w = GR/2$ and Eq.\ \eqref{eq: FD mean velocity} one arrives at:
\begin{equation}
 \label{eq: Re MRC}
 \Rey_{MRC} \;=\; \Rey \, \frac{3}{5} \, \frac{(1-\ry)^2 (5 + 6\ry + 4\ry^2)}{3 + 2\ry + \ry^2}
\end{equation}
where the expression has been scaled so that $\Rey_{MRC} = \Rey$ for $\ry = 0$.

The above expression can be manipulated into an alternative form by substituting for $(1-\ry)^2$
using the identity
\begin{equation}
 \label{eq: identity}
 \left( 3 + 2\ry + \ry^2 \right) \left( 1 - \ry \right)^2
 \;=\;
 3 - 4\ry + \ry^4
\end{equation}
to arrive at
\begin{equation}
 \label{eq: Re MRC Ookawara}
 \Rey_{MRC} \;=\; \Rey \, \frac{1}{3} \, \left( 3 - 4\ry + \ry^4 \right) \, \frac{9}{5} \,
 \frac{5 + 6\ry + 4\ry^2}{\left( 3 + 2\ry + \ry^2 \right)^2}
\end{equation}
Expression \eqref{eq: Re MRC Ookawara} is almost identical with the expression proposed by Ookawara
et al.\ \cite{Ookawara_2000}, but there is one difference: in \cite{Ookawara_2000}, the last term in
the numerator is $-11\ry^2$ instead of the $4\ry^2$ of Eq.\ \eqref{eq: Re MRC Ookawara}. Given that
the whole of the numerator of the fraction appearing at the end of expression \eqref{eq: Re MRC
Ookawara} comes from the integration of the velocity profile, it is difficult to see how any
modification to the assumptions or rationale behind $Re_{MRC}$ could bring about a change in a
single coefficient. We did not pursue further efforts to derive the formula of \cite{Ookawara_2000}.

It may be useful for purposes of generalisability to discuss another path to arrive at expression
\eqref{eq: Re MRC}. Consider a length $l$ of the pipe where the flow is fully developed; the kinetic
energy of the fluid contained in this length of pipe is:
\begin{equation}
 \label{eq: Re MRC KE}
 E \;=\; \int_A \frac{1}{2} \, \rho \, l \, \bar{u}^2 \, \mathrm{d} \! A
\end{equation}
This kinetic energy is constantly being dissipated by viscous stresses, but at the same time
replenished by the work of pressure so that it remains constant. The rate of work done by pressure
on the volume of fluid contained in the length $l$ of pipe is equal to the pressure times the
velocity at the inlet minus the pressure times the velocity at the outlet. The velocity is the same
at the inlet and outlet, while the pressure difference across the length $l$ is $G l$. Therefore,
the rate of work is:
\begin{equation}
 \label{eq: Re MRC P}
 P \;=\; \int_A G \, l \, \bar{u} \, \mathrm{d}\! A
 \;=\; G \, l \, \int_A \bar{u} \mathrm{d}\! A
 \;=\; G \, l \, U \, A
\end{equation}
As said, this is also the rate of work done by the viscous forces. The ratio $E/P$ is the amount of
time that the viscous forces would need in order to dissipate all of the fluid's kinetic energy;
during that time, the fluid travels a distance $(E/P) U$, which can be considered to be proportional
to the development length -- it is a length scale characteristic of flow changes governed by a
competition between inertia and viscosity. This would make the dimensionless development length,
$L/D$, proportional to $EU/PD$ which, from Eqs.\ \eqref{eq: Re MRC KE}, \eqref{eq: Re MRC P}, and $G
= 2 \tau_w / R$, is equal to expression \eqref{eq: Re MRC 0}.

\subsubsection*{Planar case}
% \label{sssec: Re MRC planar}

The corresponding expression in the planar case is:
\begin{equation}
 \label{eq: Re MRC planar}
 \Rey_{MRC} \;=\; \Rey \, \frac{1}{4} \, \frac{(1-\yy)^2 (8 + 7\yy)}{2 + \yy}
\end{equation}

\subsection{``Momentum gain'' Reynolds number}
\label{ssec: Re MG}

$\Rey_{MRC}$ can be viewed as a dimensionless length scale for changes to occur in the flow. An idea
for further enhancement may be to account also for the amount of change that occurs in each flow.
For example, a viscoplastic flow with its blunt velocity profile undergoes less overall development
than a Newtonian flow. We can account for this by replacing the fully developed momentum flux in the
numerator of the Reynolds number by the difference between that flux and the momentum flux at the
inlet:
\begin{equation}
 \label{eq: Re MG 0}
 \Rey_{MG} \;=\; \alpha \, \frac{\frac{1}{A} \int_A \rho \bar{u}^2 \mathrm{d}\! A \;-\;
%                                 \frac{1}{A} \int_A \rho U^2 \mathrm{d}\! A}{\tau_w}
                                 \rho U^2}{\tau_w}
\end{equation}
Comparing with Eqs.\ \eqref{eq: Re MR 0}  and \eqref{eq: Re MRC 0} we can see that $\Rey_{MG}$ is a
weighted combination of $\Rey_{MR}$ and $\Rey_{MRC}$. Because $U = \frac{1}{A} \int_A \bar{u}
\mathrm{d} \! A$, the Cauchy-Schwarz inequality guarantees that the numerator of \eqref{eq: Re MG 0}
is non-negative (it is zero only if $\bar{u} = U$ everywhere).

For our particular flow, integrating \eqref{eq: FD velocity}, using $\tau_w = GR/2$ and Eq.\
\eqref{eq: FD mean velocity}, and scaling, we arrive at the following expression:
\begin{equation}
 \label{eq: Re MG}
 \Rey_{MG} \;=\; \Rey \, \frac{1}{5} \, (1 - \ry)^3 \,
 \frac{15 + 27 \ry + 25\ry^2 + 5\ry^3}{3 + 2\ry + \ry^2}
\end{equation}

Again, it may be somewhat more illuminating and conducive to the generalisability of this Reynolds
number to consider an alternative path to its derivation. Let $L_a$ be the length travelled by the
fluid until the pressure gradient can accelerate it from the flat $U$ profile to the fully
developed $\bar{u}(r)$ profile, neglecting the viscous resistance. Omitting the viscous forces, the
momentum balance on the volume of fluid from the inlet to a distance $L_a$ is (using the Reynolds
transport theorem):
\begin{equation}
 \label{eq: Re MG deriv}
 \int_A \rho \bar{u}^2 \mathrm{d}\! A \;-\; \rho U^2 A \;=\; G \, L_a \, A
\end{equation}
where we have approximated the pressure difference between the inlet and $z=L_a$ by $G L_a$. We can
then rearrange Eq.\ \eqref{eq: Re MG deriv} into an expression for the dimensionless length
$L_a/D$, which, using $G = 2\tau_w/R$, is the same as \eqref{eq: Re MG 0}. So, the hypothesis for a
successful $\Rey_{MG}$ is that the development length is proportional to $L_a$.

\subsubsection*{Planar case}

In planar flow, instead of Eq.\ \eqref{eq: Re MG} we arrive at:
\begin{equation}
 \label{eq: Re MG planar}
 \Rey_{MG} \;=\; \Rey \, \frac{1}{2} \, (1 - \yy)^3 \,
 \frac{4 + 5\yy}{2 + \yy}
\end{equation}

\subsection{``Energy gain'' Reynolds number}
\label{ssec: Re EG}

Finally, we could not resist the temptation to, instead of considering the pressure force and the
gain in momentum, consider the pressure work and the gain in kinetic energy. This leads to the
following equation, instead of Eq.\ \eqref{eq: Re MG deriv}:
\begin{equation}
 \label{eq: Re EG deriv}
 \int_A \frac{1}{2} \rho \bar{u}^3 \mathrm{d}\! A \;-\; \frac{1}{2} \rho U^3 A
 \;=\;
 \int_A G \, L_a \, \bar{u} \, \mathrm{d}\! A
 \;=\;
 G \, L_a \, U \, A
\end{equation}
which, for our flow, gives rise to the following Reynolds number:
\begin{equation}
 \label{eq: Re EG}
 \Rey_{EG} \;=\; \Rey \, \frac{1}{105} \, (1 - \ry)^3 \,
 \frac{945 + 2187 \ry + 2520\ry^2 + 980\ry^3 + 245\ry^4 + 35\ry^5}
      {\left( 3 + 2\ry + \ry^2 \right)^2}
\end{equation}

\subsubsection*{Planar case}
% \label{sssec: Re EG planar}

The corresponding expression for planar flow is:
\begin{equation}
 \label{eq: Re EG planar}
 \Rey_{EG} \;=\; \Rey \, \frac{1}{38} \, (1 - \yy)^3 \,
 \frac{152 + 245\yy + 35\yy^2}
      {\left( 2 + \yy \right)^2}
\end{equation}

\section{Results}
\label{sec: results}

% \subsection{General description of the flow}

Most of the results to be presented were obtained on a single graded structured mesh with a length
of 300 diameters (600 radii), consisting of 206000 triangular elements. The mesh was more refined
near the entrance and close to the wall, with the smallest element being a right-angled triangle of
sides $\Delta r = 0.005R$ and $\Delta z = 0.01R$ at the entrance corner. The same mesh was used for
both the axisymmetric and planar calculations. To ensure that the mesh length suffices even for the
higher Reynolds number cases, we repeated the calculations for selected high-$\Rey$ cases on a
longer mesh of length $500D$ ($1000R$), but the discrepancy in the solutions was found to not be
significant. A shorter, very fine, graded mesh length $20D$ ($40R$), consisting of 160000 elements
was employed for a set of creeping flow ($\Rey = 0$) calculations where the Bingham number was
varied in the range $\Bin \in [0,50]$ (to be discussed towards the end of this section). The
smallest element (at the entrance corner) had sides $\Delta r = 0.001R$ and $\Delta z = 0.002R$.

\begin{figure}[tb]
    \centering

    \begin{subfigure}[b]{1.00\textwidth}
        \centering
        \includegraphics[width=1.00\linewidth]{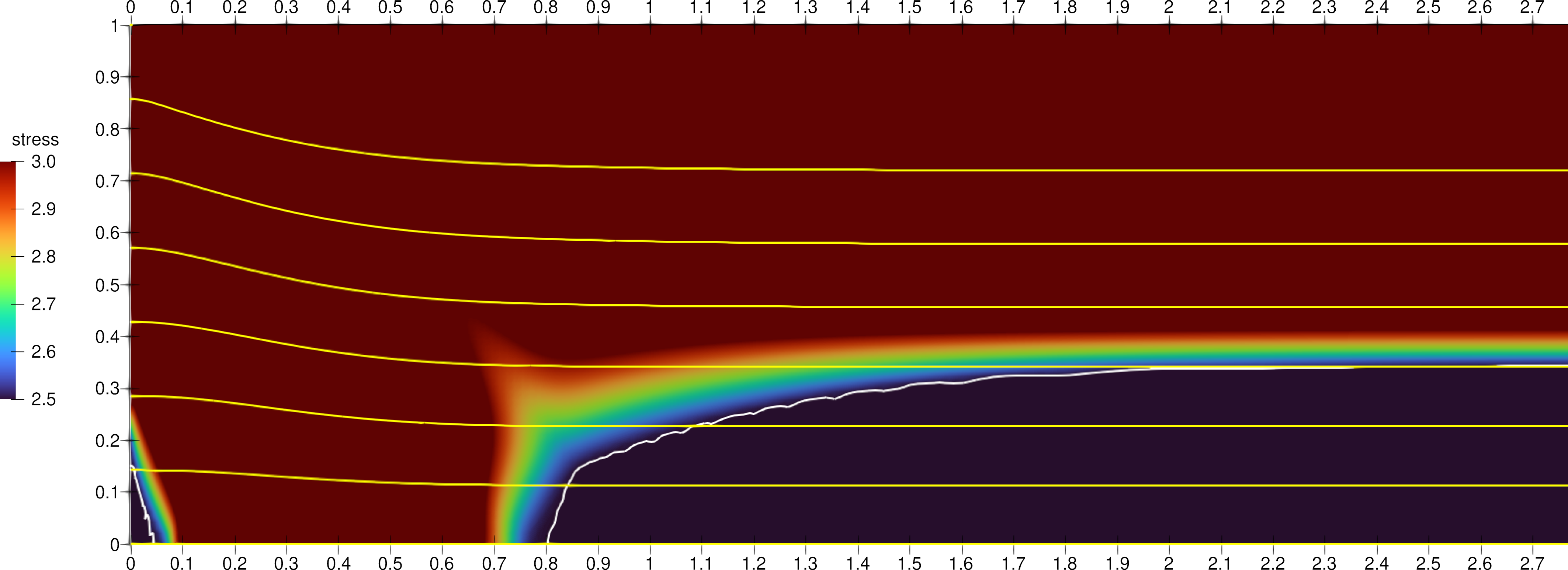}
        \caption{pipe flow}
        \label{sfig: streamlines axisymmetric}
    \end{subfigure}
    \\[0.5cm]

    \begin{subfigure}[b]{1.00\textwidth}
        \centering
        \includegraphics[width=1.00\linewidth]{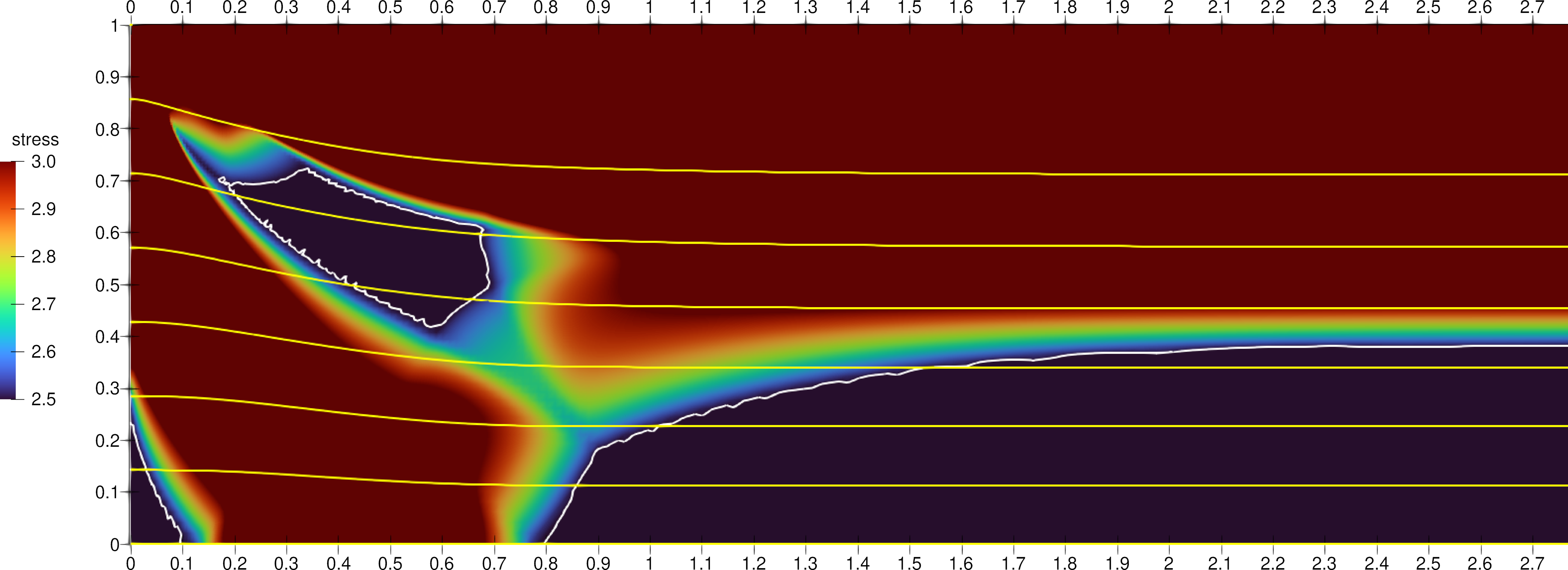}
        \caption{channel flow}
        \label{sfig: streamlines planar}
    \end{subfigure}

    \caption{Contours of dimensionless stress magnitude for \subref{sfig: streamlines axisymmetric}
pipe and \subref{sfig: streamlines planar} channel flow at $\Bin = 5$ and $\Rey = 0$, calculated
with $M = 10000$. The dimensionless yield stress for this case is $\tilde{\tau}_0 = \Bin/2 = 2.5$,
and the yield lines are drawn in white. Unyielded material ($\tilde{\tau} < 2.5$) is shown in
uniform dark blue, and regions of $\tilde{\tau} > 3$ are shown in uniform dark red. Superimposed are
streamlines, drawn in yellow.}
  \label{fig: streamlines}
\end{figure}

Before proceeding to the comparison of the correlations of the development lengths with the various
proposed Reynolds numbers, it is useful to start with a general description of the flow field, and
in particular of the pattern of yielded and unyielded regions. Although, as discussed in Section
\ref{sec: introduction}, there already exist quite a few studies on viscoplastic flow development, a
detailed description of the yield state seems to be lacking. Figure \ref{fig: streamlines} shows the
regions close to the entrance for pipe (Fig.\ \ref{sfig: streamlines axisymmetric}) and channel
(Fig.\ \ref{sfig: streamlines planar}) flow at $\Bin = 5$ and $\Rey = 0$, calculated with $M =
10000$. Unyielded zones are drawn in dark blue, and their boundaries (yield surfaces) are
delineated in white. The apparent roughness of the yield lines (the $\tilde{\tau} = \tilde{\tau}_0 =
\Bin/2$ contours) is an artefact of the high value of the regularisation parameter; a lower value
produces smoother contours, but at a cost of a loss of accuracy, as will be discussed later. Smooth
yield lines are difficult to capture also with methods other than Papanastasiou regularisation, and
a common workaround is to plot instead contours of $\tilde{\tau} = (1+\epsilon) \tilde{\tau}_0$,
where $\epsilon$ is a very small number (e.g.\ \cite{Dimakopoulos_2018, Ferrari_2024}), which are
usually notably smoother. However, in the present work no such margin is used ($\epsilon = 0$).

The plots of Fig.\ \ref{fig: streamlines} also include some streamlines. Close to the entrance, the
streamlines are sloped away from the walls, where the flow decelerates due to the no-slip condition,
and towards the centre of the pipe or channel, where the flow accelerates to compensate and satisfy
continuity. The yielded/unyielded fluid patterns look qualitatively quite similar for the
axisymmetric and planar cases, with one notable difference: while the former has two plug regions,
the latter exhibits three. Streamlines cross the boundaries of all of these unyielded zones. As
these are steady-state results, the location and size of the unyielded zones do not vary with time,
yet the material contained in them moves as a solid body. This may seem perplexing at first glance,
but what is actually taking place is that yielded material flows into these zones through their
boundaries (where streamlines cross into them) and upon entry it solidifies due to the lowering of
the stress levels. Thenceforth, all fluid particles in the zone move together as a rigid solid; but
upon reaching the boundaries where streamlines cross out of the zone, stress levels increase, and
the material liquefies again. Thus, there is a constant influx and outflux of material through the
boundaries of the unyielded plugs, with the state of the material switching between yielded and
unyielded upon traversing these boundaries, as discussed also, for example, in \cite{Syrakos_2013,
Varchanis_2020}.

\begin{table}
\begin{center}
\begin{tabular}[c]{r|cc}\toprule
 $\Bin$  &  $\ry$ & $\yy$ \\ \midrule
  0  &  0.00000 & 0.00000 \\
  1  &  0.10715 & 0.13349 \\
  2  &  0.18758 & 0.22346 \\
  5  &  0.34247 & 0.38058 \\
 10  &  0.47683 & 0.50727 \\
 20  &  0.60222 & 0.62259 \\\bottomrule
\end{tabular}
\caption{Dimensionless plug radii $\ry$ and thicknesses $\yy$ at various values of the Bingham
number, calculated from Eqs.\ \eqref{eq: Bn vs r0} and \eqref{eq: Bn vs y0}.}
\label{table: r0 values}
\end{center}
\end{table}

In pipe flow (Fig.\ \ref{sfig: streamlines axisymmetric}), the unyielded zones comprise of a a small
cone right at the central region of the inlet and a large cylindrical core at the centre of the pipe
that begins at approximately $z/R \approx 0.8$ and grows downstream, eventually reaching the
fully-developed radius $\tilde{r}_0 \approx 0.3425$ (Table \ref{table: r0 values}). In the channel
case (Fig.\ \ref{sfig: streamlines planar}) the picture looks similar, but one has to keep in mind
that the three-dimensional geometry of the plugs is entirely different. Instead of a conical zone
at the inlet there is a wedge, and instead of a cylindrical core there is a slab extending
infinitely in the direction perpendicular to the page; the latter also begins at $x/H \approx 0.8$
and its semi-height grows to the fully-developed value $\tilde{y}_0 \approx 0.38$ (Table \ref{table:
r0 values}).

In the channel flow there is an additional plug zone with no analogue in the axisymmetric case: it
is an oblong slanted zone near the top-left corner. This zone is also apparent in the results of
Dimakopoulos et al.\ \cite{Dimakopoulos_2018}. As seen from the curvature of the streamlines inside
this zone, its motion involves solid-body rotation. Its existence seems explicable by the fact that,
since the flow decelerates near the wall and accelerates near the centre, there must be an
intermediate region where the velocity does not change much, and where therefore stresses remain
low, giving rise to this zone. On the other hand, in the axisymmetric case the slanting of the
streamlines in that region makes it impossible for such a zone to exist. Had such a zone existed,
it would actually be an annulus of unyielded material. If we then consider a ring of particles
inside this zone all at the same initial radial coordinate $r_A$, then as they move along their
streamlines their radial distance from the centreline must decrease since the streamlines are
slanted inwards (towards the centreline). This means that at a later time, while still inside the
zone, the same ring of particles will have a smaller radius $r_B < r_A$. But if the state of the
material is that of a rigid solid, then it is impossible for a ring of such material to deform so
as to reduce its radius. Hence the inwards slanting of the streamlines in that region precludes the
possibility of existence of such an unyielded zone in the axisymmetric case.

\begin{figure}[!tb]
    \centering

    \begin{subfigure}[b]{0.99\textwidth}
        \centering
        \includegraphics[width=0.99\linewidth]{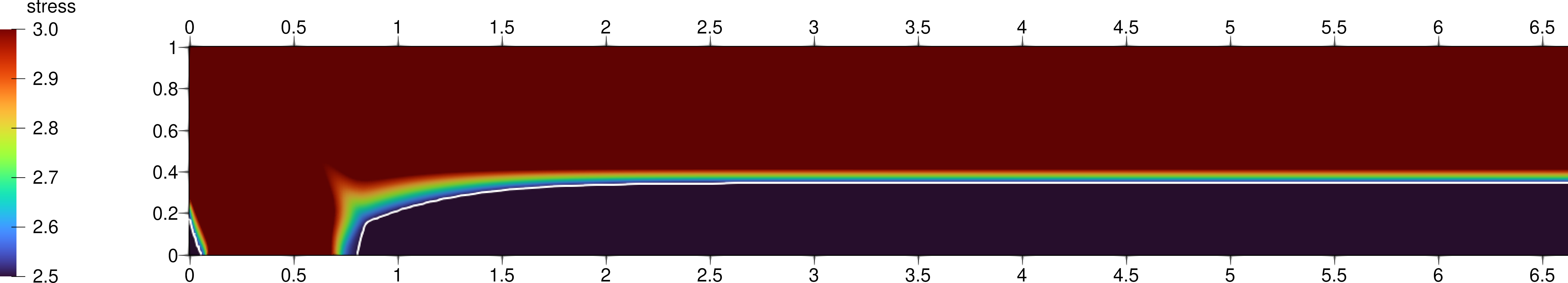}
        \caption{$M$ = 500}
        \label{sfig: stress contours Bn5 Re0 M500}
    \end{subfigure}
    \\[0.5cm]

    \begin{subfigure}[b]{0.99\textwidth}
        \centering
        \includegraphics[width=0.99\linewidth]{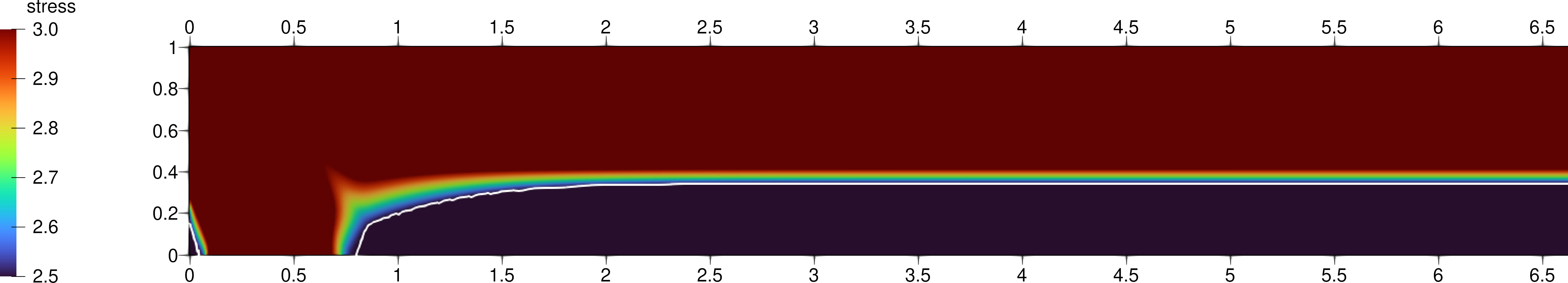}
        \caption{$M$ = 10000}
        \label{sfig: stress contours Bn5 Re0 M10000}
    \end{subfigure}

    \caption{Pipe flow, $\Bin = 5$, $\Rey = 0$: Contours of dimensionless stress magnitude,
calculated with \subref{sfig: stress contours Bn5 Re0 M500} $M = 500$ and \subref{sfig: stress
contours Bn5 Re0 M10000} $M = 10000$. Yield lines ($\tilde{\tau} = \tilde{\tau}_0 = \Bin/2 = 2.5$)
are drawn in white. Unyielded material ($\tilde{\tau} < 2.5$) is shown in uniform dark blue, and
regions of $\tilde{\tau} > 3$ are shown in uniform dark red.}
  \label{fig: stress contours Bn5 Re0}
% \end{figure}
%
%
%
% \begin{figure}[!h]
%     \centering

    \vspace{1cm}

    \begin{subfigure}[b]{0.99\textwidth}
        \centering
        \includegraphics[width=0.99\linewidth]{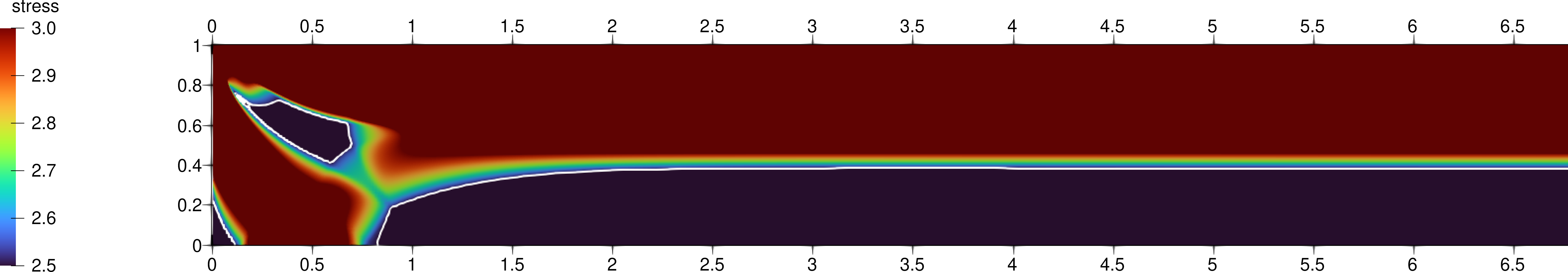}
        \caption{$M$ = 500}
        \label{sfig: stress contours Bn5 Re0 M500 planar}
    \end{subfigure}
    \\[0.5cm]

    \begin{subfigure}[b]{0.99\textwidth}
        \centering
        \includegraphics[width=0.99\linewidth]{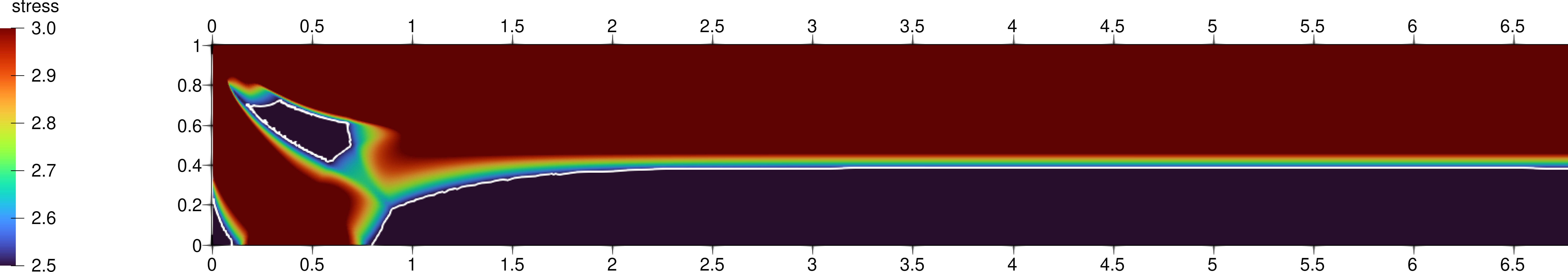}
        \caption{$M$ = 10000}
        \label{sfig: stress contours Bn5 Re0 M10000 planar}
    \end{subfigure}

    \caption{Channel flow, $\Bin = 5$, $\Rey = 0$: Contours of dimensionless stress magnitude,
calculated with \subref{sfig: stress contours Bn5 Re0 M500 planar} $M = 500$ and \subref{sfig:
stress contours Bn5 Re0 M10000 planar} $M = 10000$. Yield lines ($\tilde{\tau} = \tilde{\tau}_0 =
\Bin/2 = 2.5$) are drawn in white. Unyielded material ($\tilde{\tau} < 2.5$) is shown in uniform
dark blue, and regions of $\tilde{\tau} > 3$ are shown in uniform dark red.}
  \label{fig: stress contours Bn5 Re0 planar}
\end{figure}

\begin{figure}[!tb]
    \centering

    \begin{subfigure}[b]{0.99\textwidth}
        \centering
        \includegraphics[width=0.99\linewidth]{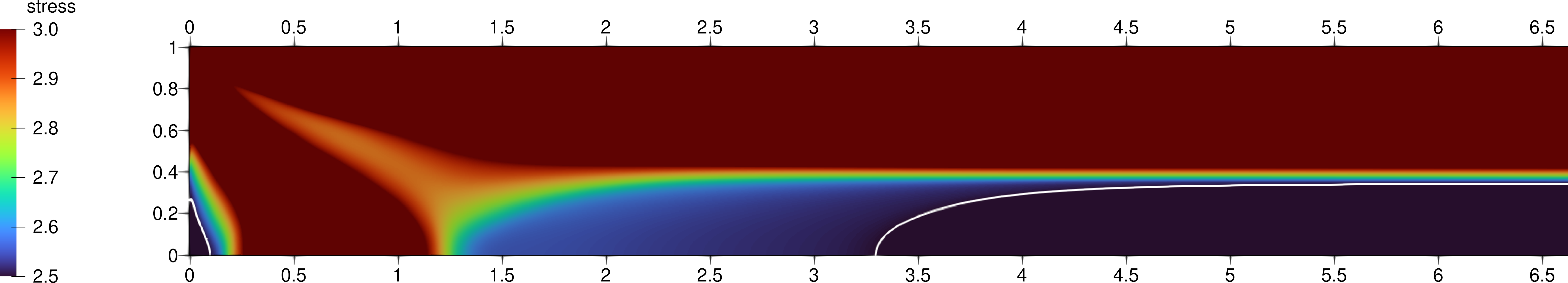}
        \caption{$M$ = 500}
        \label{sfig: stress contours Bn5 Re61p44 M500}
    \end{subfigure}
    \\[0.5cm]

    \begin{subfigure}[b]{0.99\textwidth}
        \centering
        \includegraphics[width=0.99\linewidth]{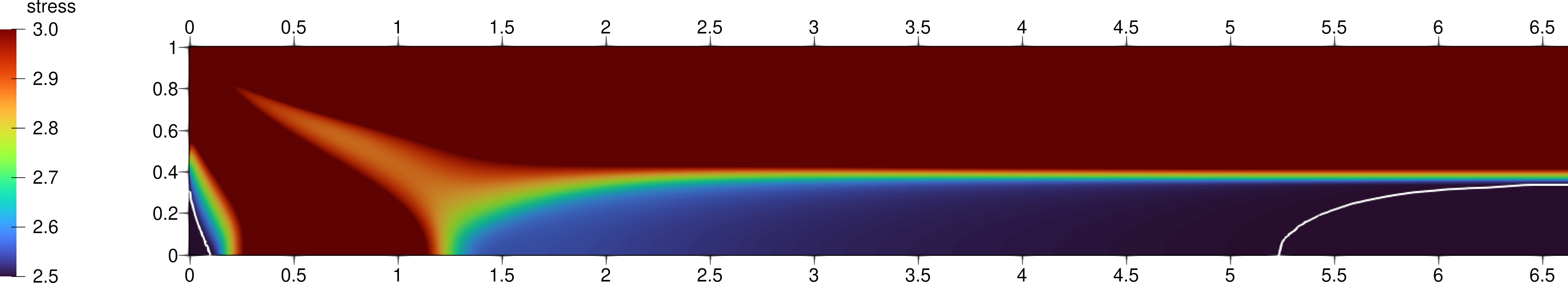}
        \caption{$M$ = 10000}
        \label{sfig: stress contours Bn5 Re61p44 M10000}
    \end{subfigure}

    \caption{Pipe flow, $\Bin = 5$, $\Rey = 61$: Contours of dimensionless stress magnitude,
calculated with \subref{sfig: stress contours Bn5 Re61p44 M500} $M = 500$ and \subref{sfig: stress
contours Bn5 Re61p44 M10000} $M = 10000$. Yield lines ($\tilde{\tau} = \tilde{\tau}_0 = \Bin/2 =
2.5$) are drawn in white. Unyielded material ($\tilde{\tau} < 2.5$) is shown in uniform dark blue,
and regions of $\tilde{\tau} > 3$ are shown in uniform dark red.}
  \label{fig: stress contours Bn5 Re61p44}
% \end{figure}
%
%
%
% \begin{figure}[!h]
%     \centering

    \vspace{1cm}

    \begin{subfigure}[b]{0.99\textwidth}
        \centering
        \includegraphics[width=0.99\linewidth]{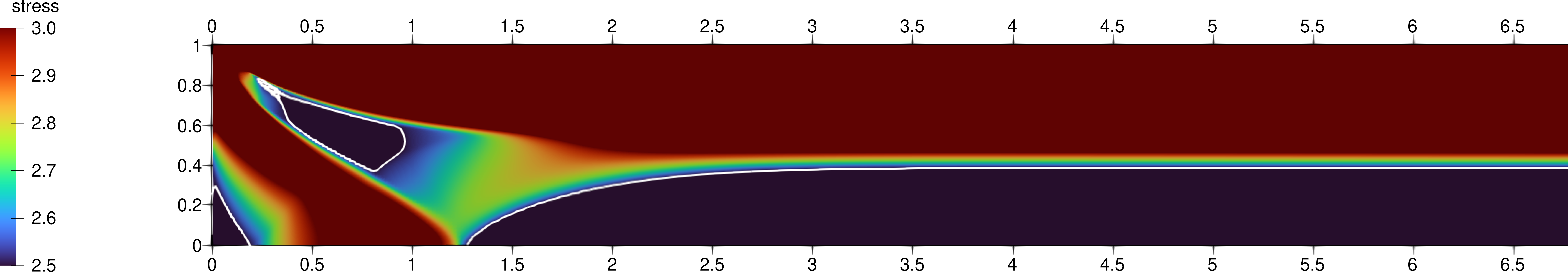}
        \caption{$M$ = 500}
        \label{sfig: stress contours Bn5 Re61p44 M500 planar}
    \end{subfigure}
    \\[0.5cm]

    \begin{subfigure}[b]{0.99\textwidth}
        \centering
        \includegraphics[width=0.99\linewidth]{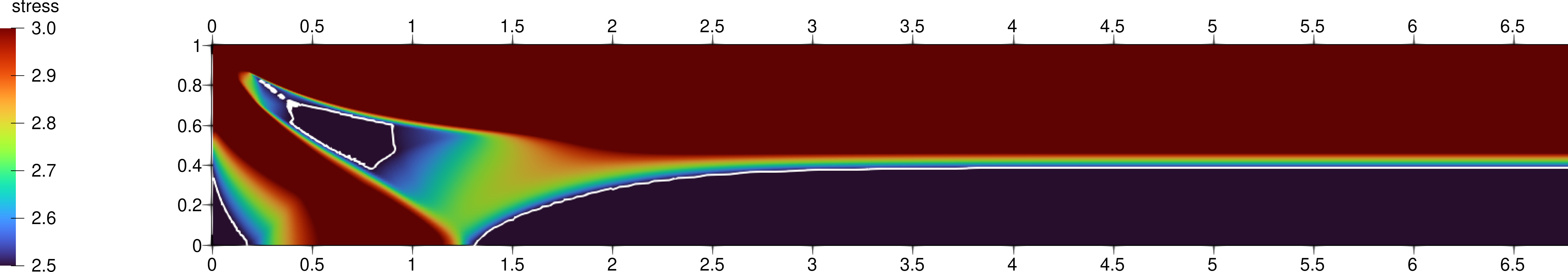}
        \caption{$M$ = 10000}
        \label{sfig: stress contours Bn5 Re61p44 M10000 planar}
    \end{subfigure}

    \caption{Channel flow, $\Bin = 5$, $\Rey = 61$: Contours of dimensionless stress magnitude,
calculated with \subref{sfig: stress contours Bn5 Re61p44 M500 planar} $M = 500$ and \subref{sfig:
stress contours Bn5 Re61p44 M10000 planar} $M = 10000$. Yield lines ($\tilde{\tau} = \tilde{\tau}_0
= \Bin/2 = 2.5$) are drawn in white. Unyielded material ($\tilde{\tau} < 2.5$) is shown in uniform
dark blue, and regions of $\tilde{\tau} > 3$ are shown in uniform dark red.}
  \label{fig: stress contours Bn5 Re61p44 planar}
\end{figure}

\begin{figure}[!tb]
    \centering

    \begin{subfigure}[b]{1.00\textwidth}
        \centering
        \includegraphics[width=1.00\linewidth]{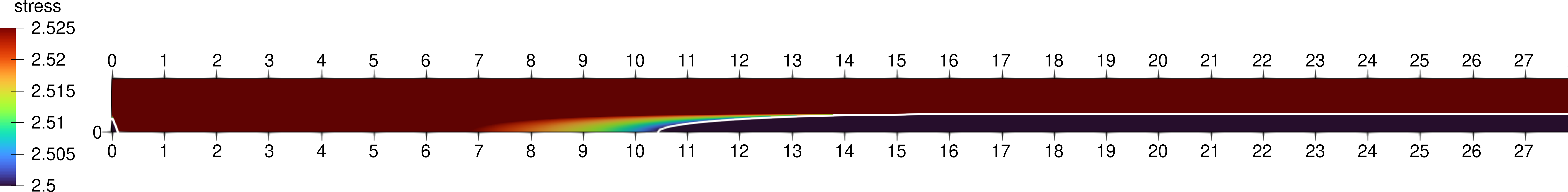}
        \caption{$M$ = 500}
        \label{sfig: stress contours Bn5 Re245p76 M500}
    \end{subfigure}
    \\[0.5cm]

    \begin{subfigure}[b]{1.00\textwidth}
        \centering
        \includegraphics[width=1.00\linewidth]{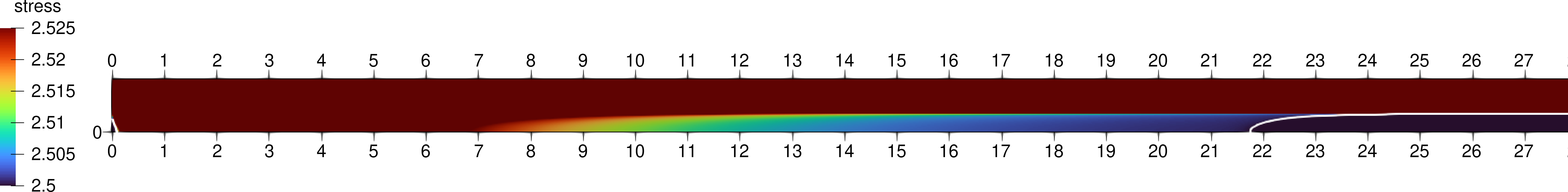}
        \caption{$M$ = 10000}
        \label{sfig: stress contours Bn5 Re245p76 M10000}
    \end{subfigure}

    \caption{Pipe flow, $\Bin = 5$, $\Rey = 246$: Contours of dimensionless stress magnitude,
calculated with \subref{sfig: stress contours Bn5 Re245p76 M500} $M = 500$ and \subref{sfig: stress
contours Bn5 Re245p76 M10000} $M = 10000$. Yield lines ($\tilde{\tau} = \tilde{\tau}_0 = \Bin/2 =
2.5$) are drawn in white. Unyielded material ($\tilde{\tau} < \tilde{\tau}_0 = 2.5$) is shown in
uniform dark blue, and regions of $\tilde{\tau} > 1.01\tilde{\tau}_0 = 2.525$ are shown in uniform
dark red.}
  \label{fig: stress contours Bn5 Re245p76}
% \end{figure}
%
%
%
% \begin{figure}[!tb]
%     \centering

    \vspace{1cm}

    \begin{subfigure}[b]{1.00\textwidth}
        \centering
        \includegraphics[width=1.00\linewidth]{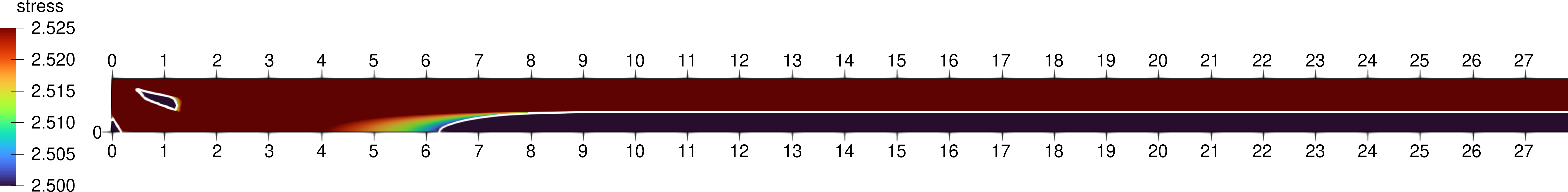}
        \caption{$M$ = 500}
        \label{sfig: stress contours Bn5 Re245p76 M500 planar}
    \end{subfigure}
    \\[0.5cm]

    \begin{subfigure}[b]{1.00\textwidth}
        \centering
        \includegraphics[width=1.00\linewidth]{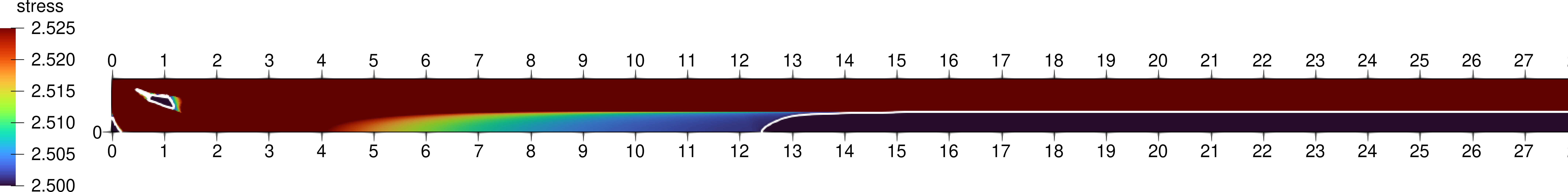}
        \caption{$M$ = 10000}
        \label{sfig: stress contours Bn5 Re245p76 M10000 planar}
    \end{subfigure}

    \caption{Channel flow, $\Bin = 5$, $\Rey = 246$: Contours of dimensionless stress magnitude,
calculated with \subref{sfig: stress contours Bn5 Re245p76 M500 planar} $M = 500$ and \subref{sfig:
stress contours Bn5 Re245p76 M10000 planar} $M = 10000$. Yield lines ($\tilde{\tau} = \tilde{\tau}_0
= \Bin/2 = 2.5$) are drawn in white. Unyielded material ($\tilde{\tau} < \tilde{\tau}_0 = 2.5$) is
shown in uniform dark blue, and regions of $\tilde{\tau} > 1.01\tilde{\tau}_0 = 2.525$ are shown in
uniform dark red.}
  \label{fig: stress contours Bn5 Re245p76 planar}
\end{figure}

The accuracy of the calculation of the yield lines can be quite sensitive to the regularisation
parameter $M$ \cite{Syrakos_2015, Dimakopoulos_2013}. Figures \ref{fig: stress contours Bn5 Re0} and
\ref{fig: stress contours Bn5 Re0 planar} compare the yield lines and stress contours computed with
$M = 500$ and $M = 10000$ for the creeping flow ($\Rey = 0$) axisymmetric and planar flows with
$\Bin = 5$. There are no noticeable differences, and adequate accuracy is already provided by $M =
500$. However, the situation changes if the Reynolds number is increased. Figures \ref{fig: stress
contours Bn5 Re61p44} and \ref{fig: stress contours Bn5 Re61p44 planar} make the same comparison,
for pipe and channel flow, respectively, at a higher Reynolds number of $61$ while $\Bin$ is kept at
$5$. While again the differences are marginal in the planar case (Fig.\ \ref{fig: stress contours
Bn5 Re61p44 planar}), in the axisymmetric case there is a very significant discrepancy between the
results obtained with $M = 500$ (Fig.\ \ref{sfig: stress contours Bn5 Re61p44 M500}) and those
obtained with $M = 10000$ (Fig.\ \ref{sfig: stress contours Bn5 Re61p44 M10000}). The $M = 500$
results predict a significantly faster evolution of stress along the axial direction in a region
upstream of the core plug, and a significantly earlier formation of the core plug by almost two pipe
radii. The discrepancy becomes more pronounced with further increase of $\Rey$. Figures \ref{fig:
stress contours Bn5 Re245p76} and \ref{fig: stress contours Bn5 Re245p76 planar} are for $\Rey =
246$. Now $M = 500$ is inadequate even for the planar case. For the axisymmetric case, $M = 500$
predicts the core plug formation nearly 11.5 pipe radii upstream from where $M = 10000$ predicts it.

These results are in line with past experience according to which small values of $M$ suffice when
the yield lines are located in regions of rapid variations of stress but high values of $M$ are
required when the stress varies gradually. For example, for the lid-driven cavity benchmark problem
where the flow occurs in a confined space, the unyielded regions can be calculated with satisfactory
accuracy using small values of $M$, e.g.\ $M=400$ \cite{Syrakos_2013, Syrakos_2014}. For unconfined
flow over a cylinder \cite{Syrakos_2015} low values of $M$ can predict satisfactorily the unyielded
regions that arise close to the cylinder, where the stresses exhibit large spatial gradients, but
can be completely off the mark in predicting the outer extent of the yield zone caused by the
presence of the cylinder, as the stresses die out very gradually in the far field. Similar
observations have been made also for the rising bubble problem \cite{Dimakopoulos_2013}. The present
results exhibit the same pattern. At low Reynolds numbers the flow development, determined by a
competition among viscoplastic stresses themselves, occurs within a short distance from the inlet;
the flow field varies rapidly, and $M = 500$ is entirely adequate to provide a good estimate of the
yield lines. However, at higher Reynolds numbers the stresses are not strong enough to rein in the
momentum of the incoming fluid quickly, and the flow development is gradual. These are not
favourable conditions for regularisation to capture the yield lines and thus we see that high values
of $M$ are required. In Fig.\ \ref{sfig: stress contours Bn5 Re245p76 M10000} the colour contours
indicate that the stress magnitude upstream of the unyielded core varies by only 1\% over a length
of 15 pipe radii, something that can be captured with $M = 10000$ but not with $M = 500$.

\begin{figure}[!tb]
    \centering

    \begin{subfigure}[b]{0.49\textwidth}
        \centering
        \includegraphics[width=0.99\linewidth]{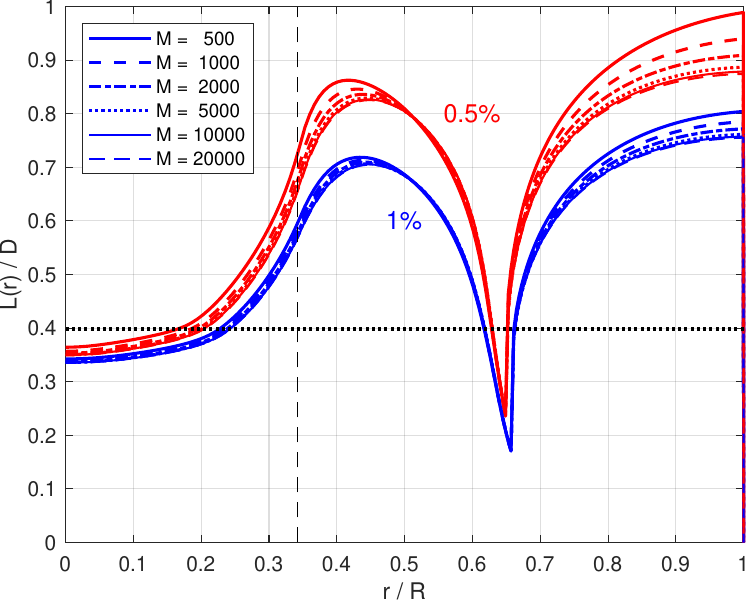}
        \caption{Pipe flow, $\Rey = 0$, $\Bin = 5$}
        \label{sfig: L vs r Bn5 Re0}
    \end{subfigure}
    \begin{subfigure}[b]{0.49\textwidth}
        \centering
        \includegraphics[width=0.99\linewidth]{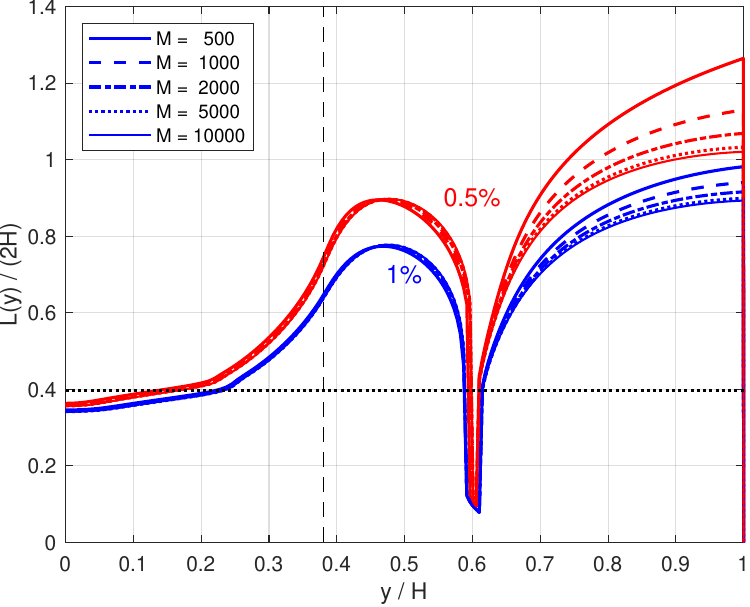}
        \caption{Channel flow, $\Rey = 0$, $\Bin = 5$}
        \label{sfig: L vs r Bn5 Re0 planar}
    \end{subfigure}
    \\[0.25cm]

    \begin{subfigure}[b]{0.49\textwidth}
        \centering
        \includegraphics[width=0.99\linewidth]{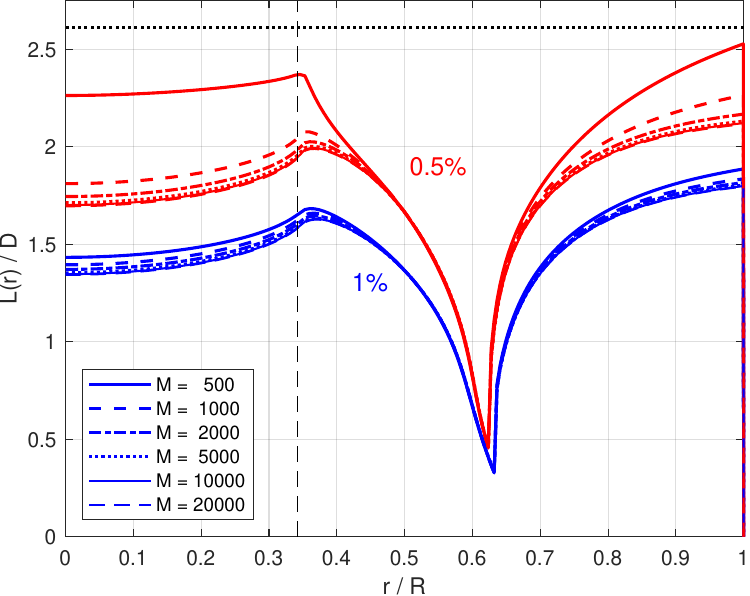}
        \caption{Pipe flow, $\Rey = 61$, $\Bin = 5$}
        \label{sfig: L vs r Bn5 Re61p44}
    \end{subfigure}
    \begin{subfigure}[b]{0.49\textwidth}
        \centering
        \includegraphics[width=0.99\linewidth]{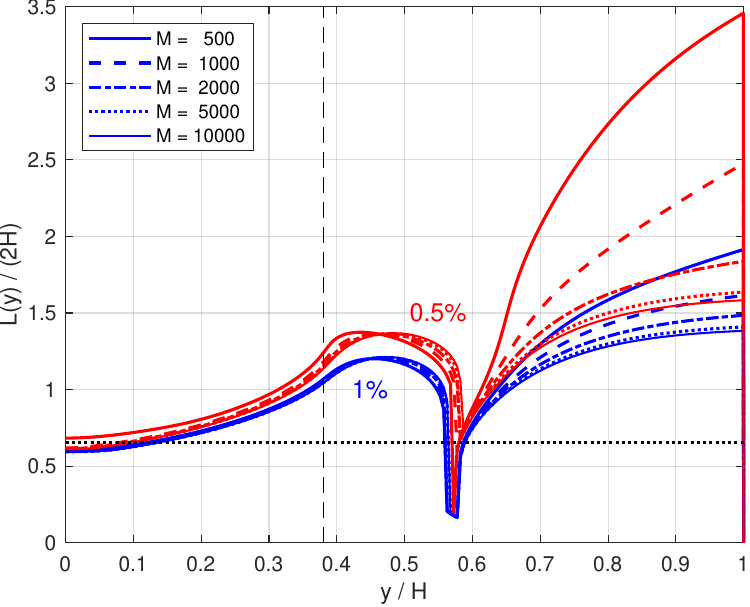}
        \caption{Channel flow, $\Rey = 61$, $\Bin = 5$}
        \label{sfig: L vs r Bn5 Re61p44 planar}
    \end{subfigure}
    \\[0.25cm]

    \begin{subfigure}[b]{0.49\textwidth}
        \centering
        \includegraphics[width=0.99\linewidth]{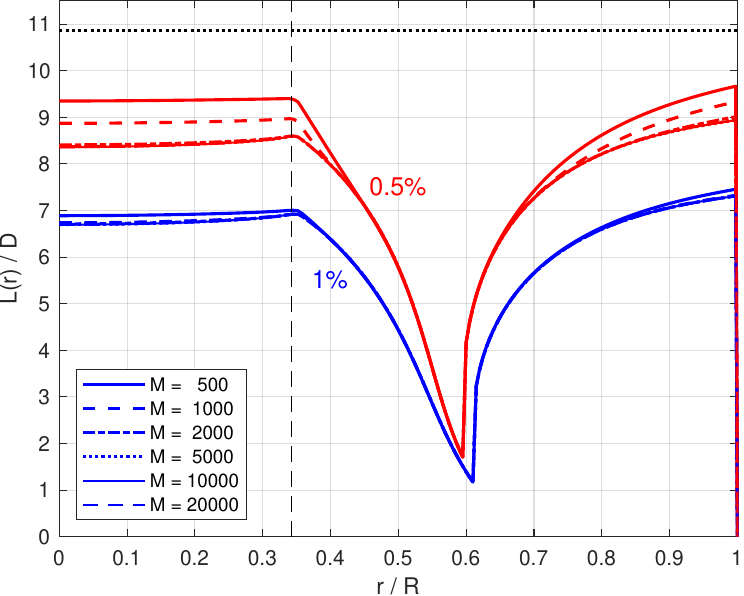}
        \caption{Pipe flow, $\Rey = 246$, $\Bin = 5$}
        \label{sfig: L vs r Bn5 Re245p76}
    \end{subfigure}
    \begin{subfigure}[b]{0.49\textwidth}
        \centering
        \includegraphics[width=0.99\linewidth]{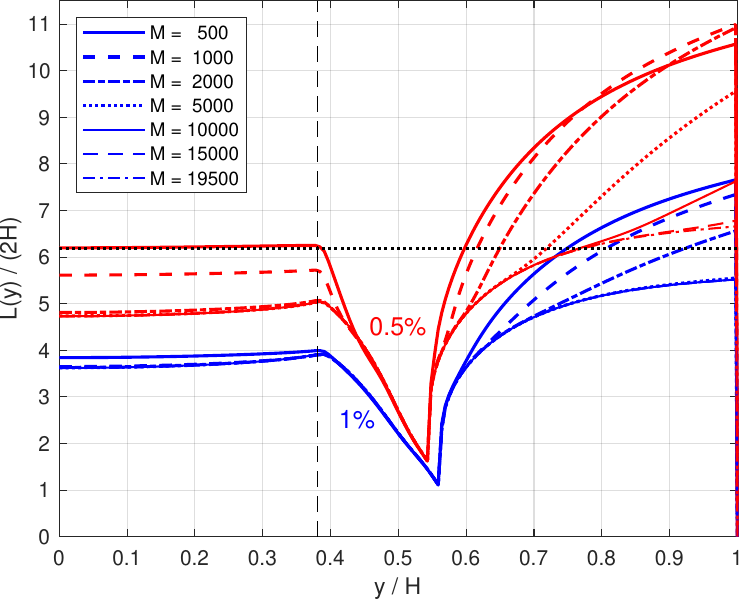}
        \caption{Channel flow, $\Rey = 246$, $\Bin = 5$}
        \label{sfig: L vs r Bn5 Re245p76 planar}
    \end{subfigure}

    \caption{Variation of the development length $L(r)$ or $L(y)$ across the pipe radius (left
column) or channel width (right column), for $\Bin = 5$ and $\Rey = 0$ (top row), $\Rey = 61$
(middle row), and $\Rey = 246$ (bottom row). Results obtained with different values of $M$ are
compared. Blue and red lines denote the development lengths based on 1\% and 0.5\% margins from the
fully developed velocity, respectively. The black vertical dashed line marks the fully developed
plug radius (calculated analytically). The black horizontal dotted line marks the tip of the plug
(obtained from the $M = 10000$ results).}
  \label{fig: L vs r-y Bn5}
\end{figure}

These results may raise concern about the ability of regularisation methods to accurately compute
the development length. However, typically other flow features (e.g.\ the velocity and pressure
fields) are much more accurately predicted compared to the yield lines. This turns out to be true in
the present case as well. Figure \ref{fig: L vs r-y Bn5} shows the variation of the local
development length $L(r)$ or $L(y)$ across the pipe radius or channel width, as computed for various
values of $M$, for $\Bin =5$ and $\Rey = 0$, $61$, and $246$. The blue lines depict the development
length beyond which the velocity remains within 1\% of its fully developed value, but also included
are the red lines which mark the development length for settling within 0.5\% of the fully developed
velocity. Let us denote these lengths as $L_{1\%}$ and $L_{0.5\%}$, respectively. In the
axisymmetric case (left column of diagrams) it may be seen that even $M = 500$ can be considered
adequate for accurate calculation of $L_{1\%}$ in all three cases shown, which may come as a
surprise given the discrepancy between figures \ref{sfig: stress contours Bn5 Re245p76 M500} and
\ref{sfig: stress contours Bn5 Re245p76 M10000}. The calculation of $L_{0.5\%}$ is more demanding,
with $M = 2000$ being a better choice that ensures accuracy. Interestingly, the planar case seems to
require higher values of $M$, of the order of $2000$, or even $5000$ for $\Rey = 246$, to accurately
capture $L_{1\%}$ close to the wall. The situation is worse for $L_{0.5\%}$, for which values of $M$
in excess of $10000$ are required in the $\Rey = 246$ case in order to achieve accuracy close to
the wall.

In general, Fig.\ \ref{fig: L vs r-y Bn5} shows that increasing $M$ makes the development length
shorter. On the contrary, Figs.\ \ref{fig: stress contours Bn5 Re61p44}, \ref{fig: stress contours
Bn5 Re245p76}, and \ref{fig: stress contours Bn5 Re245p76 planar} show that increasing $M$ moves
the predicted central plug farther into the pipe or channel, and makes the stress evolution just
upstream of it more gradual. Thus higher values of $M$ predict a faster initial flow development
near the inlet, possibly due to higher stresses, followed by a slower development afterwards,
compared to lower values of $M$.

% \FloatBarrier
\afterpage{\clearpage}

Let us now examine what Fig.\ \ref{fig: L vs r-y Bn5} tells us about the flow. A salient feature of
all diagrams is that they exhibit a sharp minimum, at about $\tilde{r} \approx 0.6 - 0.65$ or
$\tilde{y} \approx 0.55 - 0.6$, which can be explained thus: Since the flow, initially uniform at
the inlet, decelerates near the walls and accelerates near the centre, there must be some
intermediate point where the fully developed velocity is equal to that at the inlet. It is there
that the development length is minimal -- albeit not zero, because the velocity, even there,
undergoes some fluctuation before resettling to the same value that it had at the inlet. This point
separates the accelerating from the decelerating region, both of which have local $L$ maxima. The
maximum of the decelerating region occurs exactly at the wall, and is a bit larger than that of the
accelerating region, although not by a wide margin -- especially in the axisymmetric case they are
nearly equal.

In the creeping-flow axisymmetric case (Fig.\ \ref{sfig: L vs r Bn5 Re0}), the local maximum of $L$
in the accelerating region occurs at about $\tilde{r} \approx 0.43$, which is beyond the fully
developed plug radius $\tilde{r}_0 \approx 0.34$ (see Table \ref{table: r0 values} and the
vertical dashed line in Fig.\ \ref{sfig: L vs r Bn5 Re0}). At the centre of the pipe ($\tilde{r} =
0$) the development length is slightly shorter (by about $0.05D$) than the distance where the tip of
the core plug forms (marked with a horizontal dotted line in the figure). Interestingly, from
$\tilde{r} \approx 0.2$ up to the fully developed core plug radius $\tilde{r}_0 \approx 0.34$, the
development length (either $L_{1\%}$ or $L_{0.5\%}$) extends deeper inside the pipe than the tip of
the core plug. This, of course, does not mean that flow development is completed inside the plug,
something that is impossible since the velocity inside the core plug is everywhere fully-developed
and constant. Rather, because the upstream edge of the core plug is not flat but tapered, as seen
in Fig.\ \ref{fig: stress contours Bn5 Re0}, at these radii the flow becomes developed somewhere
past the tip of the core plug but inside yielded material nonetheless. Very similar observations
can be made about the planar case, Fig.\ \ref{sfig: L vs r Bn5 Re0 planar}.

As the Reynolds number is increased (Figs.\ \ref{sfig: L vs r Bn5 Re61p44}, \ref{sfig: L vs r Bn5
Re245p76}) the local maximum of the accelerating region moves towards $r_0$, while the $L(r)$
distribution for $0 \leq r \leq r_0$ becomes flat; thus, the cylindrical volume of yielded fluid of
radius $r_0$ that extends some distance upstream from the central plug also exhibits some plug-like
characteristics. Both the 1\% and the 0.5\% developments occur prior to the formation of the central
plug. Similar observations hold also for the planar case (Figs.\ \ref{sfig: L vs r Bn5 Re61p44
planar}, \ref{sfig: L vs r Bn5 Re245p76 planar}).

\begin{figure}[!tb]
    \centering

    \begin{subfigure}[b]{0.328\textwidth}
        \centering
        \includegraphics[width=0.99\linewidth]{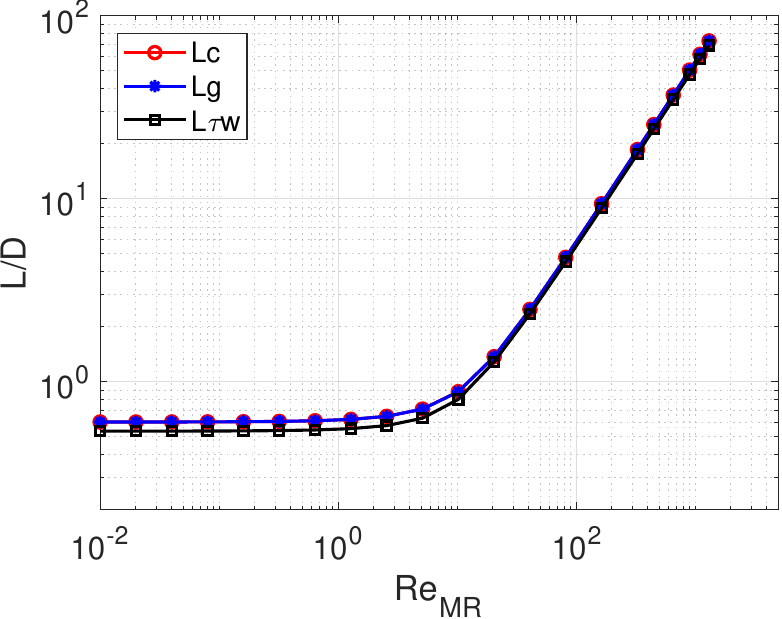}
        \caption{$\Bin = 0$}
        \label{sfig: Ls Bn0 ax}
    \end{subfigure}
    \begin{subfigure}[b]{0.328\textwidth}
        \centering
        \includegraphics[width=0.99\linewidth]{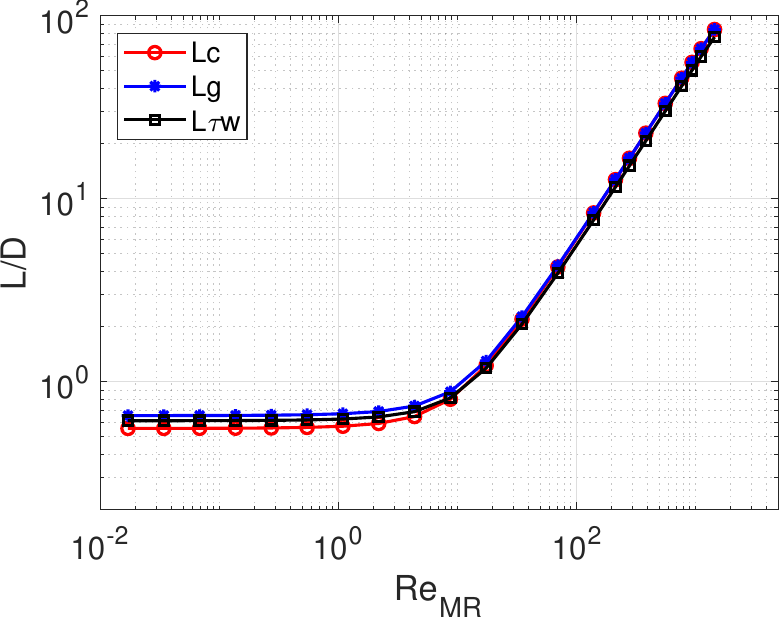}
        \caption{$\Bin = 1$}
        \label{sfig: Ls Bn1 ax}
    \end{subfigure}
    \begin{subfigure}[b]{0.328\textwidth}
        \centering
        \includegraphics[width=0.99\linewidth]{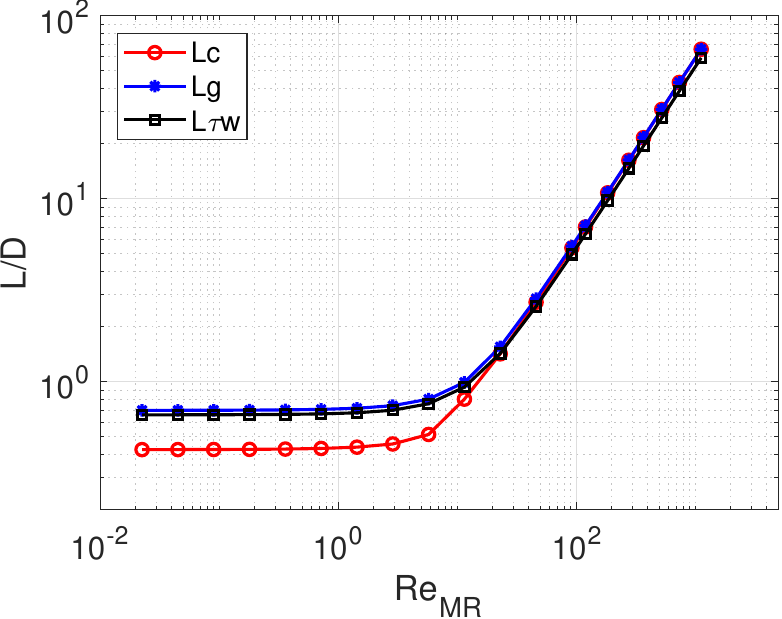}
        \caption{$\Bin = 2$}
        \label{sfig: Ls Bn2 ax}
    \end{subfigure}
    \\[0.5cm]

    \begin{subfigure}[b]{0.328\textwidth}
        \centering
        \includegraphics[width=0.99\linewidth]{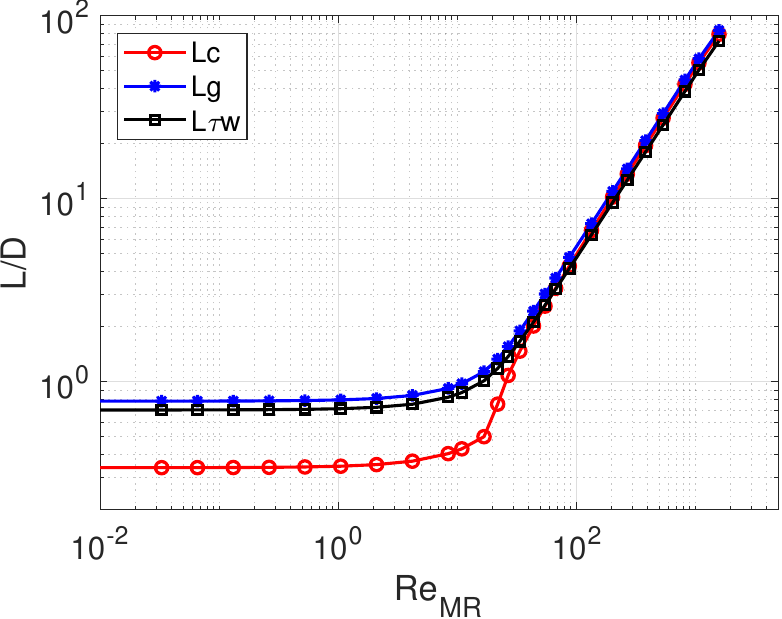}
        \caption{$\Bin = 5$}
        \label{sfig: Ls Bn5 ax}
    \end{subfigure}
    \begin{subfigure}[b]{0.328\textwidth}
        \centering
        \includegraphics[width=0.99\linewidth]{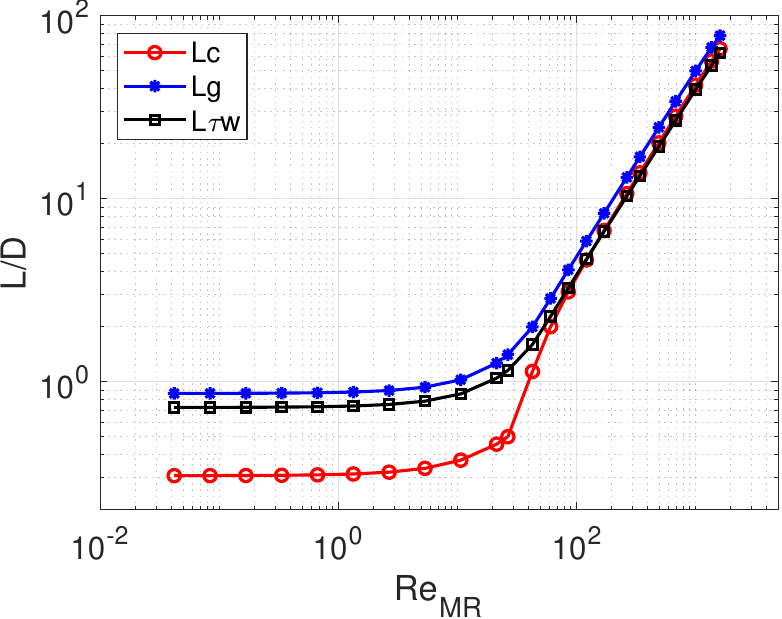}
        \caption{$\Bin = 10$}
        \label{sfig: Ls Bn10 ax}
    \end{subfigure}
    \begin{subfigure}[b]{0.328\textwidth}
        \centering
        \includegraphics[width=0.99\linewidth]{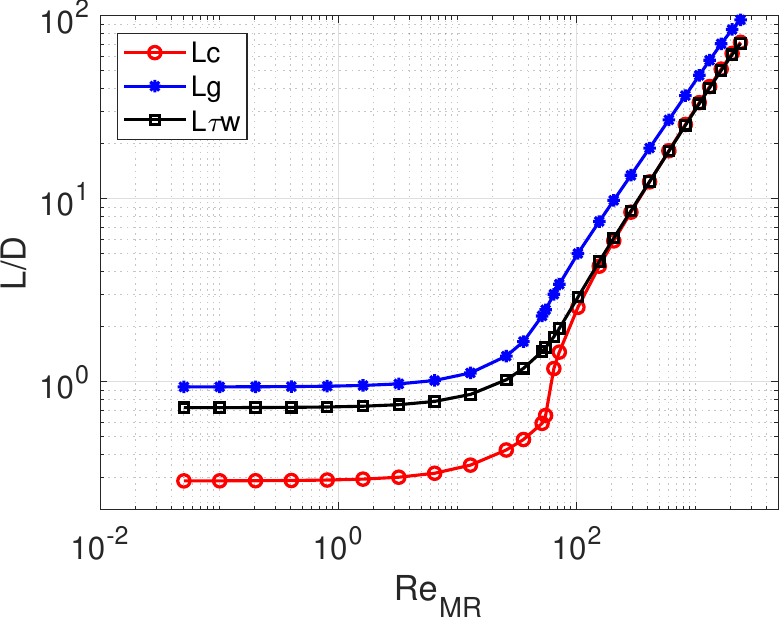}
        \caption{$\Bin = 20$}
        \label{sfig: Ls Bn20 ax}
    \end{subfigure}

    \caption{Pipe flow: Comparison of the variations of the three development lengths with the
Reynolds number for different Bingham numbers. For convenience, the Metzner-Reed Reynolds number
(Eq.\ \eqref{eq: Re MR}) is used instead of $\Rey$ so as to use the same $x$-axis range for all
plots. Computed with $M = 2000$.}
  \label{fig: Ls ax}
\end{figure}

\begin{figure}[!tb]
    \centering

    \begin{subfigure}[b]{0.328\textwidth}
        \centering
        \includegraphics[width=0.99\linewidth]{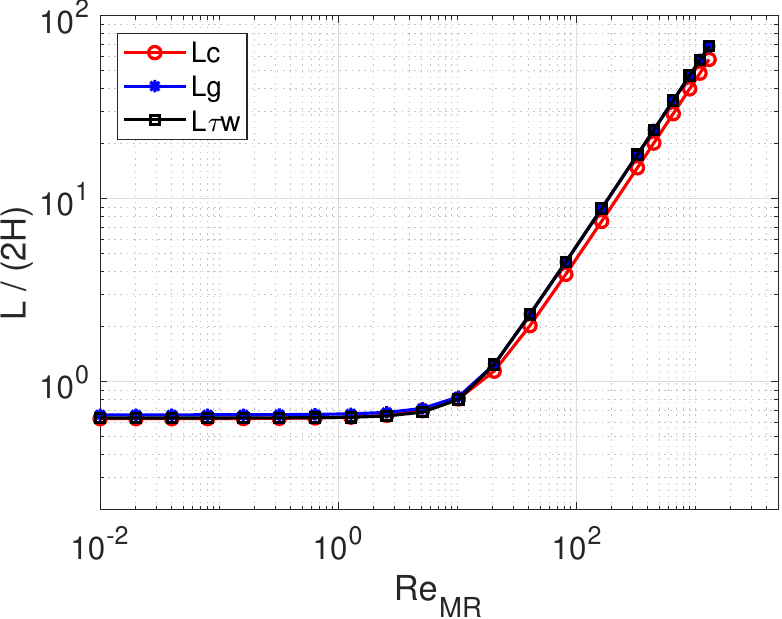}
        \caption{$\Bin = 0$}
        \label{sfig: Ls Bn0 pl}
    \end{subfigure}
    \begin{subfigure}[b]{0.328\textwidth}
        \centering
        \includegraphics[width=0.99\linewidth]{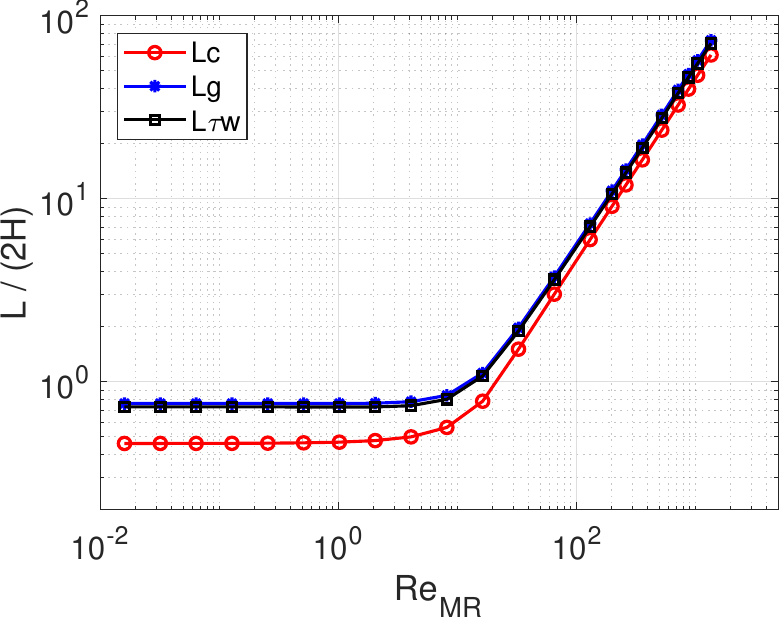}
        \caption{$\Bin = 1$}
        \label{sfig: Ls Bn1 pl}
    \end{subfigure}
    \begin{subfigure}[b]{0.328\textwidth}
        \centering
        \includegraphics[width=0.99\linewidth]{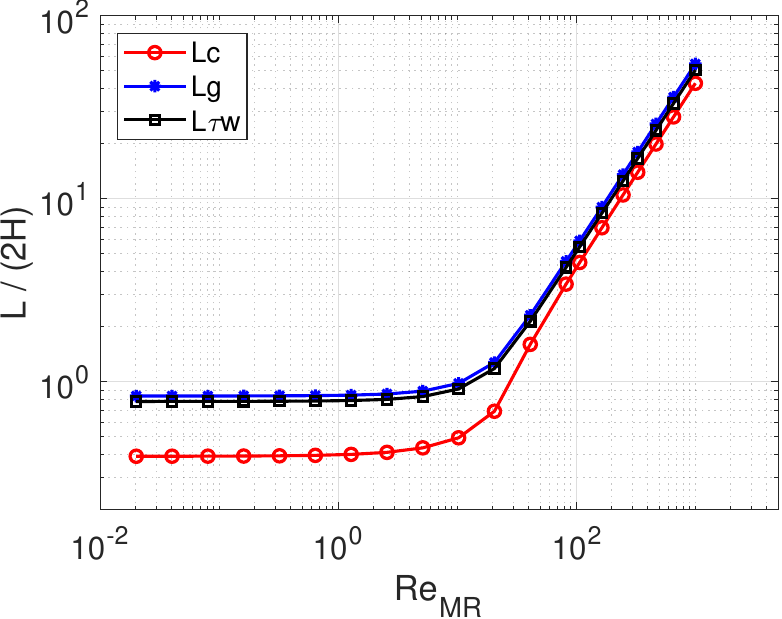}
        \caption{$\Bin = 2$}
        \label{sfig: Ls Bn2 pl}
    \end{subfigure}
    \\[0.5cm]

    \begin{subfigure}[b]{0.328\textwidth}
        \centering
        \includegraphics[width=0.99\linewidth]{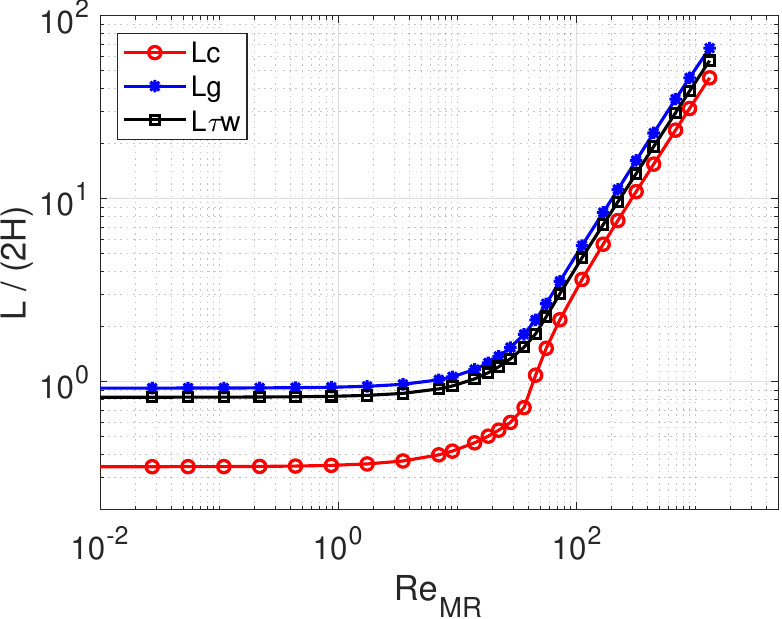}
        \caption{$\Bin = 5$}
        \label{sfig: Ls Bn5 pl}
    \end{subfigure}
    \begin{subfigure}[b]{0.328\textwidth}
        \centering
        \includegraphics[width=0.99\linewidth]{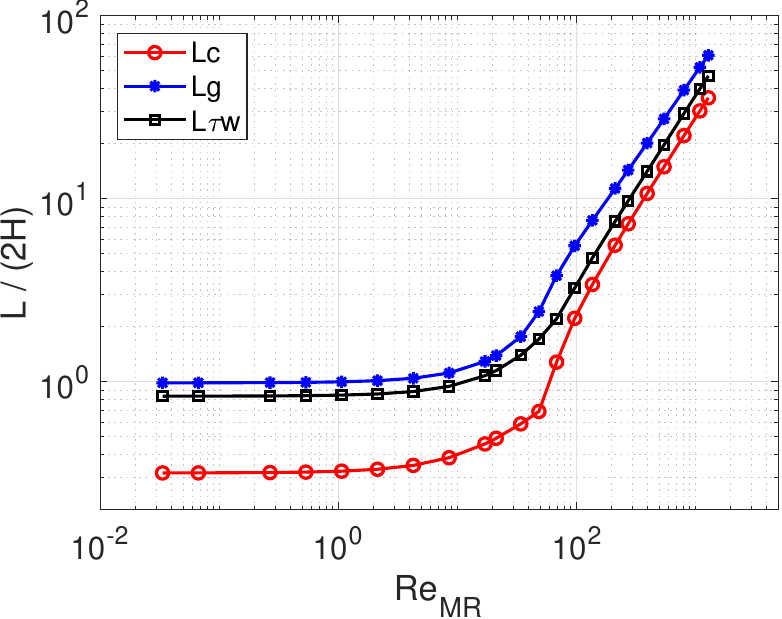}
        \caption{$\Bin = 10$}
        \label{sfig: Ls Bn10 pl}
    \end{subfigure}
    \begin{subfigure}[b]{0.328\textwidth}
        \centering
        \includegraphics[width=0.99\linewidth]{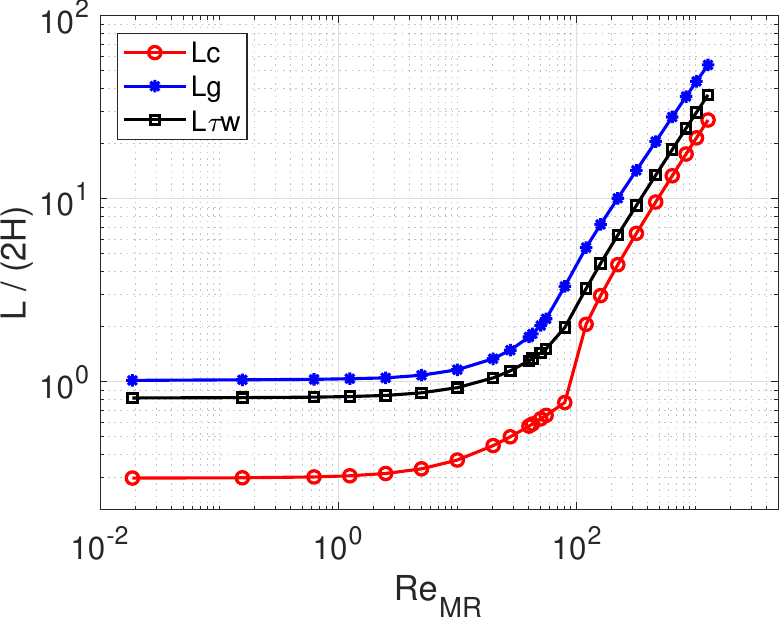}
        \caption{$\Bin = 20$}
        \label{sfig: Ls Bn20 pl}
    \end{subfigure}

    \caption{Channel flow: Comparison of the variations of the three development lengths with the
Reynolds number for different Bingham numbers. For convenience, the Metzner-Reed Reynolds number
(Eq.\ \eqref{eq: Re MR planar}) is used instead of $\Rey$ so as to use the same $x$-axis range for
all plots. The computations were performed with $M = 2000$ (for $\Bin = 1$ and $2$), $M = 5000$ (for
$\Bin = 5$ and $10$) and $M = 15000$ (for $\Bin = 20)$.}
  \label{fig: Ls pl}
\end{figure}

Figures \ref{fig: Ls ax} and \ref{fig: Ls pl} compare the three main development lengths, $L_c$,
$L_g$ and $L_{\tau w}$, for axisymmetric and planar flow, respectively. To obtain the data,
simulations were performed where the Bingham number was held fixed and the Reynolds number was
varied, and the three development lengths were recorded. The procedure was repeated for a set of
different Bingham numbers, listed in Table \ref{table: r0 values} along with the corresponding plug
radii/thicknesses.

In accordance with previous studies, the diagrams show that the development length tends to a finite
($\Bin$-dependent) value as $\Rey \rightarrow 0$; it exhibits an approximately constant plateau at
low $\Rey$ values and increases proportionally to $\Rey$ at higher $\Rey$ values. This holds for all
development lengths, including the newly-proposed $L_{\tau w}$. In Newtonian flow, $L_c$ and $L_g$
are identical and $L_{\tau w}$ is very slightly smaller. As the Bingham number is increased, the
value of $L_c$ in the low-$\Rey$ plateau decreases because the central plug expands and its flat
velocity tends to equate to that at the inlet, so that less development is required near the centre
\cite{Philippou_2016, Lambride_2023}. However, the other two development lengths, $L_g$ and $L_{\tau
w}$, actually increase with $\Bin$: the rapid development at the central region due to the expansion
of the plug is balanced by a slower flow development near the walls. Although at low $\Rey$ increase
of $\Bin$ causes divergence between $L_c$ on one hand and $L_g$ and $L_{\tau w}$ on the other,
beyond some value of $\Rey$ the length $L_c$ catches up rapidly with the other lengths -- or at
least with $L_{\tau w}$, because $L_g$ starts to pull away from the rest at high $\Bin$. More
insight is provided by Fig.\ \ref{fig: L vs r-y Bn5} (where, for $\Bin = 5$, $\Rey = 61$ translates
to $\Rey_{MR} = 33$/$28$ for pipe/channel flow, and $\Rey = 246$ translates to $\Rey_{MR} =
135$/$112$, respectively), which shows that at higher Reynolds numbers $L_c$ is quite close to the
maximum value of $L(r)$ in the accelerating region, while $L(r)$ attains even higher values (albeit
not much higher) in the decelerating region, and in particular at the wall: $L_g = L(R)$. The same
hold also for the planar case.

\begin{figure}[!tb]
    \centering

    \begin{subfigure}[b]{0.49\textwidth}
        \centering
        \includegraphics[width=0.99\linewidth]{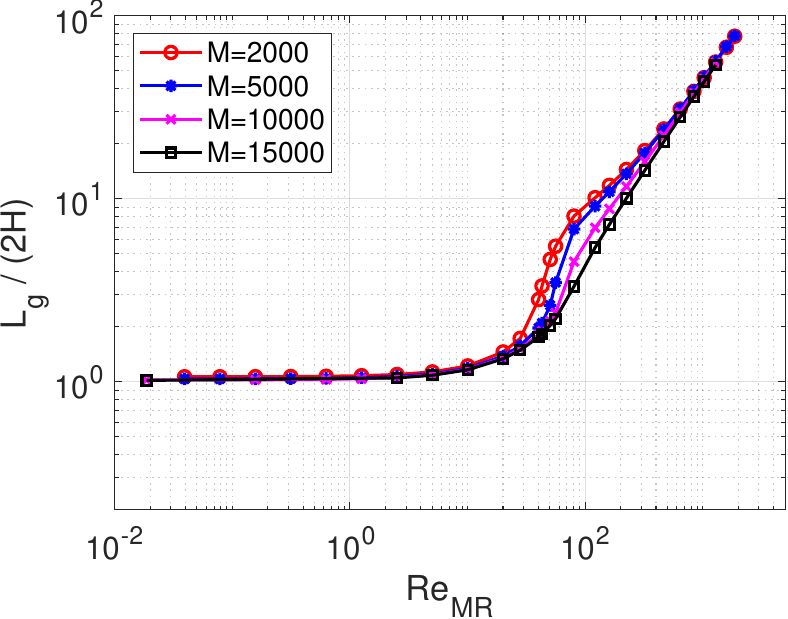}
        \caption{$L_g$}
        \label{sfig: Lg vs ReMR Bn20 planar per M}
    \end{subfigure}
    \begin{subfigure}[b]{0.49\textwidth}
        \centering
        \includegraphics[width=0.99\linewidth]{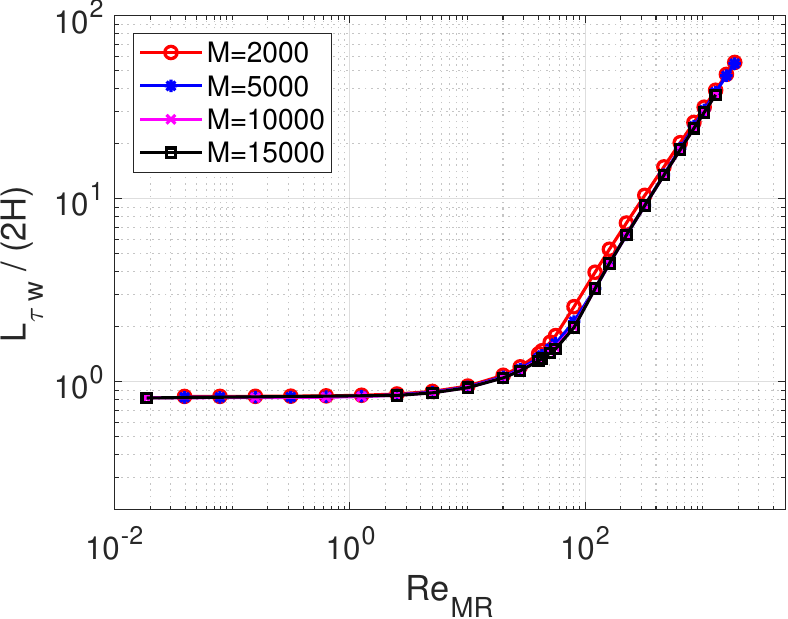}
        \caption{$L_{\tau w}$}
        \label{sfig: Lw vs ReMR Bn20 planar per M}
    \end{subfigure}

    \caption{Variations of $L_g$ \subref{sfig: Lg vs ReMR Bn20 planar per M} and $L_{\tau w}$
\subref{sfig: Lw vs ReMR Bn20 planar per M} with $\Rey_{MR}$ for the $\Bin = 20$ planar case,
obtained with various values of $M$.}
  \label{fig: Ls vs ReMR Bn20 planar per M}
\end{figure}

The accurate calculation of $L_g$ for high-$Bn$ planar flow is quite challenging, due to a tendency
to exaggerate the development length close to the wall, exhibited in Fig.\ \ref{sfig: L vs r Bn5
Re245p76 planar}. Figure \ref{sfig: Lg vs ReMR Bn20 planar per M} compares the results obtained
with different values of $M$. This exaggeration is manifested as a spurious ``hump'' exhibited by
the $L_g$ curves in the approximate range $25 \leq \Rey_{MR} \leq 500$, which decreases as $M$ is
increased. Interestingly, the difficulty appears to vanish not only at small Reynolds numbers but
also at large ones ($\Rey_{MR} > 500$), for which even $M = 2000$ seems to provide good accuracy.
Figure \ref{sfig: Lw vs ReMR Bn20 planar per M} shows the corresponding variations of $L_{\tau w}$,
for which the errors appear to be much smaller -- the results for $M = 5000$ and $M = 15000$ are
almost indistinguishable. This is a fortunate and perhaps surprising observation given that
$L_{\tau w}$ is also determined by the flow conditions at the wall.

\begin{figure}[tb]
    \centering

    \begin{subfigure}[b]{0.328\textwidth}
        \centering
        \includegraphics[width=0.99\linewidth]{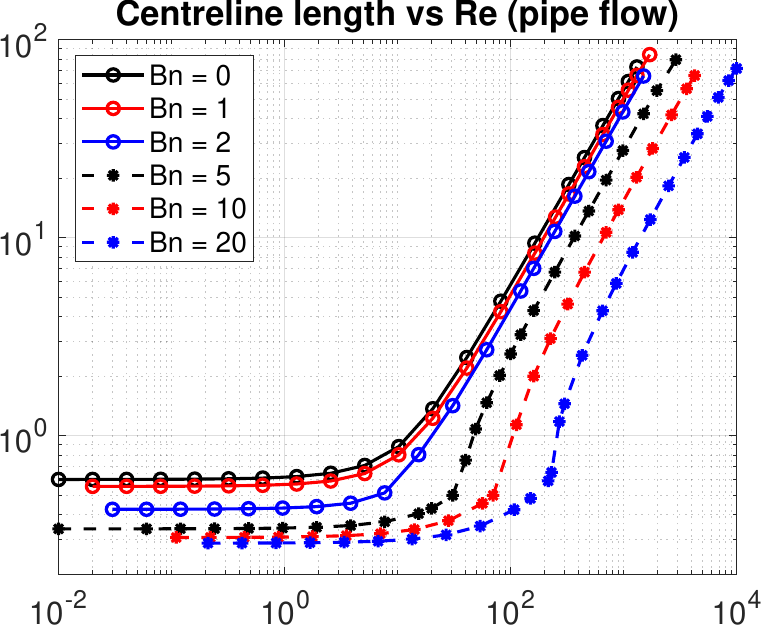}
        \caption{$\tilde{L}_c(\Rey)$}
        \label{sfig: Re Lc}
    \end{subfigure}
    \begin{subfigure}[b]{0.328\textwidth}
        \centering
        \includegraphics[width=0.99\linewidth]{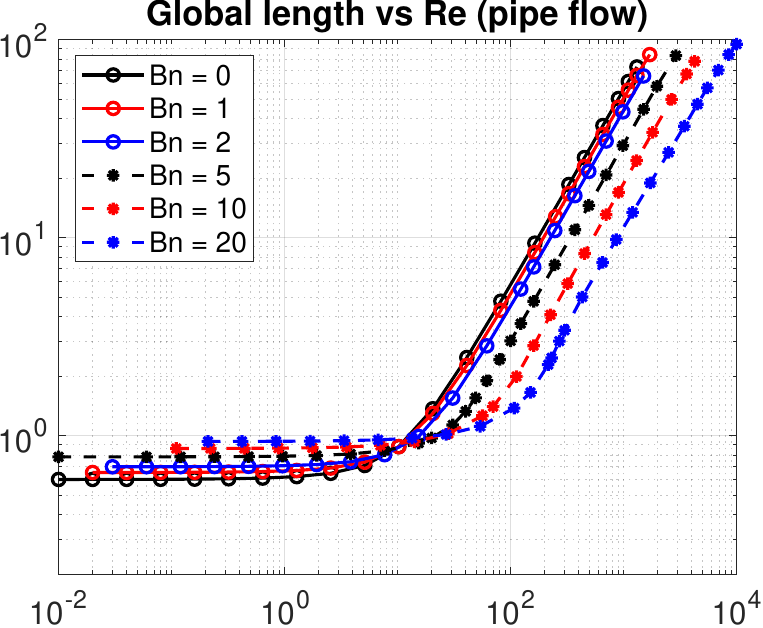}
        \caption{$\tilde{L}_g(\Rey)$}
        \label{sfig: Re Lg}
    \end{subfigure}
    \begin{subfigure}[b]{0.328\textwidth}
        \centering
        \includegraphics[width=0.99\linewidth]{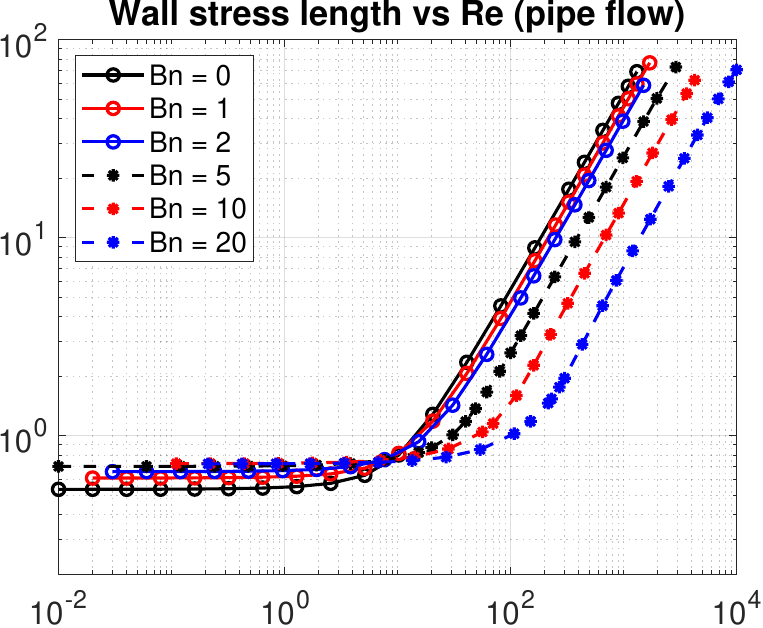}
        \caption{$\tilde{L}_{\tau w}(\Rey)$}
        \label{sfig: Re Lw}
    \end{subfigure}
    \\[0.5cm]

    \begin{subfigure}[b]{0.328\textwidth}
        \centering
        \includegraphics[width=0.99\linewidth]{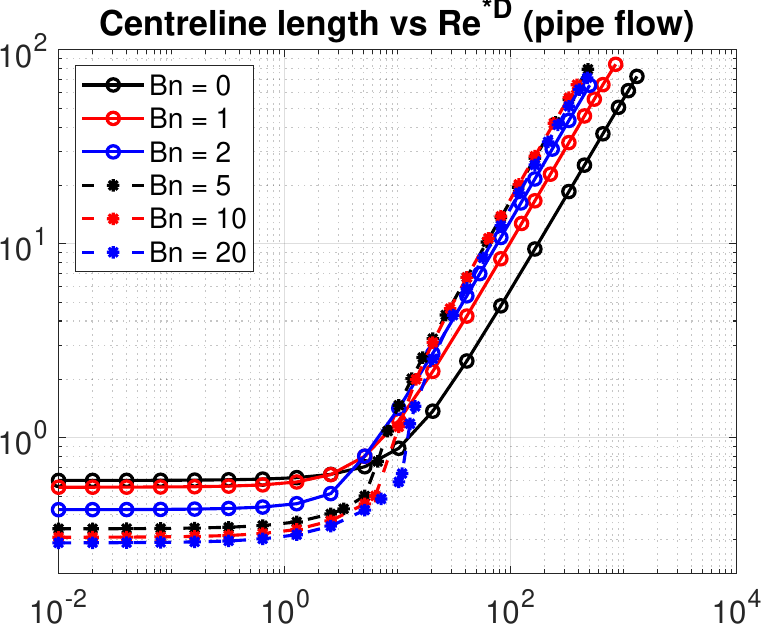}
        \caption{$\tilde{L}_c(\Rey^{*D})$}
        \label{sfig: ReeD Lc}
    \end{subfigure}
    \begin{subfigure}[b]{0.328\textwidth}
        \centering
        \includegraphics[width=0.99\linewidth]{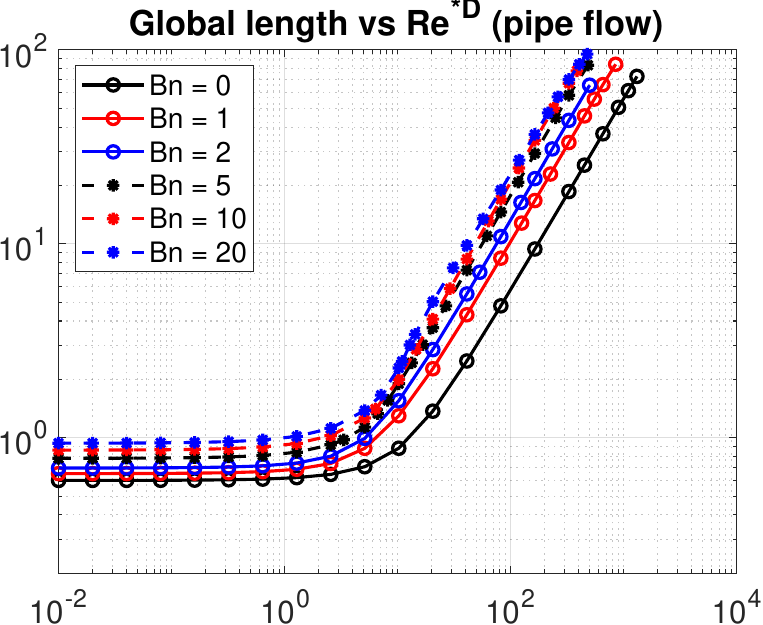}
        \caption{$\tilde{L}_g(\Rey^{*D})$}
        \label{sfig: ReeD Lg}
    \end{subfigure}
    \begin{subfigure}[b]{0.328\textwidth}
        \centering
        \includegraphics[width=0.99\linewidth]{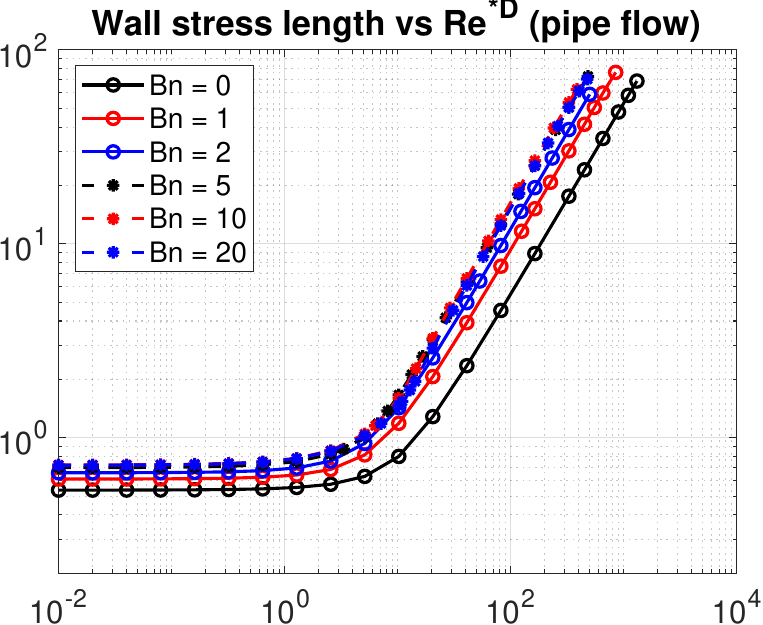}
        \caption{$\tilde{L}_{\tau w}(\Rey^{*D})$}
        \label{sfig: ReeD Lw}
    \end{subfigure}
    \\[0.5cm]

    \begin{subfigure}[b]{0.328\textwidth}
        \centering
        \includegraphics[width=0.99\linewidth]{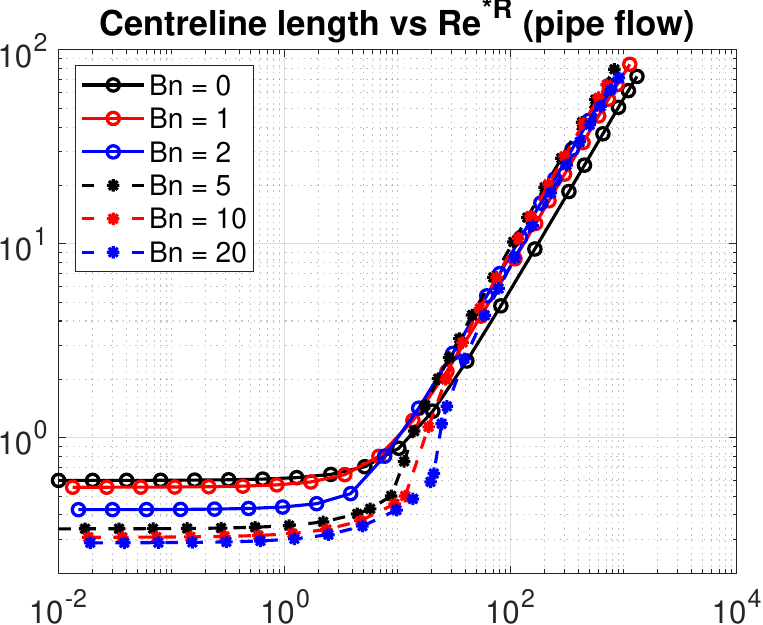}
        \caption{$\tilde{L}_c(\Rey^{*R})$}
        \label{sfig: ReeR Lc}
    \end{subfigure}
    \begin{subfigure}[b]{0.328\textwidth}
        \centering
        \includegraphics[width=0.99\linewidth]{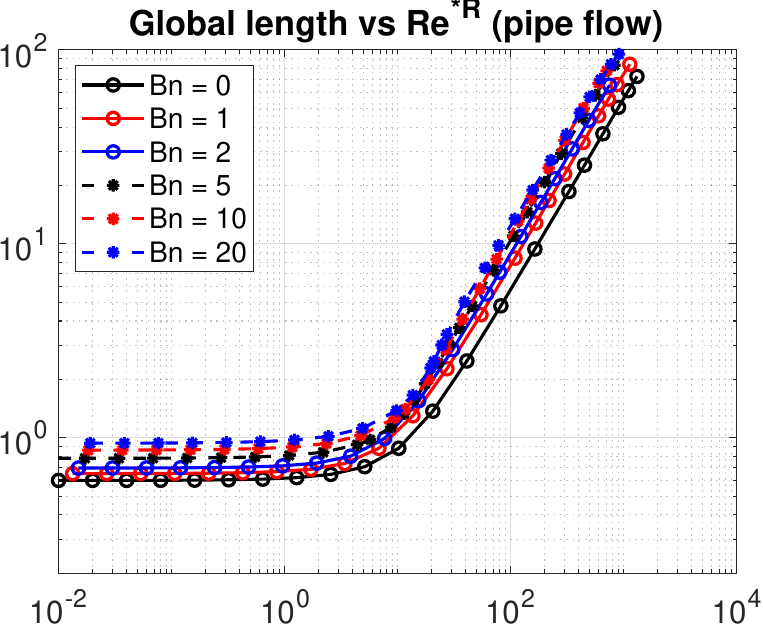}
        \caption{$\tilde{L}_g(\Rey^{*R})$}
        \label{sfig: ReeR Lg}
    \end{subfigure}
    \begin{subfigure}[b]{0.328\textwidth}
        \centering
        \includegraphics[width=0.99\linewidth]{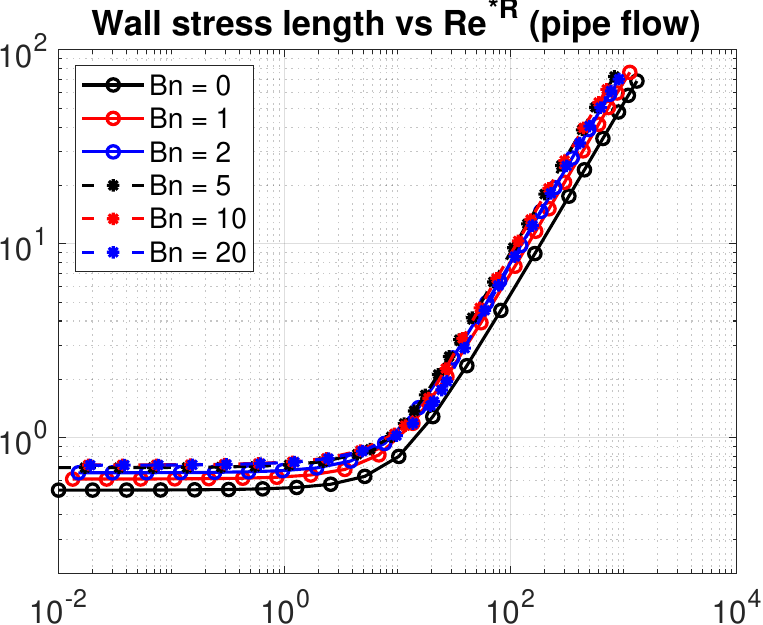}
        \caption{$\tilde{L}_{\tau w}(\Rey^{*R})$}
        \label{sfig: ReeR Lw}
    \end{subfigure}
    \\[0.5cm]

    \begin{subfigure}[b]{0.328\textwidth}
        \centering
        \includegraphics[width=0.99\linewidth]{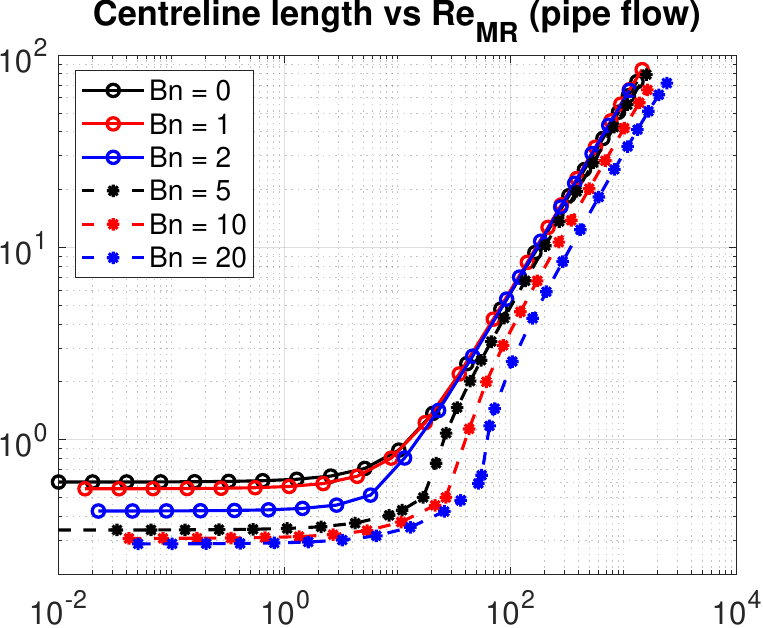}
        \caption{$\tilde{L}_c(\Rey_{MR})$}
        \label{sfig: ReMR Lc}
    \end{subfigure}
    \begin{subfigure}[b]{0.328\textwidth}
        \centering
        \includegraphics[width=0.99\linewidth]{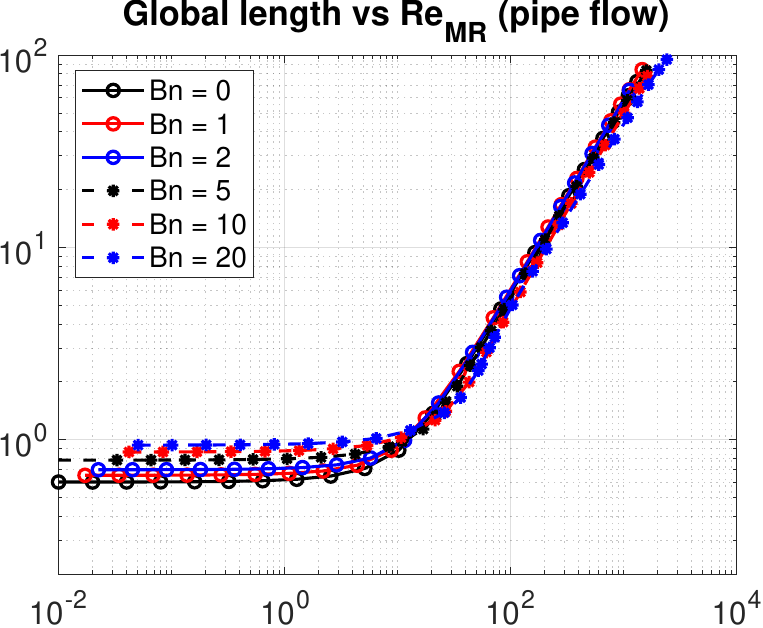}
        \caption{$\tilde{L}_g(\Rey_{MR})$}
        \label{sfig: ReMR Lg}
    \end{subfigure}
    \begin{subfigure}[b]{0.328\textwidth}
        \centering
        \includegraphics[width=0.99\linewidth]{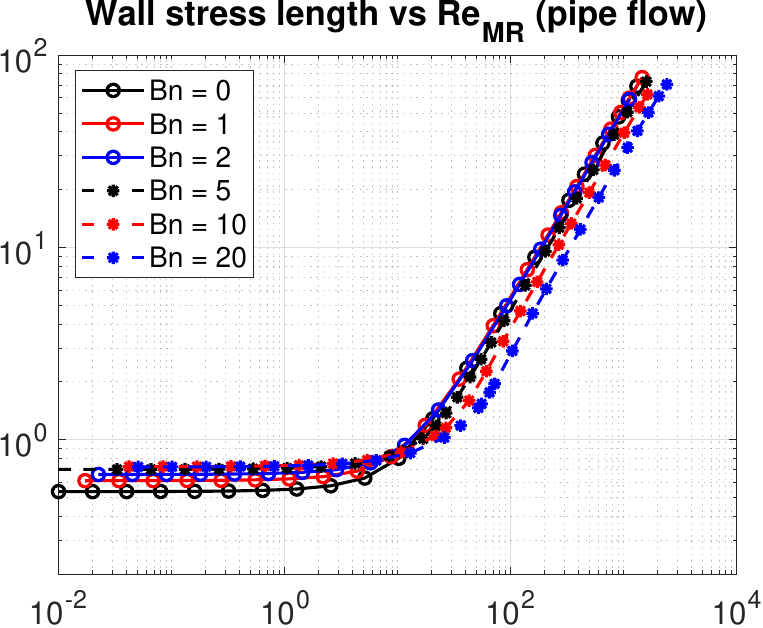}
        \caption{$\tilde{L}_{\tau w}(\Rey_{MR})$}
        \label{sfig: ReMR Lw}
    \end{subfigure}

    \caption{Pipe flow at various Bingham numbers: The three development lengths $L_c$ (left), $L_c$
(middle) and $L_{\tau_w}$ (right), scaled by the pipe diameter $D$, as a function of $\Rey$
\eqref{eq: Re} (top row), $\Rey^{*D}$ \eqref{eq: Re effective D} (2nd row), $\Rey^{*R}$ \eqref{eq:
Re effective R} (3rd row), and $\Rey_{MR}$ \eqref{eq: Re MR} (last row).}
  \label{fig: L vs Re ReeD ReeR ReMR axisymmetric}
\end{figure}

The $x$-axis of Figs.\ \ref{fig: Ls ax} and \ref{fig: Ls pl} is in terms of the Metzner-Reed
Reynolds number, which facilitates using the same $x$-axis ranges in all plots, irrespective of the
Bingham number. This brings us to the issue of examining the various Reynolds number variants of
Sec.\ \ref{sec: Renolds numbers} and how the development lengths correlate to them. The results are
plotted in Figs.\ \ref{fig: L vs Re ReeD ReeR ReMR axisymmetric}-\ref{fig: L vs ReMR ReMRC ReMG ReEG
planar}.

\begin{figure}[tb]
    \centering

    \begin{subfigure}[b]{0.328\textwidth}
        \centering
        \includegraphics[width=0.99\linewidth]{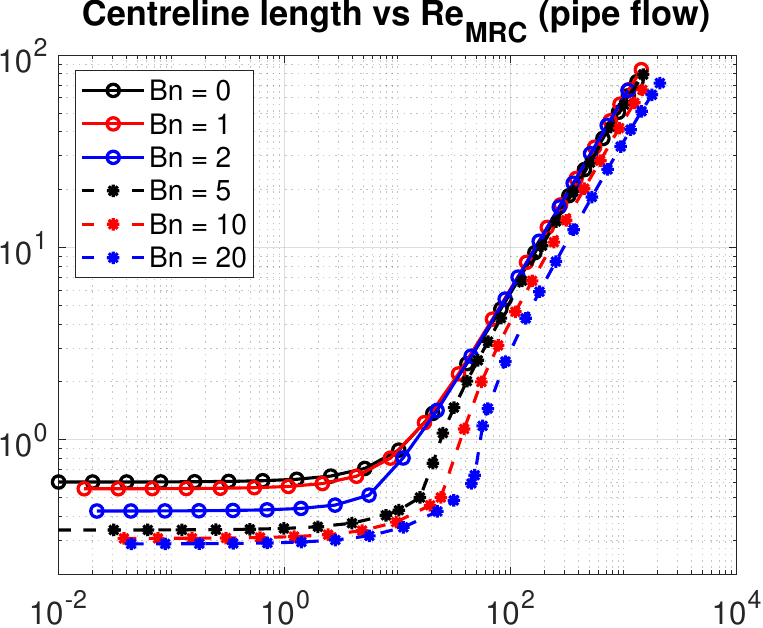}
        \caption{$\tilde{L}_c(\Rey_{MRC})$}
        \label{sfig: ReMRC Lc}
    \end{subfigure}
    \begin{subfigure}[b]{0.328\textwidth}
        \centering
        \includegraphics[width=0.99\linewidth]{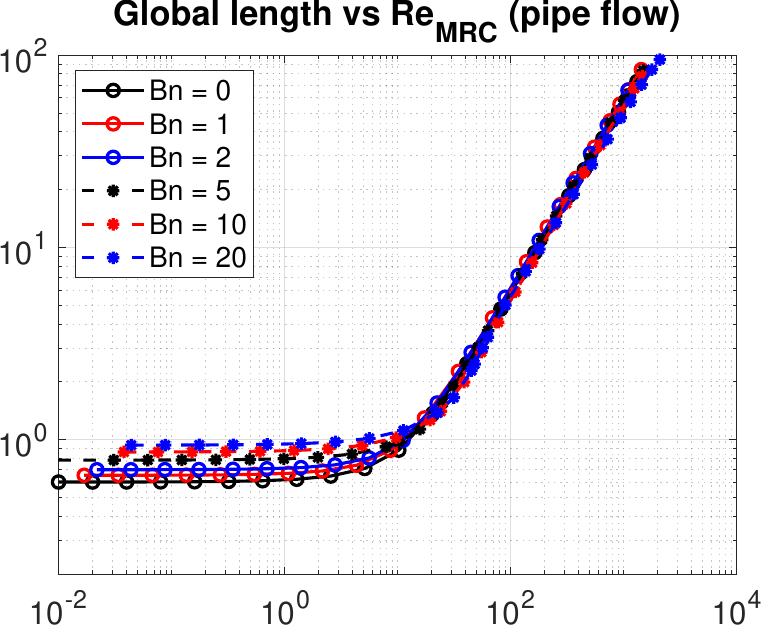}
        \caption{$\tilde{L}_g(\Rey_{MRC})$}
        \label{sfig: ReMRC Lg}
    \end{subfigure}
    \begin{subfigure}[b]{0.328\textwidth}
        \centering
        \includegraphics[width=0.99\linewidth]{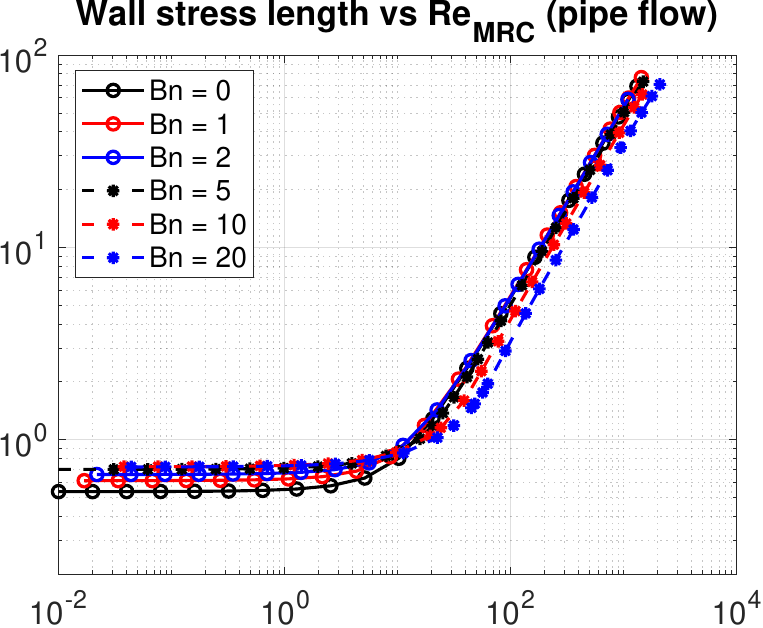}
        \caption{$\tilde{L}_{\tau w}(\Rey_{MRC})$}
        \label{sfig: ReMRC Lw}
    \end{subfigure}
    \\[0.5cm]

    \begin{subfigure}[b]{0.328\textwidth}
        \centering
        \includegraphics[width=0.99\linewidth]{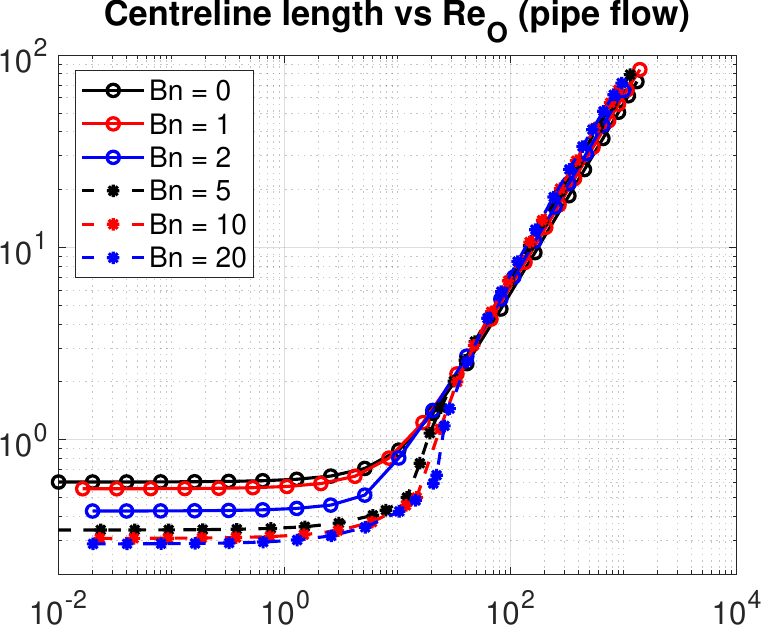}
        \caption{$\tilde{L}_c(\Rey_O)$}
        \label{sfig: ReO Lc}
    \end{subfigure}
    \begin{subfigure}[b]{0.328\textwidth}
        \centering
        \includegraphics[width=0.99\linewidth]{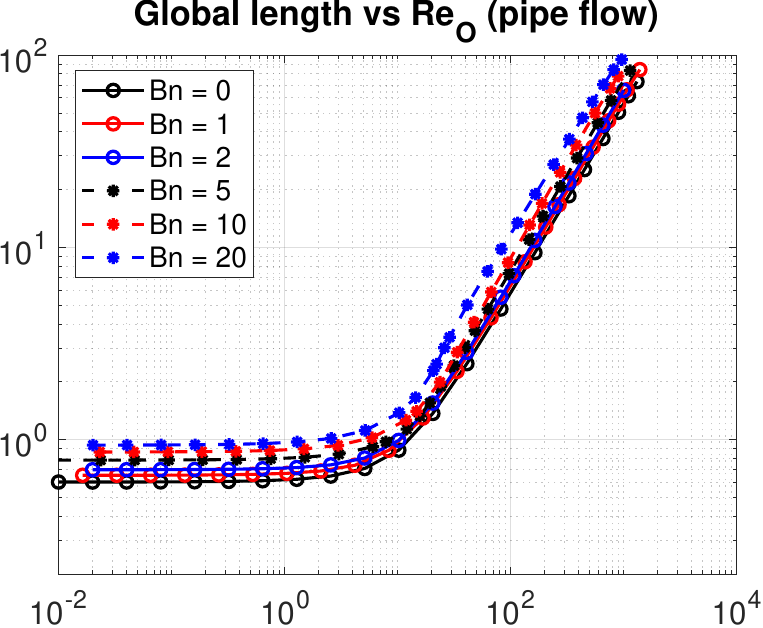}
        \caption{$\tilde{L}_g(\Rey_O)$}
        \label{sfig: ReO Lg}
    \end{subfigure}
    \begin{subfigure}[b]{0.328\textwidth}
        \centering
        \includegraphics[width=0.99\linewidth]{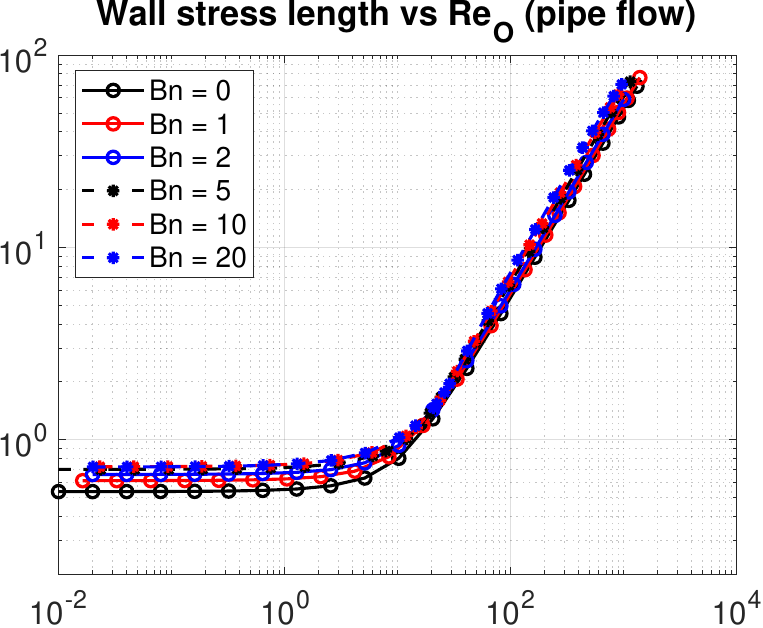}
        \caption{$\tilde{L}_{\tau w}(\Rey_O)$}
        \label{sfig: ReO Lw}
    \end{subfigure}
    \\[0.5cm]

    \begin{subfigure}[b]{0.328\textwidth}
        \centering
        \includegraphics[width=0.99\linewidth]{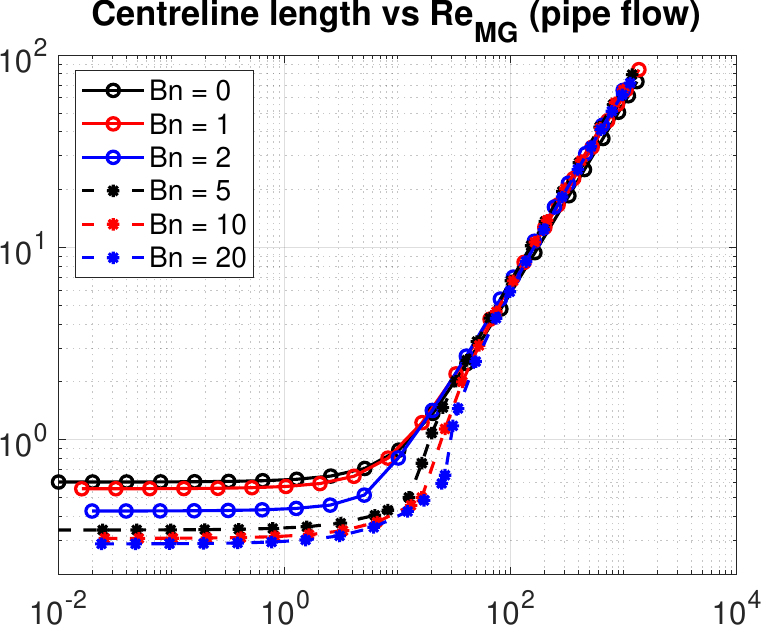}
        \caption{$\tilde{L}_c(\Rey_{MG})$}
        \label{sfig: ReMG Lc}
    \end{subfigure}
    \begin{subfigure}[b]{0.328\textwidth}
        \centering
        \includegraphics[width=0.99\linewidth]{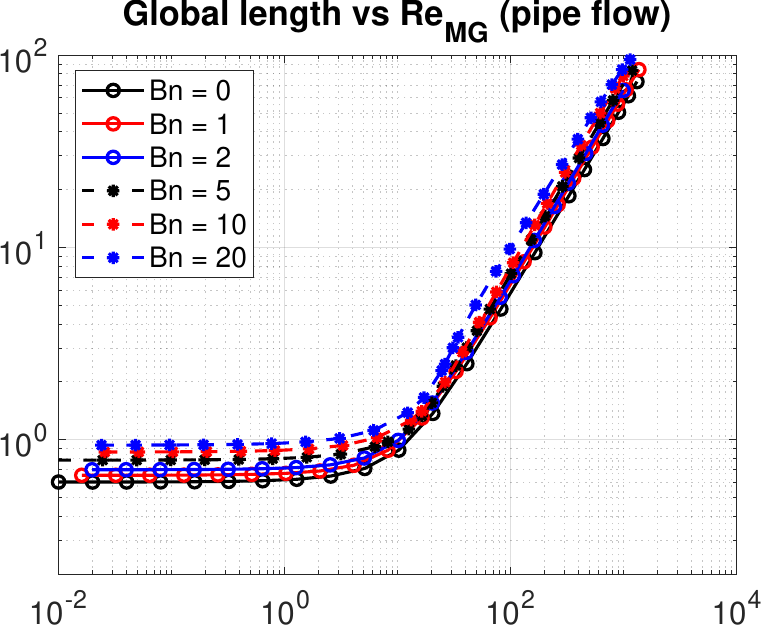}
        \caption{$\tilde{L}_g(\Rey_{MG})$}
        \label{sfig: ReMG Lg}
    \end{subfigure}
    \begin{subfigure}[b]{0.328\textwidth}
        \centering
        \includegraphics[width=0.99\linewidth]{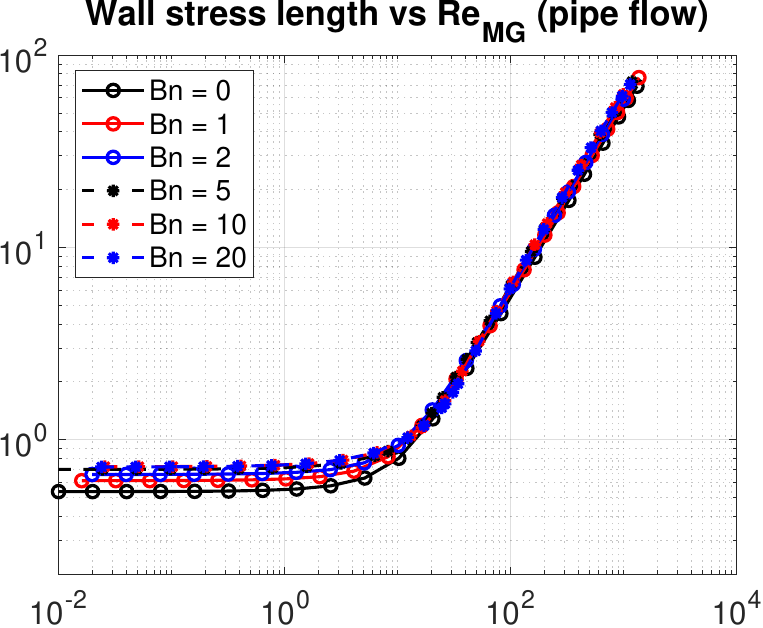}
        \caption{$\tilde{L}_{\tau w}(\Rey_{MG})$}
        \label{sfig: ReMG Lw}
    \end{subfigure}
    \\[0.5cm]

    \begin{subfigure}[b]{0.328\textwidth}
        \centering
        \includegraphics[width=0.99\linewidth]{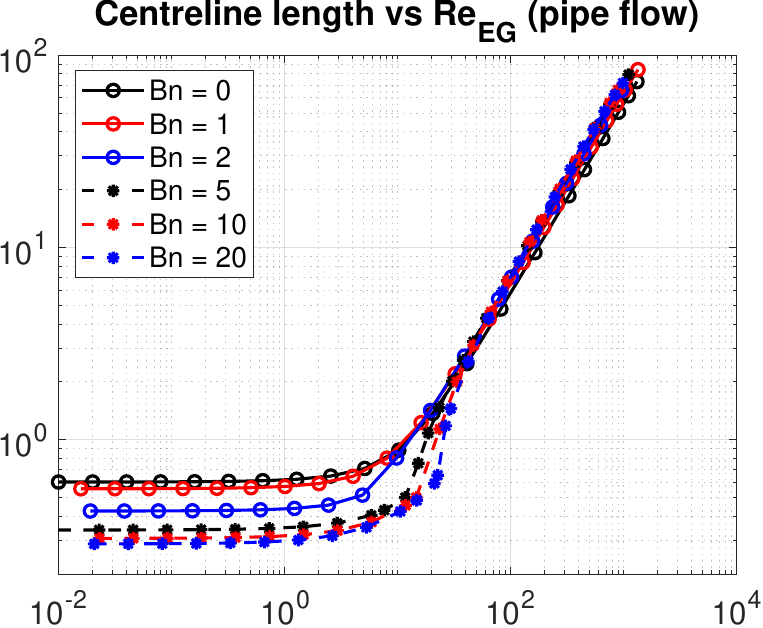}
        \caption{$\tilde{L}_c(\Rey_{EG})$}
        \label{sfig: ReEG Lc}
    \end{subfigure}
    \begin{subfigure}[b]{0.328\textwidth}
        \centering
        \includegraphics[width=0.99\linewidth]{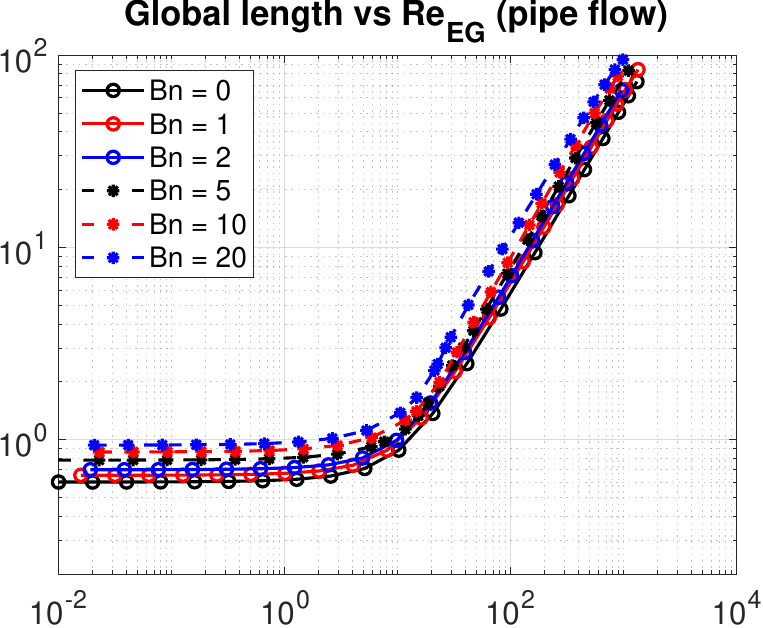}
        \caption{$\tilde{L}_g(\Rey_{EG})$}
        \label{sfig: ReEG Lg}
    \end{subfigure}
    \begin{subfigure}[b]{0.328\textwidth}
        \centering
        \includegraphics[width=0.99\linewidth]{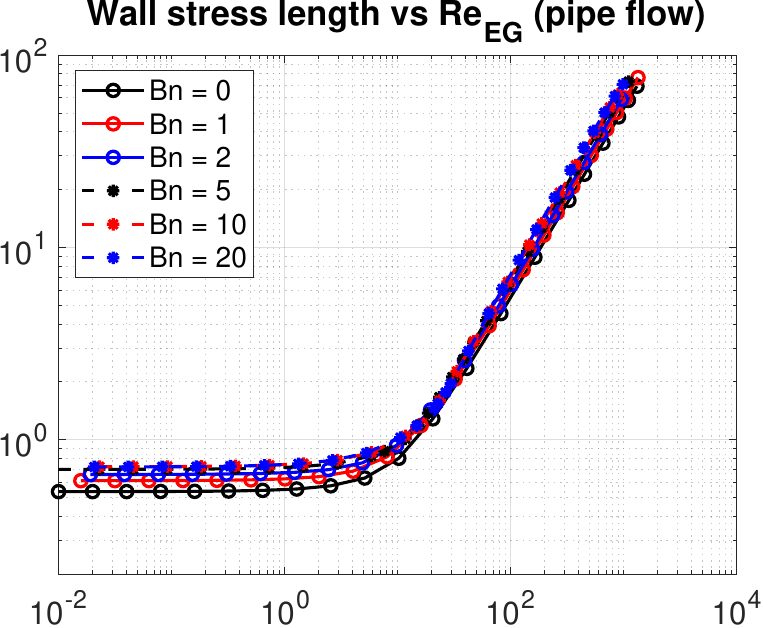}
        \caption{$\tilde{L}_{\tau w}(\Rey_{EG})$}
        \label{sfig: ReEG Lw}
    \end{subfigure}

    \caption{Pipe flow at various Bingham numbers: The three development lengths $L_c$ (left), $L_c$
(middle) and $L_{\tau_w}$ (right), scaled by the pipe diameter $D$, as a function of $\Rey_{MRC}$
\eqref{eq: Re MRC} (top row), Ookawara's Reynolds number \cite{Ookawara_2000} (2nd row, denoted
as $\Rey_O$), $\Rey_{MG}$ \eqref{eq: Re MG} (3rd row), and $\Rey_{EG}$ \eqref{eq: Re EG} (last
row).}
  \label{fig: L vs ReMRC ReO ReMG ReEG axisymmetric}
\end{figure}

\begin{figure}[tb]
    \centering

    \begin{subfigure}[b]{0.328\textwidth}
        \centering
        \includegraphics[width=0.99\linewidth]{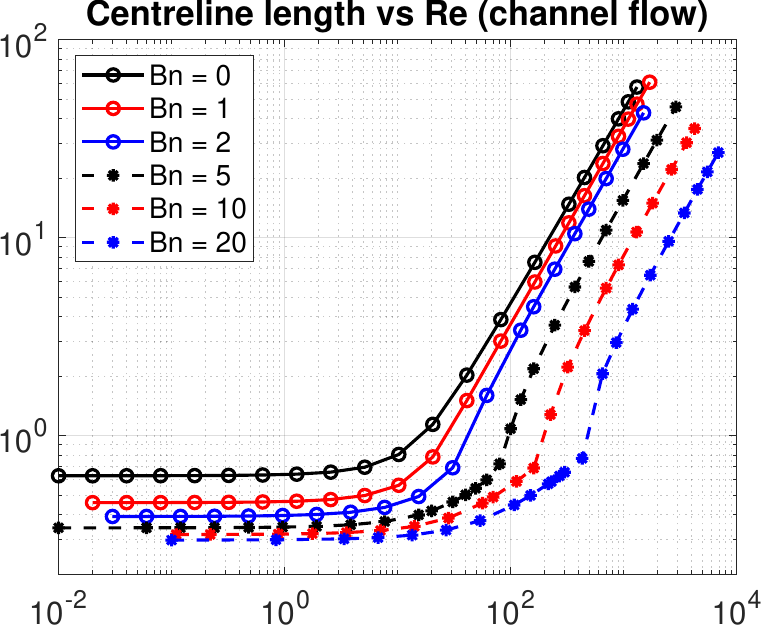}
        \caption{$\tilde{L}_c(\Rey)$}
        \label{sfig: Re Lc P}
    \end{subfigure}
    \begin{subfigure}[b]{0.328\textwidth}
        \centering
        \includegraphics[width=0.99\linewidth]{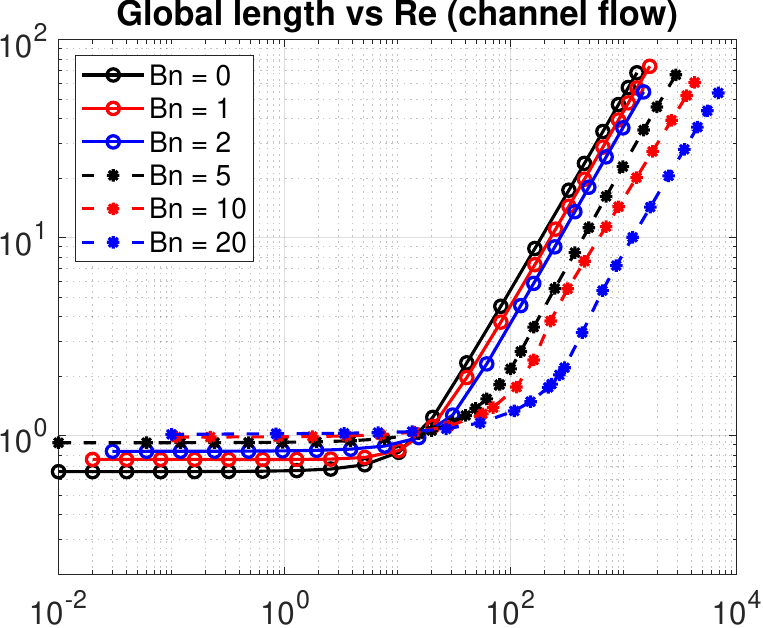}
        \caption{$\tilde{L}_g(\Rey)$}
        \label{sfig: Re Lg P}
    \end{subfigure}
    \begin{subfigure}[b]{0.328\textwidth}
        \centering
        \includegraphics[width=0.99\linewidth]{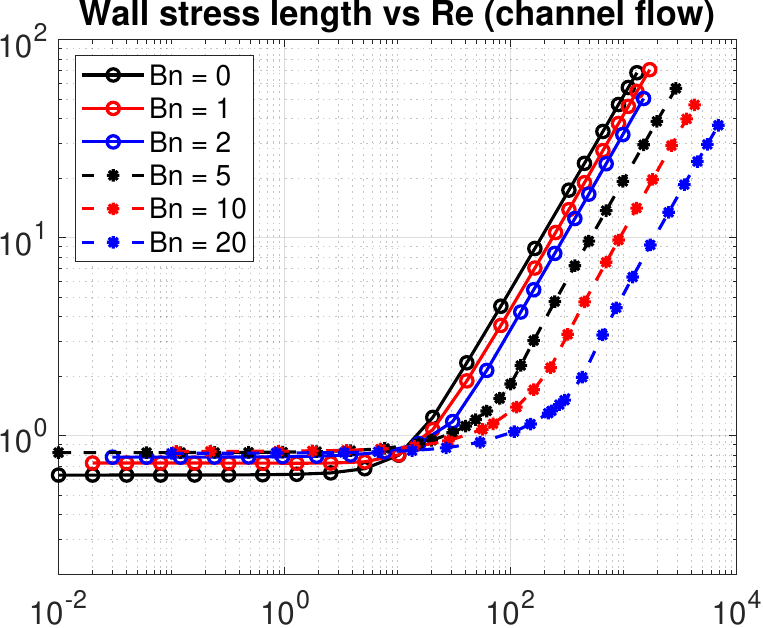}
        \caption{$\tilde{L}_{\tau w}(\Rey)$}
        \label{sfig: Re Lw P}
    \end{subfigure}
    \\[0.5cm]

    \begin{subfigure}[b]{0.328\textwidth}
        \centering
        \includegraphics[width=0.99\linewidth]{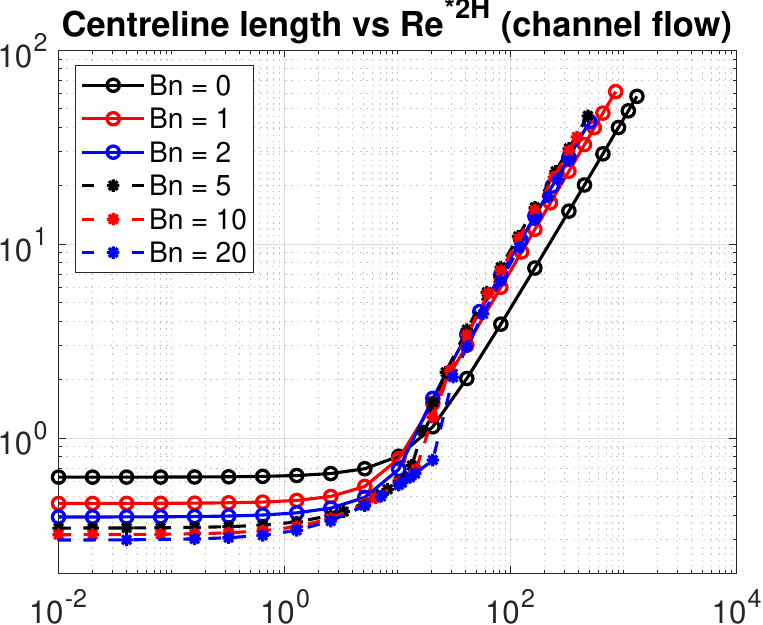}
        \caption{$\tilde{L}_c(\Rey^{*2H})$}
        \label{sfig: Ree2H Lc P}
    \end{subfigure}
    \begin{subfigure}[b]{0.328\textwidth}
        \centering
        \includegraphics[width=0.99\linewidth]{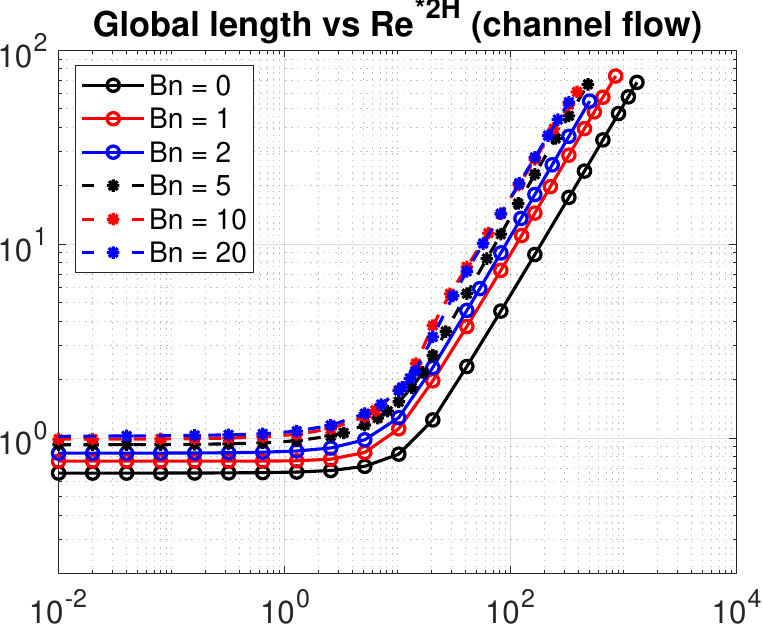}
        \caption{$\tilde{L}_g(\Rey^{*2H})$}
        \label{sfig: Ree2H Lg P}
    \end{subfigure}
    \begin{subfigure}[b]{0.328\textwidth}
        \centering
        \includegraphics[width=0.99\linewidth]{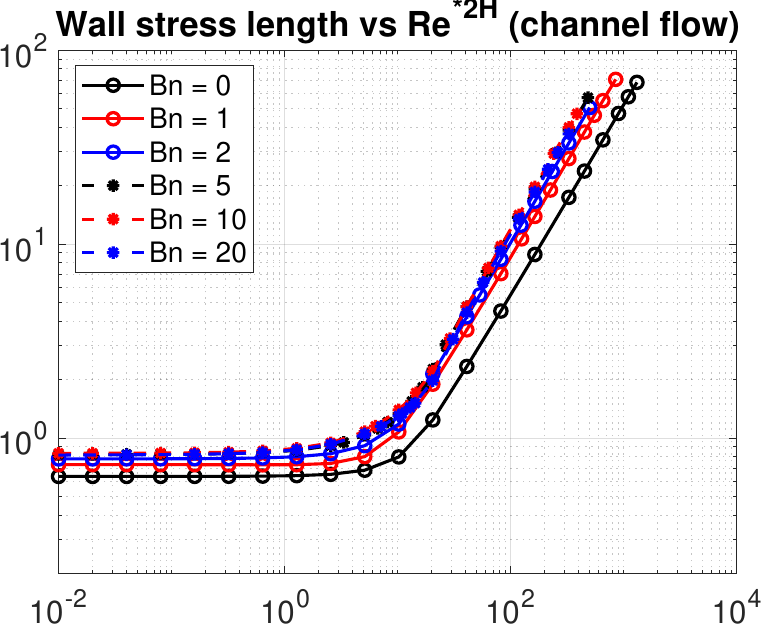}
        \caption{$\tilde{L}_{\tau w}(\Rey^{*2H})$}
        \label{sfig: Ree2H Lw P}
    \end{subfigure}
    \\[0.5cm]

    \begin{subfigure}[b]{0.328\textwidth}
        \centering
        \includegraphics[width=0.99\linewidth]{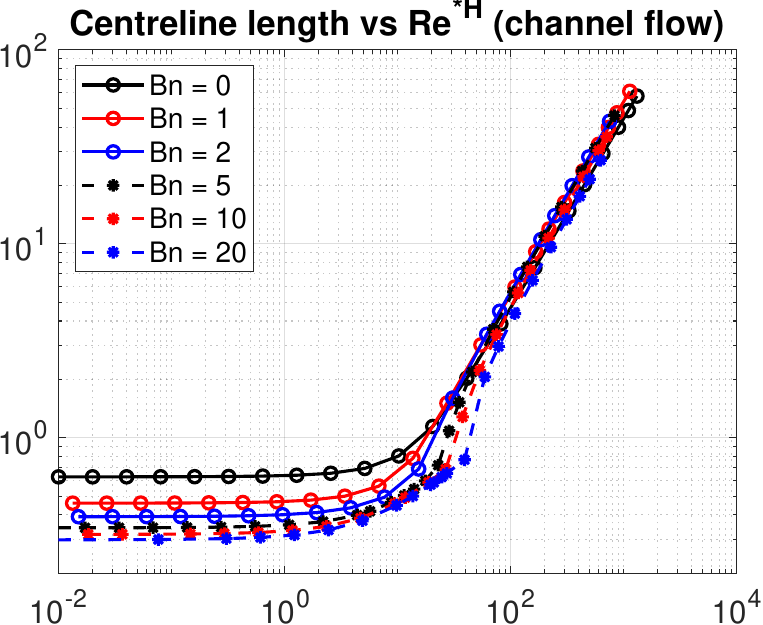}
        \caption{$\tilde{L}_c(\Rey^{*H})$}
        \label{sfig: ReeH Lc P}
    \end{subfigure}
    \begin{subfigure}[b]{0.328\textwidth}
        \centering
        \includegraphics[width=0.99\linewidth]{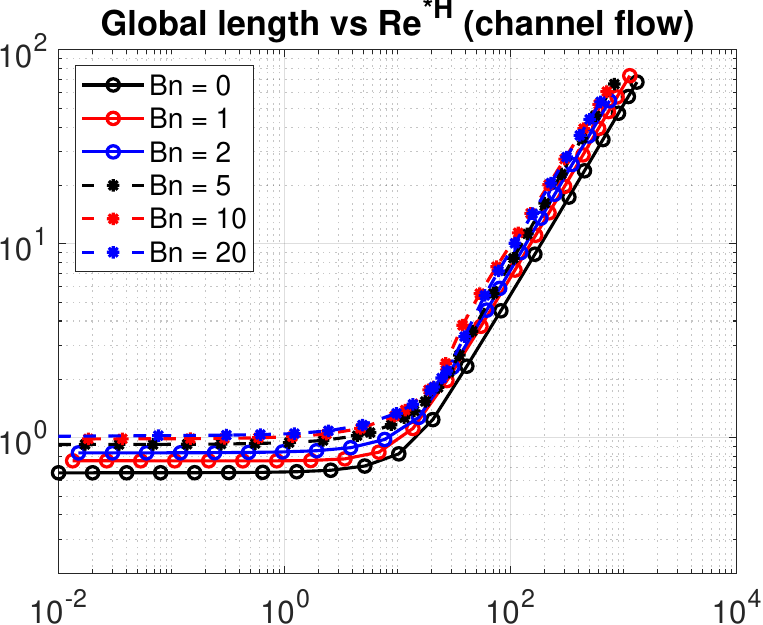}
        \caption{$\tilde{L}_g(\Rey^{*H})$}
        \label{sfig: ReeH Lg P}
    \end{subfigure}
    \begin{subfigure}[b]{0.328\textwidth}
        \centering
        \includegraphics[width=0.99\linewidth]{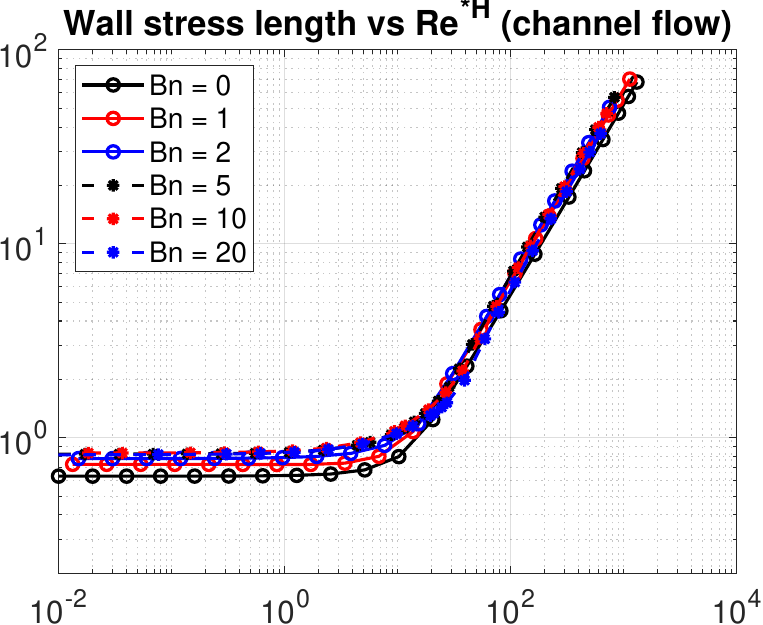}
        \caption{$\tilde{L}_{\tau w}(\Rey^{*H})$}
        \label{sfig: ReeH Lw P}
    \end{subfigure}

    \caption{Channel flow at various Bingham numbers: The three development lengths $L_c$ (left),
$L_c$ (middle) and $L_{\tau_w}$ (right), scaled by the channel height diameter $2H$, as a function
of $\Rey$ (top row), $\Rey^{*2H}$ (2nd row), and $\Rey^{*H}$ (last row).}
  \label{fig: L vs Re Ree2H ReeH planar}
\end{figure}

It is no surprise that when the development lengths are plotted against the standard Reynolds
number $\Rey$ \eqref{eq: Re} then the curves for the various Bingham numbers are spaced apart
(Figs.\ \ref{sfig: Re Lc}-\ref{sfig: Re Lw} for axisymmetric flow, \ref{sfig: Re Lc P}-\ref{sfig: Re
Lw P} for planar flow); as the Bingham number is increased, the same trends are observed as for
lower $\Bin$ but delayed to higher $\Rey$. At low $\Rey$, the $L_c$'s are more dispersed for the
various $\Bin$ compared to $L_g$ and $L_{\tau w}$.

The ``effective'' Reynolds numbers, $\Rey^{*D}$ and $\Rey^{*R}$ (or $\Rey^{*2H}$ and $\Rey^{*H}$)
go some way towards bringing the different $\Bin$ curves together. In Figs.\ \ref{sfig: ReeD Lc},
\ref{sfig: ReeD Lg} and \ref{sfig: ReeD Lw} we can see that the higher the Bingham number the more
shifted the curves are towards the left, which suggests that the characteristic stress inherent in
$\Rey^{*D}$, $\tau_0 + \mu U/D$, grows faster with $\Bin$ than the actual characteristic stress
that determines the development length. $\Rey^{*R}$, which uses the characteristic stress
$\frac{1}{2}(\tau_0 + \mu U/R)$ does a better job, as seen in Figs.\ \ref{sfig: ReeR Lc},
\ref{sfig: ReeR Lg} and \ref{sfig: ReeR Lw}; the shifting of the curves towards the left with
increase of $\Bin$ is still exhibited, but to a smaller degree. The planar case (Figs.\ \ref{sfig:
Ree2H Lc P}--\ref{sfig: ReeH Lw P}) is similar, and in fact the correlations are better than in the
axisymmetric case. Especially $\Rey^{*H}$ (Figs.\ \ref{sfig: ReeH Lc P}--\ref{sfig: ReeH Lw P}) is
quite effective in collapsing the development length curves for all $\Bin$.

\begin{figure}[tb]
    \centering

    \begin{subfigure}[b]{0.328\textwidth}
        \centering
        \includegraphics[width=0.99\linewidth]{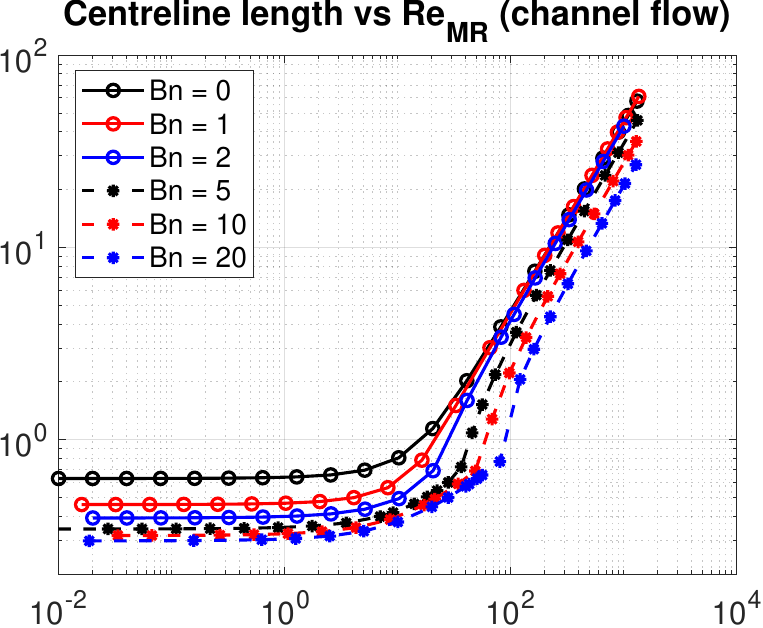}
        \caption{$\tilde{L}_c(\Rey_{MR})$}
        \label{sfig: ReMR Lc P}
    \end{subfigure}
    \begin{subfigure}[b]{0.328\textwidth}
        \centering
        \includegraphics[width=0.99\linewidth]{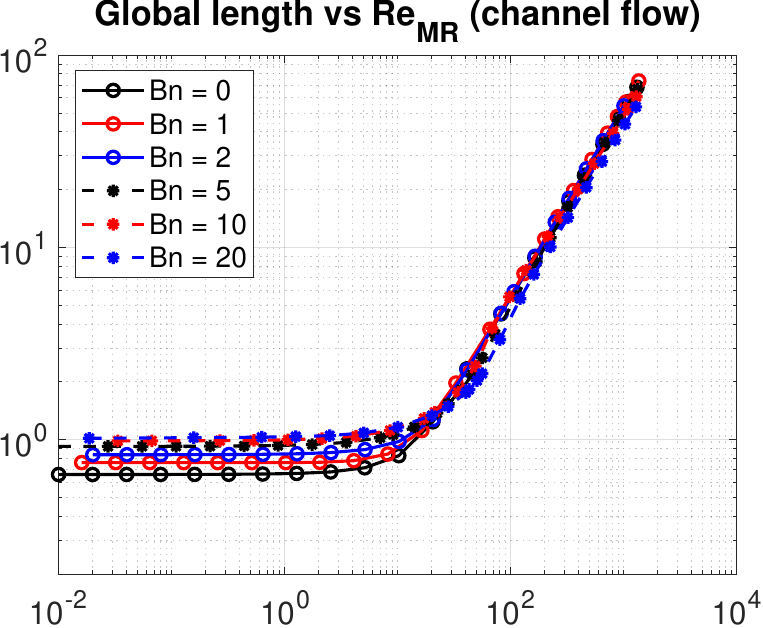}
        \caption{$\tilde{L}_g(\Rey_{MR})$}
        \label{sfig: ReMR Lg P}
    \end{subfigure}
    \begin{subfigure}[b]{0.328\textwidth}
        \centering
        \includegraphics[width=0.99\linewidth]{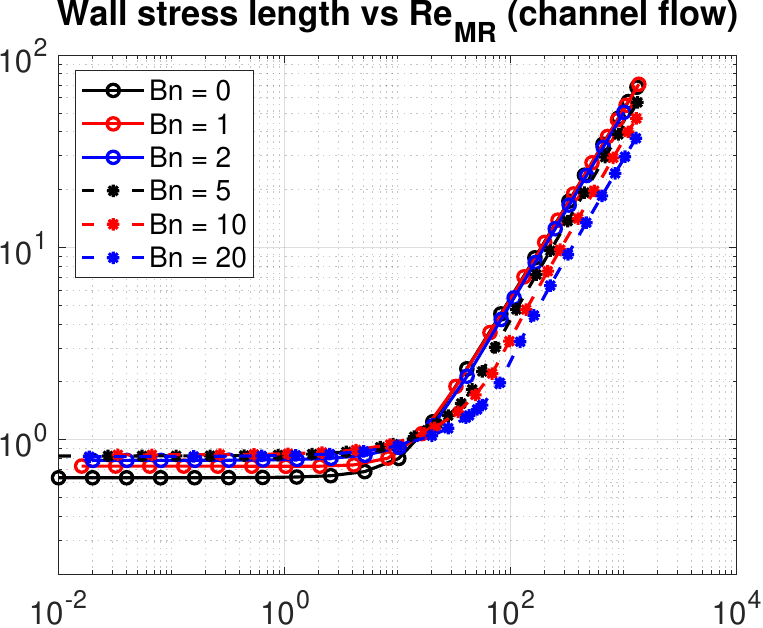}
        \caption{$\tilde{L}_{\tau w}(\Rey_{MR})$}
        \label{sfig: ReMR Lw P}
    \end{subfigure}
    \\[0.5cm]

    \begin{subfigure}[b]{0.328\textwidth}
        \centering
        \includegraphics[width=0.99\linewidth]{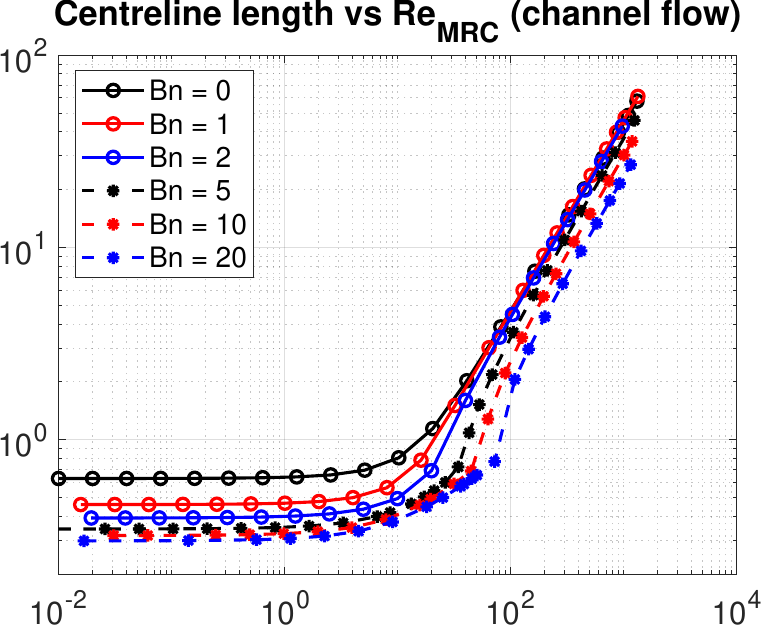}
        \caption{$\tilde{L}_c(\Rey_{MRC})$}
        \label{sfig: ReMRC Lc P}
    \end{subfigure}
    \begin{subfigure}[b]{0.328\textwidth}
        \centering
        \includegraphics[width=0.99\linewidth]{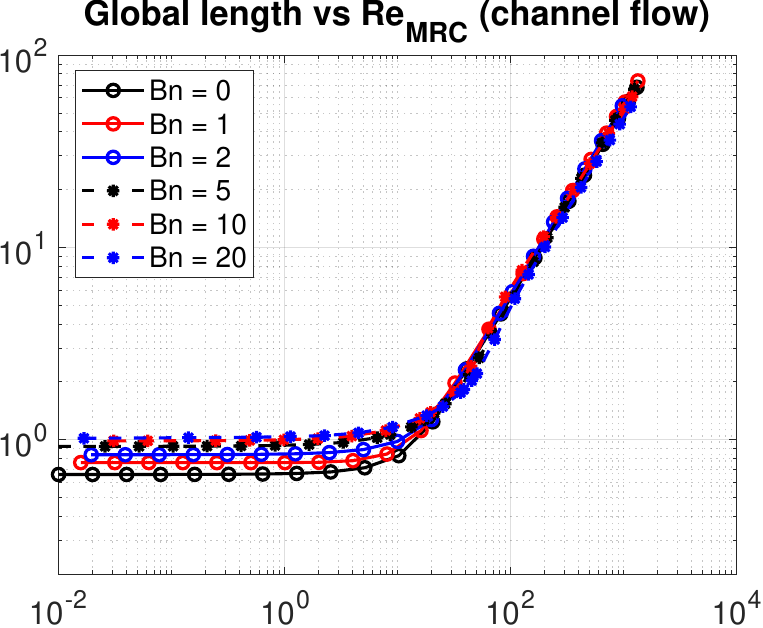}
        \caption{$\tilde{L}_g(\Rey_{MRC})$}
        \label{sfig: ReMRC Lg P}
    \end{subfigure}
    \begin{subfigure}[b]{0.328\textwidth}
        \centering
        \includegraphics[width=0.99\linewidth]{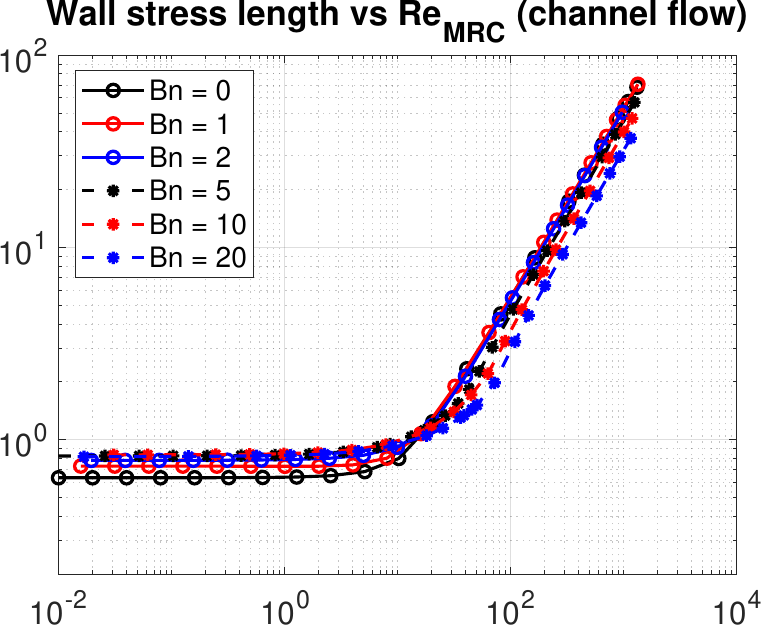}
        \caption{$\tilde{L}_{\tau w}(\Rey_{MRC})$}
        \label{sfig: ReMRC Lw P}
    \end{subfigure}
    \\[0.5cm]

    \begin{subfigure}[b]{0.328\textwidth}
        \centering
        \includegraphics[width=0.99\linewidth]{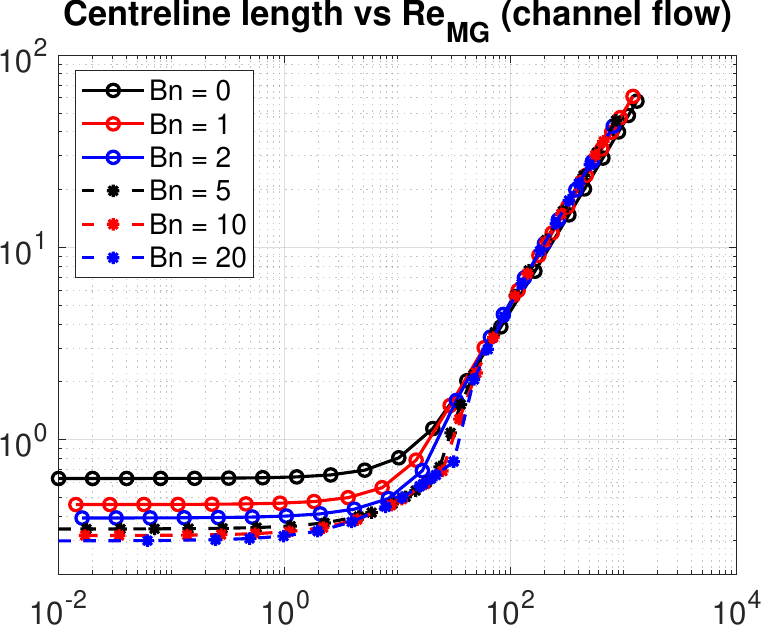}
        \caption{$\tilde{L}_c(\Rey_{MG})$}
        \label{sfig: ReMG Lc P}
    \end{subfigure}
    \begin{subfigure}[b]{0.328\textwidth}
        \centering
        \includegraphics[width=0.99\linewidth]{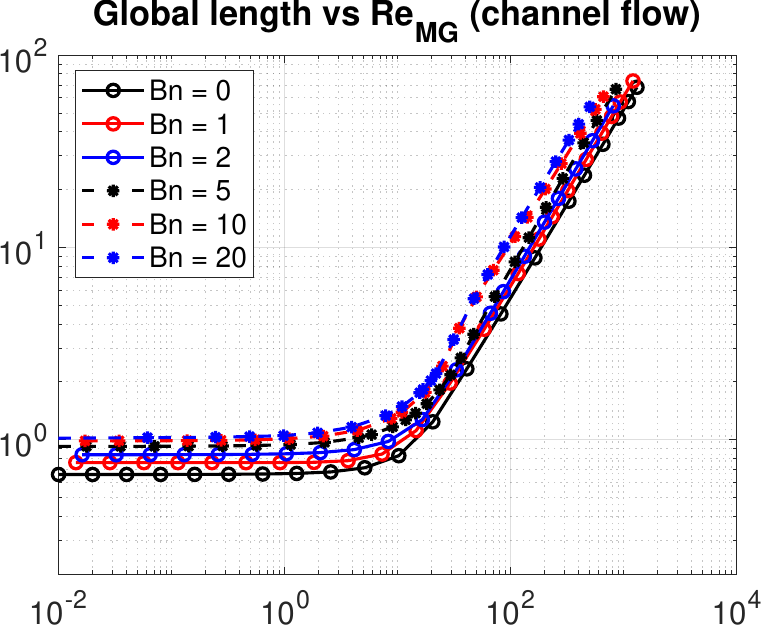}
        \caption{$\tilde{L}_g(\Rey_{MG})$}
        \label{sfig: ReMG Lg P}
    \end{subfigure}
    \begin{subfigure}[b]{0.328\textwidth}
        \centering
        \includegraphics[width=0.99\linewidth]{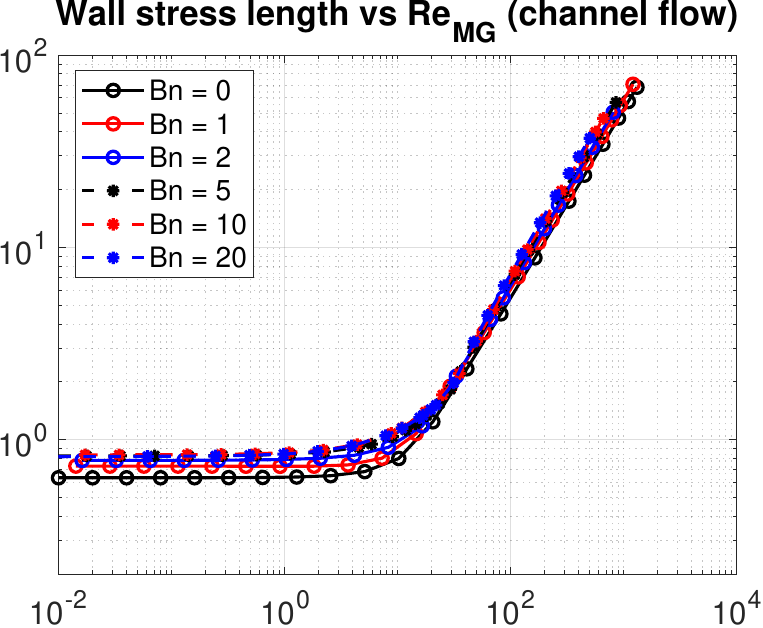}
        \caption{$\tilde{L}_{\tau w}(\Rey_{MG})$}
        \label{sfig: ReMG Lw P}
    \end{subfigure}
    \\[0.5cm]

    \begin{subfigure}[b]{0.328\textwidth}
        \centering
        \includegraphics[width=0.99\linewidth]{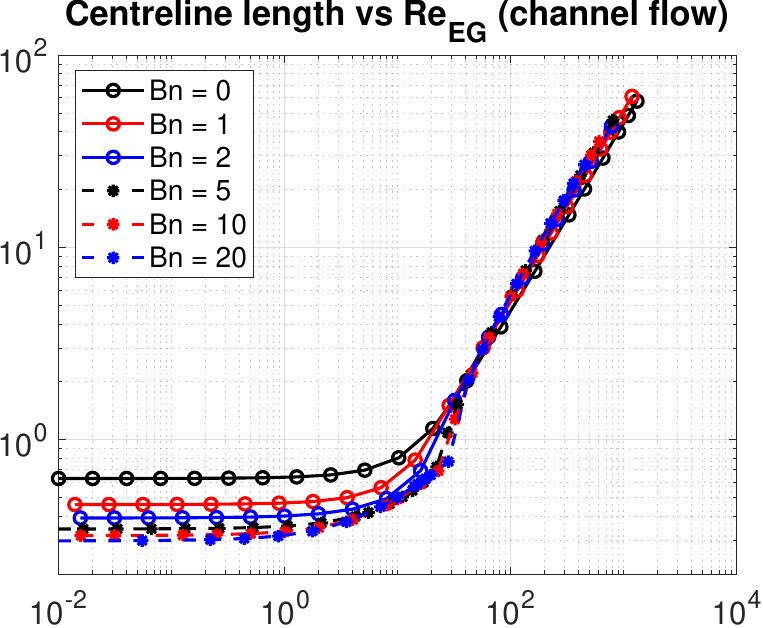}
        \caption{$\tilde{L}_c(\Rey_{EG})$}
        \label{sfig: ReEG Lc P}
    \end{subfigure}
    \begin{subfigure}[b]{0.328\textwidth}
        \centering
        \includegraphics[width=0.99\linewidth]{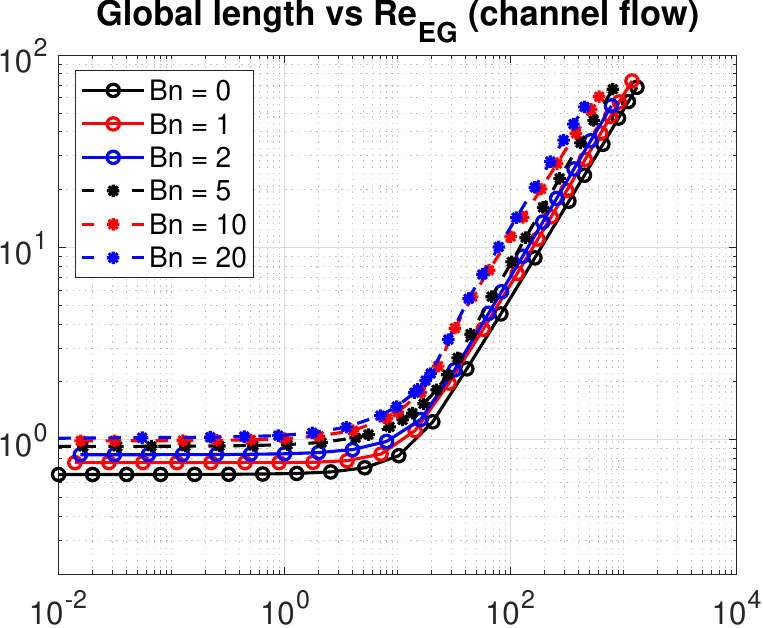}
        \caption{$\tilde{L}_g(\Rey_{EG})$}
        \label{sfig: ReEG Lg P}
    \end{subfigure}
    \begin{subfigure}[b]{0.328\textwidth}
        \centering
        \includegraphics[width=0.99\linewidth]{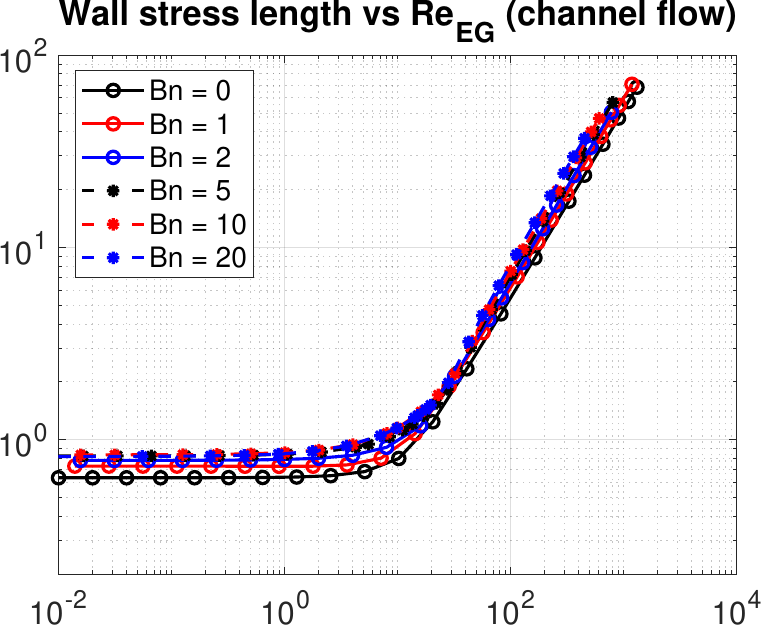}
        \caption{$\tilde{L}_{\tau w}(\Rey_{EG})$}
        \label{sfig: ReEG Lw P}
    \end{subfigure}

    \caption{Channel flow at various Bingham numbers: The three development lengths $L_c$ (left),
$L_c$ (middle) and $L_{\tau_w}$ (right), scaled by the channel height diameter $2H$, as a function
of $\Rey_{MR}$ (top row), $\Rey_{MRC}$ (2nd row), $\Rey_{MG}$ (3rd row), and $\Rey_{EG}$ (last
row).}
  \label{fig: L vs ReMR ReMRC ReMG ReEG planar}
\end{figure}

Let us now turn to Reynolds numbers that employ the wall shear stress as a characteristic stress.
Figures \ref{sfig: ReMR Lc}--\ref{sfig: ReMR Lw} plot the development lengths with respect to
$\Rey_{MR}$, for pipe flow. What is immediately striking is the near-perfect collapse of the $L_g$
curves (Fig.\ \ref{sfig: ReMR Lg}). On the other hand, $L_c$ and $L_{\tau w}$ seem somewhat more
dispersed than when plotted against $\Rey^{*R}$. In fact the $L_c$ and $L_{\tau w}$ curves are now
displaced towards the right as $\Bin$ is increased, which means that $\Rey_{MR}$ overestimates the
ratio of inertia to viscoplastic resistance that is determining these development lengths. If then
we apply a correction to the momentum flux in the numerator as in $\Rey_{MRC}$, then we observe a
slight improvement in Figs.\ \ref{sfig: ReMRC Lc}--\ref{sfig: ReMRC Lw}. The $L_g(\Rey_{MRC})$
correlations are almost perfectly matched, while the $L_c(\Rey_{MRC})$ and $L_{\tau w}(\Rey_{MRC})$
curves are slightly closer together compared to $L_c(\Rey_{MR})$ and $L_{\tau w}(\Rey_{MR})$ but
some noticeable dispersion persists. Interestingly, the Reynolds number of Ookawara et al.,
$\Rey_O$, (Figs.\ \ref{sfig: ReO Lc}--\ref{sfig: ReO Lw}) does a markedly better job collapsing the
$L_c$ and $L_{\tau w}$ curves, albeit at the expense of the $L_g$ curves. Similar observations hold
true also for the planar case (Figs.\ \ref{sfig: ReMR Lc P}--\ref{sfig: ReMRC Lw P}).

Finally, we consider the remaining Reynolds numbers, which account only for the increase in the
momentum or energy of the fluid rather than the total amount. As seen in Fig.\ \ref{sfig: ReMG Lc},
$\Rey_{MG}$ is the most effective Reynolds number for collapsing the $L_c$ curves. This is not
surprising given that the the flow development dynamics near the axis is governed to a large extent
by the fact that higher viscoplasticity (Bingham number) brings the fully developed plug velocity
closer to the inlet velocity thereby reducing the required development length. Consequently, near
the pipe centre the change in momentum between the inlet and the fully developed state is more
relevant to the determination of the development length than the fully developed momentum as a
whole. However, this is not the case close to the walls, which explains why $\Rey_{MG}$ performs
less effectively regarding $L_g$ (Fig.\ \ref{sfig: ReMG Lg}); nevertheless, the $L_g(\Rey_{MG})$
correlation is quite decent, surpassed only by $L_g(\Rey_{MR})$ and $L_g(\Rey_{MRC})$. On the other
hand, perhaps surprisingly, $\Rey_{MG}$ is also the most effective Reynolds number for collapsing
the $L_{\tau w}$ curves, despite $L_{\tau w}$ referring to conditions at the wall. In fact,
returning to Fig.\ \ref{fig: Ls ax}, we notice that while in the creeping flow regime $L_c$ and
$L_{\tau w}$ differ substantially, in the inertia-dominated regime they converge onto a single
curve, for all values of $\Bin$ considered, while $L_g$ is larger. So, once inertia begins to become
prominent, the development of the core flow determines also the development of the wall shear stress
($L_{\tau w} = L_c$); while the flow near the walls continues to develop for a larger distance ($L_g
> L_{\tau_w}$), this development seems to occur in a way that does not impact the wall shear stress.
As for the last Reynolds number, $\Rey_{EG}$, a comparison between Figs.\ \ref{sfig: ReEG
Lc}--\ref{sfig: ReEG Lw} and Figs.\ \ref{sfig: ReMG Lc}--\ref{sfig: ReMG Lw} shows that it exhibits
the same traits as $\Rey_{MG}$, but it is slightly less effective than the latter.

Turning to the channel flow case, we can make similar observations although the $L_{\tau
w}(\Rey_{MG})$ and especially the $L_g(\Rey_{MG})$ correlations are not as good as in the pipe flow
case. Now in Fig.\ \ref{fig: Ls pl} one can see that $L_c$ and $L_{\tau w}$ do not coincide even in
the region of prominent inertia, with $L_{\tau w}$ being consistently greater. Therefore the
equality of $L_c$ and $L_{\tau w}$ is particular to the pipe flow case and cannot be generalised to
all conduit geometries.

% \FloatBarrier
\afterpage{\clearpage}

% \begin{figure}[tb]
%     \centering
%
%     \begin{subfigure}[b]{8cm}
%         \centering
%         \includegraphics[width=0.99\linewidth]{Bn5planar_Re0_L_vs_y.pdf}
%         \caption{}
%         \label{sfig: L vs r Bn5 Re0 planar}
%     \end{subfigure}
%     \\[0.25cm]
%
%     \begin{subfigure}[b]{8cm}
%         \centering
%         \includegraphics[width=0.99\linewidth]{Bn5planar_Re61p44_L_vs_y.pdf}
%         \caption{}
%         \label{sfig: L vs r Bn5 Re61p44 planar}
%     \end{subfigure}
%     \\[0.25cm]
%
%     \begin{subfigure}[b]{8cm}
%         \centering
%         \includegraphics[width=0.99\linewidth]{Bn5planar_Re245p76_L_vs_y.pdf}
%         \caption{}
%         \label{sfig: L vs r Bn5 Re245p76 planar}
%     \end{subfigure}
%
%     \caption{Variation of the development length $L(r)$ across the channel width, for $\Bin = 5$ and
% \subref{sfig: L vs r Bn5 Re0 planar} $\Rey = 0$, \subref{sfig: L vs r Bn5 Re61p44 planar} $\Rey =
% 61$, and \subref{sfig: L vs r Bn5 Re245p76 planar} $\Rey = 246$. Results obtained with different
% values of the regularisation parameter $M$ are compared. Blue and red lines denote the development
% lengths based on 1\% and 0.5\% margins from the fully developed velocity, respectively. The black
% vertical dashed line marks the fully developed plug width. The black horizontal dotted line marks
% the tip of the plug.}
%   \label{fig: L vs r Bn5}
% \end{figure}

\begin{figure}[tb]
    \centering

    \begin{subfigure}[b]{8cm}
        \centering
        \includegraphics[width=0.99\linewidth]{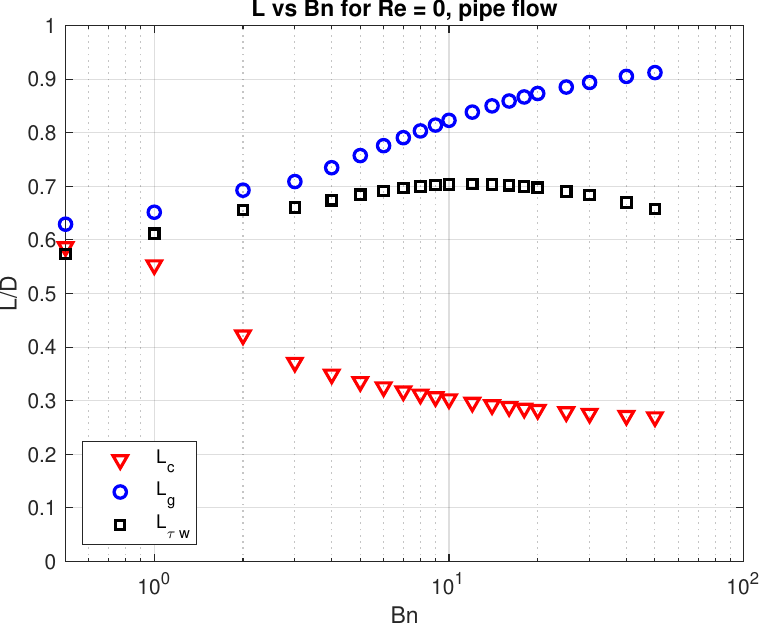}
        \caption{}
        \label{sfig: L vs Bn Re0 axisymmetric}
    \end{subfigure}
    \\[0.25cm]

    \begin{subfigure}[b]{8cm}
        \centering
        \includegraphics[width=0.99\linewidth]{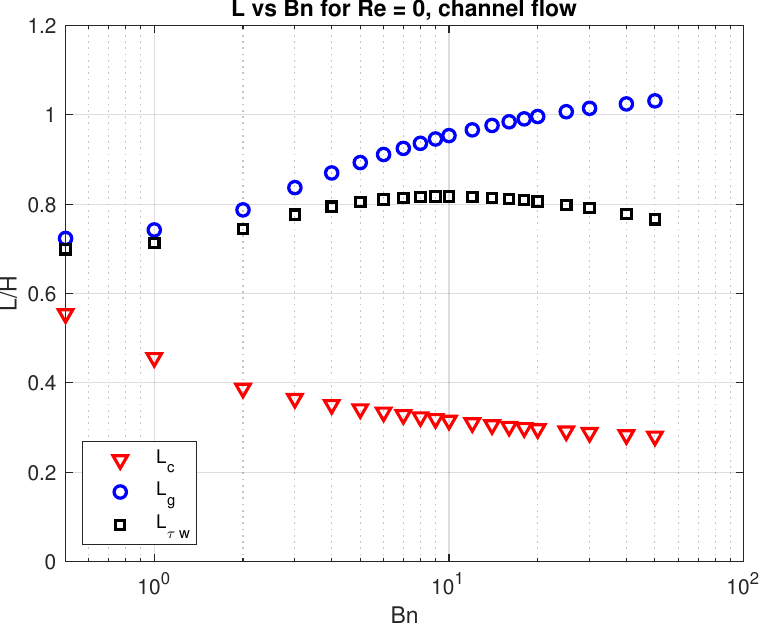}
        \caption{}
        \label{sfig: L vs Bn Re0 planar}
    \end{subfigure}
    \\[0.25cm]

    \begin{subfigure}[b]{8cm}
        \centering
        \includegraphics[width=0.99\linewidth]{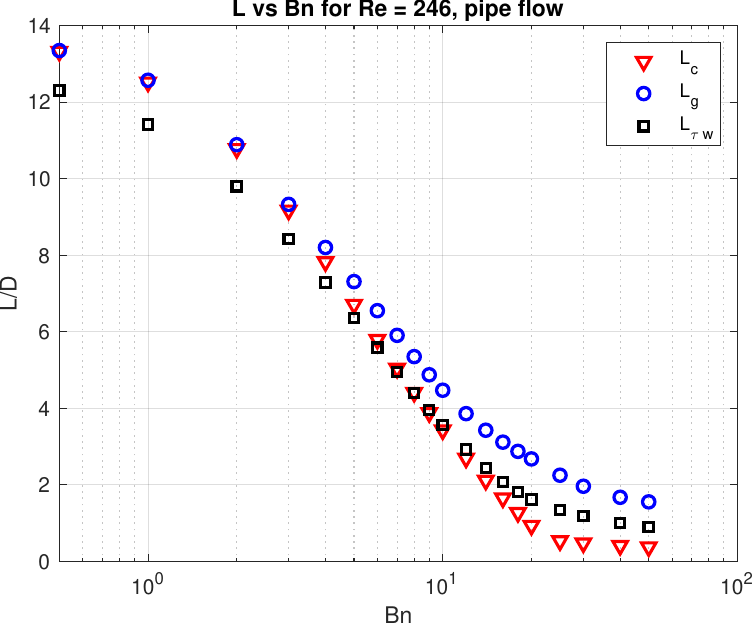}
        \caption{}
        \label{sfig: L vs Bn Re246 axisymmetric}
    \end{subfigure}

    \caption{Variation of the development lengths with the Bingham number: \subref{sfig: L vs Bn
Re0 axisymmetric} creeping ($\Rey = 0$) pipe flow with $M=10000$; \subref{sfig: L vs Bn Re0 planar}
creeping ($\Rey = 0$) channel flow with $M=10000$; \subref{sfig: L vs Bn Re246 axisymmetric}
pipe flow at $\Rey = 246$ with $M=2000$.}
  \label{fig: L vs Bn}
\end{figure}

The last set of results to be discussed here concern the variation of the development lengths with
the Bingham number, plotted in Fig.\ \ref{fig: L vs Bn}. The focus is on the inertialess case
(Figs.\ \ref{sfig: L vs Bn Re0 axisymmetric} and \ref{sfig: L vs Bn Re0 planar}) because when
inertia is important then the development length is determined by an appropriate effective Reynolds
number, and not by the Bingham number per se. In that case, increasing the Bingham number is
equivalent to lowering the effective Reynolds number, thus causing a reduction of the development
lengths as studied extensively in the previous discussion of the correlation of the lengths with the
several Reynolds number variants. An indicative plot is Fig.\ \ref{sfig: L vs Bn Re246 axisymmetric}
where the standard Reynolds number is held at $\Rey = 246$ and the reduction of all development
lengths as the Bingham number is increased (and the effective Reynolds numbers are therefore
decreased) is apparent. But in the low inertia regime the Reynolds number becomes irrelevant and the
development lengths are determined by the Bingham number alone. The behaviour of the development
length in this regime is summarised in Figs.\ \ref{sfig: L vs Bn Re0 axisymmetric} and \ref{sfig: L
vs Bn Re0 planar} for pipe and channel flow, respectively. In both cases, $L_c$ decreases and $L_g$
increases with $\Bin$, as already noted in \cite{Philippou_2016}. The new piece of data in Figs.\
\ref{sfig: L vs Bn Re0 axisymmetric} and \ref{sfig: L vs Bn Re0 planar} is the variation of $L_{\tau
w}$ which, interestingly, is non-monotonic. Up to a Bingham number of about 10, $L_{\tau w}$
increases with $\Bin$, similarly to $L_g$, but thereafter it decreases, similarly to $L_c$ (and in
fact at a faster rate). Due to this non-monotonous behaviour, $L_{\tau w}$ varies less in the range
considered ($0 \leq \Bin \leq 50$) compared to $L_g$ and $L_c$: $L_{\tau w} \in (0.57,0.72)$ in the
axisymmetric case, and $L_{\tau w} \in (0.70,0.83)$ in the planar case.

\section{Conclusions}
\label{sec: conclusions}

We have revisited the problem of laminar flow development of viscoplastic fluids by means of finite
element simulations, employing the Papanastasiou regularisation of the Bingham plastic equation. We
have focused on three definitions of the development length, based on the centreline velocity
($L_c$), the global velocity field ($L_g$), and the wall shear stress ($L_{\tau w}$). Usually, it
was found that $L_c < L_{\tau w} <  L_g$, although at very low Bingham numbers $L_c$ and $L_g$ are
identical or nearly identical and $L_{\tau w}$ is slightly smaller than them.

In an effort to collapse the development length versus Reynolds number curves onto a single master
curve, we tested several alternative definitions of the Reynolds number. The results showed that the
``momentum gain'' Reynolds number, $\Rey_{MG}$, (Eq.\ \eqref{eq: Re MG 0}) is most effective in
accomplishing this for $L_c$ and $L_{\tau w}$, both for pipe and channel flow, while for $L_g$ the
most effective choice is $\Rey_{MRC}$, the corrected ``Metzner-Reed'' Reynolds number (Eq.\
\eqref{eq: Re MRC 0}). The uncorrected ``Metzner-Reed'' Reynolds number (Eq.\ \eqref{eq: Re MR 0})
is almost as effective, and does not require knowledge of the fully-developed velocity profile (it
requires knowledge of the wall shear stress, but in an experimental setting this is easy to obtain
from the pressure gradient). An even simpler choice is the effective Reynolds number based on the
pipe diameter or channel half-width, $\Rey^{*R}$ or $\Rey^{*H}$, (Eq.\ \eqref{eq: Re effective R})
which is not as effective but still has a decent performance.

In the low-inertia regime the development length curves are impossible to collapse as they are
independent of the Reynolds number and depend only on the Bingham number. In this regime, $L_c$
decreases with the Bingham number, $L_g$ exhibits the opposite behaviour, while $L_{\tau w}$ is
non-monotonic, increasing with $\Bin$ up to a value of $\Bin \approx 10$ and decreasing thereafter.

Near the entrance, there are three unyielded regions in planar flow, one of which develops
downstream into the fully developed unyielded core plug. In axisymmetric flow, one of these regions
is absent due to physical constraints. As the Reynolds number is increased, the formation of the
core plug is pushed farther into the pipe or channel, and the flow development is more gradual,
necessitating higher values of the regularisation parameter to maintain accuracy.

It will be interesting to extend the current study to cases with shear-thinning (or thickening), by
employing the Herschel-Bulkley model, and to cases with wall slip, both of which are common in
actual viscoplastic flows. Such a study may be undertaken in the near future.

% \end{linenumbers}

\section*{Acknowledgements}

This study was funded by the Cohesion Policy Programme ``THALIA 2021-2027'', co-funded by the
European Union, through the Research and Innovation Foundation of Cyprus (Project ``CRaFTC'' --
SMALL SCALE INFRASTRUCTURES/1222/0181).

% ~~~ BIBLIOGRAPHY ~~~ > ~~~ > ~~~ > ~~~ > ~~~ > ~~~ > ~~~ > ~~~ > ~~~ > ~~~ >

\nocite{}
\bibliographystyle{ieeetr}
\bibliography{entrance}
%\printbibliography

\end{document}